\documentclass[12pt]{iopart}
\input{epsf.sty}
\input{epsfig.sty}

\begin{document}

\def\simg{\mathrel{\hbox{\rlap{\lower.55ex \hbox {$\sim$}}
                   \kern-.3em \raise.4ex \hbox{$>$}}}}
\def\siml{\mathrel{\hbox{\rlap{\lower.55ex \hbox {$\sim$}}
                   \kern-.3em \raise.4ex \hbox{$<$}}}}
\def\Mesz{M\'esz\'aros~}
\def\Pacz{Paczy\'nski~}
\def\beq{\begin{equation}}
\def\enq{\end{equation}}
\def\bea{\begin{eqnarray}}
\def\ena{\end{eqnarray}}
\def\bec{\begin{center}}
\def\enc{\end{center}}
\def\etal{{\it et~al.}}
\def\msun{M_\odot}
\def\eps{\epsilon}
\def\vareps{\varepsilon}
\def\cmsqi{{\rm cm}^{-2} {\rm s}^{-1}}
\def\barnu{\bar\nu}
\def\barnue{\bar \nu_e}
\def\barnum{\bar \nu_\mu}
\def\epm{\hbox{e}^\pm}
\def\nue{\nu_e}
\def\num{\nu_\mu}
\newcommand{\figuresize}{0.41\textwidth}
\newcommand{\boxsize}{0.89\textwidth}
\newcommand{\smallboxsize}{0.8\textwidth}
\def\bitm{\bibitem}
\def\etal{{\it et~al.\ }}
\def\s{{\rm ~s}}
\def\lbr{\linebreak}
\def\hl{\hline}
\def\tl{\tableline}
\def\etal{et al}
\def\bverb{\begin{verbatim}}
\def\everb{\end{verbatim}}

\title[Gamma-Ray Bursts]{Gamma-Ray Bursts}

\author{P. \Mesz}

\address{Dept. of Astronomy \& Astrophysics and Dept. of Physics,
Pennsylvania State University, 
525 Davey Laboratory, University Park, PA 16802, USA}

\bec {\footnotesize \it May 1, 2006 }\enc
\bec {\footnotesize \it To appear in Reports on Progress in Physics 
      \copyright 2006 IOP Publishing Ltd., http://www.iop.org} \enc

\begin{abstract}
Gamma-ray bursts are the most luminous explosions in the Universe, 
and their origin and mechanism are the focus of intense research and debate. 
More than three decades after their discovery, and after pioneering 
breakthroughs from space and ground experiments, their study is entering 
a new phase with the recently launched Swift satellite. The interplay 
between these observations and theoretical models of the prompt gamma ray 
burst and its afterglow is reviewed.

\end{abstract}


\tableofcontents

\maketitle

\section{Introduction}
\label{sec:intro}

Gamma-ray bursts (GRB) are brief events occurring at an average rate of a few
per day throughout the universe, which for a brief period of seconds 
completely flood with their radiation an otherwise almost dark gamma-ray sky.
While they are on, they outshine every other source of gamma-rays in the 
sky, including the Sun. In fact, they are the most concentrated and brightest 
electromagnetic explosions in the Universe. Until recently, they were
undetected at any wavelengths other than gamma-rays, which provided poor 
directional information and hence no direct clues about their site of origin.

This changed in early 1997 when the Beppo-SAX  satellite succeeded in detecting 
them in X-rays, which after a delay of some hours yielded sufficiently accurate 
positions for large ground-based telescope follow-up observations. These proved
that they were at cosmological distances, comparable to those of the most distant 
galaxies and quasars known in the Universe. Since even at these extreme distances
(up to Gigaparsecs, or $\sim 10^{28}$ cm) they outshine galaxies and quasars
by a very large factor, albeit briefly, their energy needs must be far greater. 
Their electromagnetic energy output during tens of seconds is comparable to 
that of the Sun over $\sim {\rm few} \times 10^{10}$ years, the approximate age 
of the universe, or to that of our entire Milky Way over a few years. 
The current interpretation of how this prodigious energy release is produced 
is that a correspondingly large amount of gravitational energy (roughly 
a solar rest mass) is released in a very short time (seconds or less) in a very 
small region (tens of kilometers or so) by a cataclysmic stellar event (the collapse 
of the core of a massive star, or the subsequent mergers of two remnant compact 
cores). Most of the energy would escape in the first few seconds as thermal
neutrinos, while another substantial fraction may be emitted as gravitational
waves. This sudden energy liberation would result in a very high temperature 
fireball expanding at highly relativistic speeds, which undergoes internal 
dissipation leading to gamma-rays, and it would later develop into a blast wave 
as it decelerates against the external medium, producing an afterglow which gets 
progressively weaker. The resulting electromagnetic energy emitted appears to
be of the order of a percent or less of the total energy output, but even this
photon output (in $\gamma$-rays) is comparable to the total kinetic energy 
output leading to optical photons by a supernova over weeks. 
The remarkable thing about this theoretical scenario is that it successfully 
predicts many of the observed properties of the bursts. This fireball shock 
scenario and the blast wave model of the ensuing afterglow have been extensively 
tested against observations, and have become the leading paradigms for the 
current understanding of GRB.

Historically, GRBs were first discovered in 1967 by the Vela satellites, 
although they were not publicly announced until 1973 \cite{kleb73_vela}. 
These spacecraft, carrying omnidirectional gamma-ray detectors, were flown 
by the U.S. Department of Defense to monitor for nuclear explosions which 
might violate the Nuclear Test Ban Treaty.  When these mysterious gamma-ray 
flashes were first detected, and it was determined that they did not come 
from the Earth's direction, the first suspicion (quickly abandoned) was 
that they might be the product of an advanced extraterrestrial civilization.
Soon, however, it was realized that this was a new and extremely puzzling
cosmic phenomenon \cite{kleb73_vela}. For the next 25 years, only these brief 
gamma-ray flashes were observed, which could be only roughly localized, and 
which vanished too soon, leaving no traces, or so it seemed. Gamma-rays are 
notoriously hard to focus, so no sharp gamma-ray ``images" exist to this day: 
they are just diffuse pin-pricks of gamma-ray light.  This mysterious phenomenon 
led to a huge interest and to numerous conferences and publications on the 
subject, as well as to a proliferation of theories. In one famous  review 
article at the 1975 Texas Symposium on Relativistic Astrophysics, no fewer 
than 100 different possible theoretical models of GRB were listed 
\cite{ruderman75}, most of which could not be ruled out by the observations 
then available.

The first significant steps in understanding GRBs started with the 1991 
launch of the Compton Gamma-Ray Observatory, whose results were summarized 
in \cite{fm95}. The all-sky survey from the BATSE instrument showed 
that bursts were isotropically distributed, strongly suggesting a cosmological, 
or possibly an extended galactic halo distribution, with essentially zero 
dipole and quadrupole components \cite{fen93}. At cosmological distances the 
observed GRB fluxes imply enormous energies, which, from the fast time variability,
must arise in a small volume in a very short time. This must lead to the
formation of an $e^\pm -\gamma$ fireball \cite{pac86,goo86,shepi90}, which will
expand relativistically. The main difficulty with this scenario was that a 
smoothly  expanding fireball would  convert most of its energy into kinetic 
energy of accelerated baryons (rather than into photon energy), and would 
produce a quasi-thermal spectrum, while the typical timescales would not 
explain events much longer than milliseconds. This difficulty was addressed
by the ``fireball shock scenario" \cite{rm92,mr93a}, based on the realization 
that shocks are likely to arise, e.g. when the fireball ejecta runs into the 
external medium, after the fireball has become optically thin, thus 
reconverting the expansion kinetic energy into non-thermal radiation. 
The complicated light curves can also be understood, e.g. in terms of internal 
shocks \cite{rm94,sapi97,kps97} in the outflow itself, before it runs into the 
external medium, caused by velocity variations in the outflow from the source.

The next major developments came after 1997, when the Italian-Dutch satellite
Beppo-SAX succeeded in detecting fading X-ray images which, after a 
delay of 4-6 hours for processing, led to positions \cite{cos97}, allowing 
follow-ups at optical and other wavelengths, e.g. \cite{jvp97}. This paved
the way for the measurement of redshift distances, the identification of 
candidate host galaxies, and the confirmation that they were indeed at 
cosmological distances \cite{metz97,djo98_0703,kul98a_hiz,kul99b}. The detection 
of other GRB afterglows followed in rapid succession, sometimes extending to radio 
\cite{fra97,fra00} and over timescales of many months \cite{jvp00}, and in a 
number of cases resulted in the identification of candidate host galaxies, 
e.g. \cite{sah97,bloo98_0508,ode98_1214}, etc.  The study of afterglows has 
provided strong confirmation for the generic fireball shock model of GRB.  
This model led to a correct prediction \cite{mr97a}, in advance of
the observations, of the quantitative nature of afterglows at wavelengths 
longer than $\gamma$-rays, which were in substantial agreement with the data
\cite{vie97a,tav97,wax97a,rei97,wrm97}.

A consolidation of the progress made by Beppo-SAX was made possible through
the HETE-2 satellite \cite{hete2}, after the demise of CGRO and Beppo-SAX.
It provided a continuing stream of comparable quality afterglow positions, 
after typical delays of hours, and contributed to the characterization of a 
new class of sources called X-ray flashes or XRF \cite{heise01} resembling 
softer GRBs, which had been earlier identified with Beppo-SAX. It also
localized GRB 030329, which resulted in the first unambiguous association 
with a supernova (SN2003dh) \cite{sta03,hjorth03}.

The third wave of significant advances in the field is due to the Swift multi-wavelength 
afterglow satellite, launched in November 2004, which achieved the long-awaited 
goal of accurately localized afterglows starting a minute or so after the burst 
trigger, at gamma-ray, X-ray and optical wavelenghts \cite{swift_site,gehr05_md05}.
This revealed the hitherto unexplored afterglow behavior between minutes
to hours, enabling a study of the transition from the prompt emission and
the subsequent long term afterglow, and revealing a rich range of X-ray early 
behavior. It also achieved the long-awaited discovery of the afterglows of
``short" gamma-ray bursts (whose hard gamma-ray emission is briefer than 2 s).
It furthermore broke through the symbolic redshift $z=6$ barrier, beyond which
very few objects of any kind have been measured. 

On the theoretical side, a major issue raised by the large redshifts, e.g. 
\cite{kul98a_hiz,kul99}, is that the measured $\gamma$-ray fluences (the flux
integrated over time) imply a total energy of order a solar rest mass, 
$M_\odot c^2 \sim 2\times 10^{54}$ ergs, if it is emitted isotropically.
By contrast, the total radiant (and the associated kinetic expansion energy) 
of supernovae (SN), which is detected over timescales of weeks to months, is 
of the order of a thousandth of a solar rest mass, $10^{51}$ ergs. A GRB  
emission which is concentrated in a jet, rather than isotropically, alleviates 
significantly the energy requirements. There is now extensive observational 
evidence for such collimated emission from GRBs, provided by breaks in the 
optical/IR light curves of their afterglows \cite{kul99,fru99,cas99}. The 
inferred total amount of radiant and kinetic energy involved in the explosion 
is in this case comparable to that of supernovae (except that in GRBs the 
energy is mostly emitted in a jet in $\gamma$-rays over tens of seconds, 
whereas in supernovae it is emitted isotropically in the optical over weeks).
While the luminous (electromagnetic) energy output of a GRB is thus ``only" 
of the same order of magnitude as that of supernovae, the explosion is much 
more concentrated, both in time and in direction, so its specific brightness 
for an observer aligned with the jet is many orders of magnitude more 
intense, and appears at much higher characteristic photon energies.
Including the collimation correction, the GRB electromagnetic emission is 
energetically quite compatible with an origin in, say, either compact mergers 
of neutron star-neutron star (NS-NS) or black hole-neutron star 
(BH-NS) binaries \cite{pac86,eic89,napapi92,mr92}, or with a core 
collapse (hypernova or collapsar) model of a massive stellar progenitor 
\cite{woo93,pac98,pop99,macwoo99,woo05_md05}, which would  
be related to but much rarer than core-collapse supernovae. While in both
cases the outcome could be, at least temporarily, a massive fast-rotating 
ultra-high magnetic field neutron star (a magnetar), the high mass involved 
is expected to lead inevitably to the formation of a central black hole,
fed through a brief accretion episode from the surrounding disrupted 
core stellar matter, which provides the energy source for the ejection of 
relativistic matter responsible for the radiation.

A stellar origin of  GRB leads to two predictions which are similar to
those for core-collapse supernovae, albeit in so far unobserved aspects. 
In both GRB (whether from compact mergers or from collapsar scenarios)  and 
in core-collapse SN, the central material is compressed to nuclear densities 
and heated to virial temperatures characteristically in the multi-MeV range,
leading to 5-30 MeV thermal neutrinos. And in both cases, the merging or
collapsing core material acquires a time-varying quadrupole mass moment
(which may be smaller in SN not related to GRB), which leads to gravitational 
wave emission. In both GRB and supernovae, the total neutrino emission is of 
the order of a fraction of a solar rest mass, $\sim {\rm several} \times 
10^{53}$ ergs. The gravitational wave emission is of the same order for 
compact mergers, probably less than that for collapsars, and much less 
in normal core collapse SNe.  Experiments currently planned or under 
construction will be able to probe these new channels.

\section{Observational Progress up to 2005 }
\label{sec:obs}

Before reviewing, in the next section, the latest observational advances 
achieved with Swift, the observational progress made up to that time is 
briefly surveyed. More extensive discussion and references on observations 
previous to 2004 are, e.g. in \cite{fm95,jvp00,zhames04,piran05}.

\begin{figure}[ht]
\begin{center}
\centerline{\epsfxsize=5.in \epsfbox{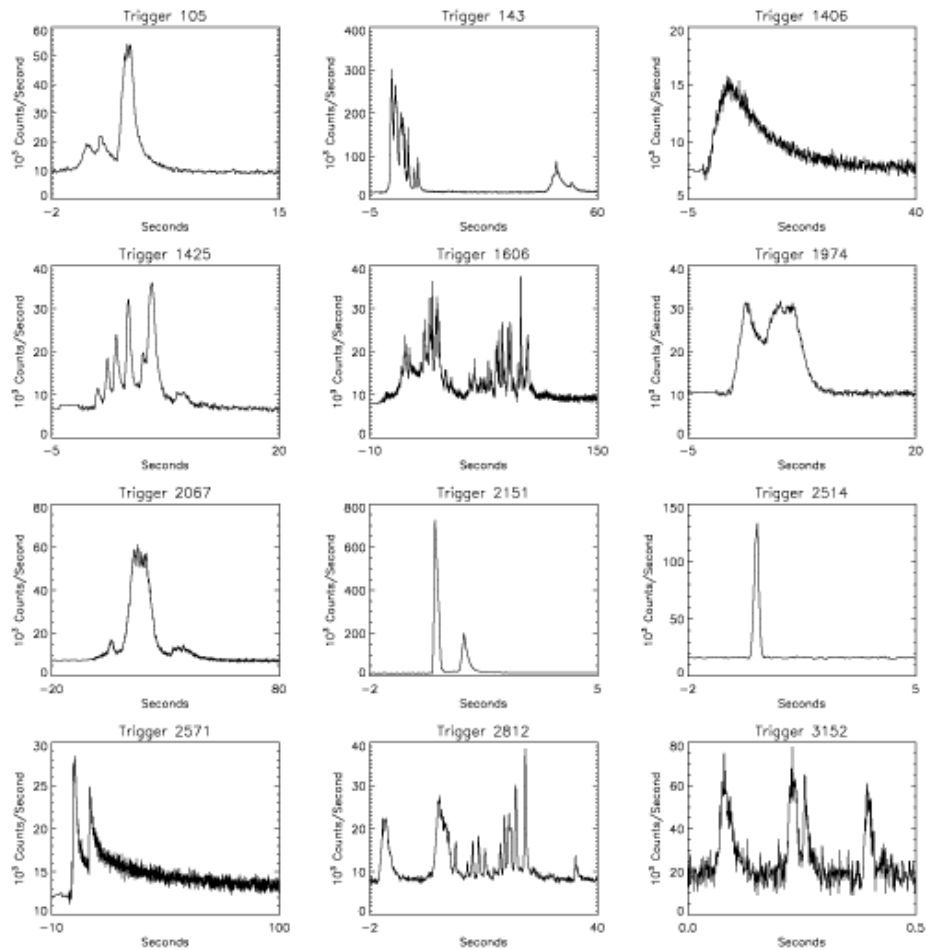}}
\end{center}
\caption{Diversity of gamma-ray light curves observed by BATSE \cite{fm95}
\label{fig:grblc} }
\end{figure}

The $\gamma$-ray phenomenology of GRB was extensively studied and
characterized by the BATSE instrument on the Compton GRO satellite 
\cite{fm95}. The $\gamma$-ray spectra are non-thermal, 
typically fitted in the MeV range by broken power-laws whose energy per 
decade peak is in the range 50-500 KeV \cite{band93}, sometimes extending 
to GeV energies \cite{hur94}. GRB appeared to leave no detectable
traces at other wavelengths, except in some cases briefly in X-rays.
The gamma-ray durations range from $10^{-3}$ s to about $10^3$ s, with a 
roughly bimodal distribution of long bursts of $t_b \simg 2$ s and short 
bursts of $t_b \siml 2$s \cite{kou93}, and substructure sometimes down 
to milliseconds. The gamma-ray light curves range from smooth, fast-rise 
and quasi-exponential decay (FREDs), through curves with several peaks, 
to highly variable curves with  many peaks \cite{fm95,kou98} 
(Figure \ref{fig:grblc}). The pulse distribution is complex \cite{pen96}, 
and the time histories can provide clues for the geometry of the emitting 
regions \cite{fen96,fen98}.

\begin{figure}[ht]
\begin{center}
\centerline{\epsfxsize=4.in \epsfbox{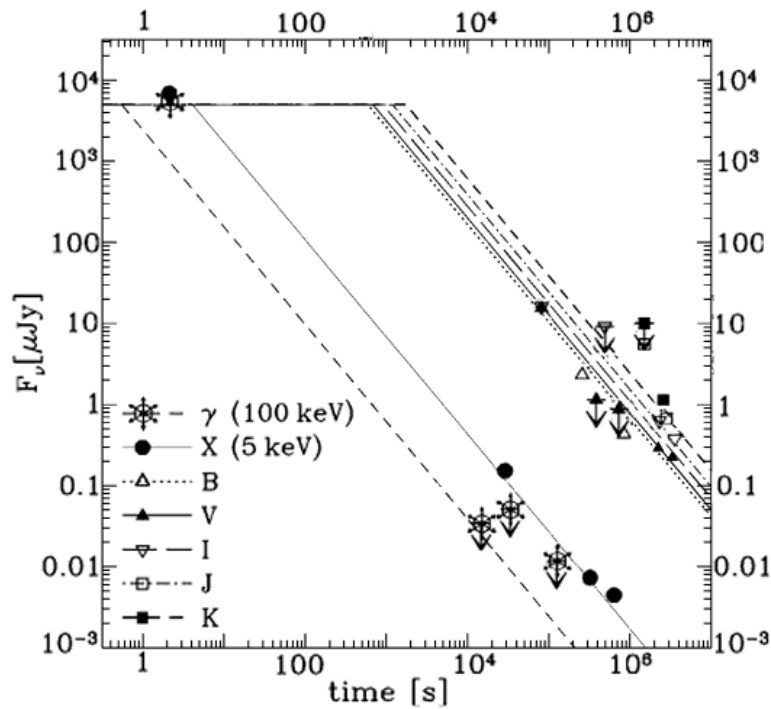}}
\end{center}
\caption{The afterglow light curves of GRB 970228 \cite{wrm97}.
\label{fig:970228lc} }
\end{figure}

GRB were conclusively shown to be at cosmological distances following 
Beppo-SAX localizations of their X-ray afterglows in 1997 \cite{cos97},
followed by optical host galaxy identification and redshift determinations 
\cite{jvp97}. The afterglows decay as a power law in time in a manner 
predicted by pre-existing models \cite{mr97a}, softening in time from 
X-rays to optical to radio (e.g. Figure \ref{fig:970228lc}).
The energy needed to explain the total (mainly gamma-ray) energy fluence
can be as large as $10^{54}(\Omega_\gamma /4\pi)$ ergs, where $\Delta 
\Omega_\gamma$ is the solid angle into which the gamma-rays are beamed. 
This is for the highest fluences seen in some of the most distant bursts,
although for many bursts the energy budget problem is not as extreme.  If 
the emission is assumed to be emitted isotropic (isotropic equivalent luminosity 
or energy) this energy ranges up to a solar rest mass in gamma-rays. This would 
strain a stellar origin interpretation, since from basic principles and experience 
it is known that, even for the most efficient radiation conversion schemes,
a dominant fraction of the energy should escape in the form of thermal neutrinos 
and gravitational waves. The energy requirements, however, are much less severe
in the case when the emission is collimated (\S \ref{sec:jetobs}).

GRB afterglow light curves such as those shown in Figure \ref{fig:970228lc}
have been followed up starting several hours after the trigger  in X-rays by 
Beppo-SAX and subsequently HETE-2, and in the optical/IR from ground-based 
telescopes (or in some case with HST), and have been explained in terms of 
forward shock emission (for discussions of the pre-Swift data interpretation
see, e.g. \cite{jvp00,mesz02,zhames04,piran05}). Afterglows have been 
followed up at radio wavelengths in some cases over months, and the analysis 
and interpretation  of the radio spectra and light curves
\cite{vanhorst05_radio,berger04_radio,fra03} provides important clues for the 
calorimetry and the multi-waveband model fits discussed in \S \ref{sec:standardag}.

\subsection{Progenitor candidates}
\label{sec:progobs}

There is now strong observational evidence \cite{jvp00} that GRB result from a 
small fraction ($\sim 10^{-6}$) of stars which undergo a catastrophic energy 
release event toward the end of their evolution. For the class of long GRB
the candidates are massive stars whose core collapses \cite{woo93,pac98,fryerwh99}
to a black hole, either directly or after a brief accretion episode, possibly
in the course of merging with a companion. This scenario is referred to as the
collapsar or hypernova scenario, which received strong support through the 
secure spectroscopic detection in some cases of an associated supernova event 
(e.g. \cite{gal98_sn,sta03,hjorth03}; see also \S\S \ref{sec:snobs}, \ref{sec:sn}). 
For short bursts the most widely speculated candidates are mergers of neutron 
star (NS) binaries or neutron star-black hole (BH) binaries 
\cite{pac86,goo86,eic89,mr92,mr97b,leeklu99,rossrr03,rossrrd03,leerrp04}, which 
lose orbital angular momentum by gravitational wave radiation and undergo a merger. 
This second progenitor scenario has only now begun to be tested thanks to the Swift 
detection of short burst afterglows (see \S \ref{sec:swift}, \S \ref{sec:short}).
Both of these progenitor types are expected to have as an end result the formation
of a few solar mass black hole, surrounded by a temporary debris torus whose
accretion can provide a sudden release of gravitational energy, 
sufficient to power a burst. An important point is that the overall
energetics from these various progenitors need not differ by more than about
one order of magnitude \cite{mrw99}. The duration of the burst in this model is
related to the fall-back time of matter to form an accretion torus around the
BH \cite{fryerwh99,pop99} or the accretion time of the torus \cite{narayan01}.
Other related scenarios include the formation from a stellar collapse of a 
fast-rotating ultra-high magnetic field neutron star 
\cite{usov94,tho94,spru99,whee00,ruderman00}.

\subsection{Light curve breaks and jets}
\label{sec:jetobs}

An important subsequent development was the observation, in many of the 
well-sampled afterglows, of a break or steepening of the X-ray and optical 
light curves \cite{kul99,fra01},  which can be interpreted as being due 
to the outflow being jet-like and the break occurs when the edge of the jet 
becomes visible, as the jet slows down \cite{rho97,rho99,sapiha99,mr99}.
The typical (long burst )inferred jet opening angles are $\theta_j \sim 5-20$ 
degrees, which reduces the total energy requirements from $10^{53}-10^{54}$ 
erg to $\sim 10^{51}$ ergs \cite{fra01,pankum01}, with a dispersion of a 
factor $\sim 10$. The details are dependent on assumptions about the jet 
geometry, and whether one addresses the gamma-ray energy \cite{fra01} or 
the jet kinetic energy \cite{pankum01}, with a somewhat larger dispersion in
the latter. Variable optical linear polarization is expected at the time of 
a jet break \cite{sari99,ghislaz99,rossi04}, which can provide additional 
constraints, also on the jet structure.
Light curve break determinations continued through the Beppo-SAX and HETE-2 
afterglow observation periods, mainly in the optical. Break observations
in the Swift era are discussed in \S \ref{sec:jet}.

\subsection{Optical flashes}
\label{sec:flashobs}

Prompt optical flashes (starting within tens of seconds after the gamma-ray 
trigger) have been reported from ground-based small robotic telescopes in 
a few  bursts \cite{ak99,and_flash_rev04}. These arise much earlier, are 
initially brighter and decay more steeply than more ubiquitous long-term, 
slow decaying optical afterglows generally detected since 1997. They are also
rare: between 1999 and 2004 there were only a handful of prompt optical flashes 
detected with robotic ground telescopes \cite{and_flash_rev04}. Since the Swift 
launch, more than twenty prompt UVOT (or ground-based robotic) optical flashes 
have been seen, mostly at times starting several hundreds of seconds after
the trigger. None have been as bright as the first one detected in GRB 990123, 
except for the notable most distant GRB 050904, whose optical brightness is 
comparable to that of GRB 990123 \cite{boer05}. This is discussed further in 
\S \ref{sec:promptopt}.

\subsection{Association with supernovae}
\label{sec:snobs}

At least some long GRBs are associated with supernova explosions. The first
reported example was the GRB980425/SN1998bw association\cite{gal98_sn,kul98b_sn,jvp99}.
SN 1998bw was a peculiar, energetic Type Ib/c supernova. Using it as a
template, other possible associations have been claimed through identifying 
a so-called red supernova bump on the optical afterglow light curves of 
GRB 980326\cite{bloom99}, GRB 970228\cite{rei99,galama00}, GRB 000911\cite{laz01}, 
GRB 991208\cite{castro01}, GRB 990712\cite{sahu00}, GRB 011121\cite{bloom02a}, 
GRB 020405\cite{price03} and GRB 031203 \cite{cobb04,malesani04}. The first
unambiguous supernova signature (SN 2003dh) was detected in the $z=0.168$ 
GRB 030329, firmly establishing the GRB-SN associations\cite{sta03,hjorth03}.
Another GRB/SN event, GRB060218/SN2006aj, is discussed in \S \ref{sec:swift},
and the supernova connection is discussed further in \S \ref{sec:sn}.

\subsection{X-ray flashes}
\label{sec:xrfobs}

X-ray flashes (XRFs) are a class of bursts whose light curves and
spectra resemble typical GRB, except for the fact that their spectra
are much softer, their spectral peaks $E_{pk}$ being typically tens of 
keV or less \cite{heise01,kippen02}. XRFs were first identified with the 
Beppo-SAX satellite,  and have been studied in greater detail and numbers 
with the HETE-2 satellite \cite{barraud04,lamb05a}.  Their fluxes and 
isotropic equivalent luminosities tend to be smaller than for GRB, which
makes afterglow searches more difficult. Nonetheless, several afterglows 
have been detected, and redshifts have been measured in some of them 
(\cite{sod04}, also \S\S \ref{sec:swift}, \ref{sec:newag}).

\subsection{Empirical correlations and distance estimators}
\label{sec:correl}

The collimation-corrected total burst energy clustering around $10^{51}$ 
ergs, while making a stellar origin quite plausible, is unfortunately
not sufficiently well defined to use as standard luminosity candles, 
whose apparent brightness would provide a distance determination.
There are other possible distance measures, based on empirical 
correlations between burst observables. One of these is an apparent
gamma-ray light curve variability correlation with the isotropic
equivalent luminosity \cite{fen00,rei00}. 
Another is the time lag (between higher energy versus lower energy)
gamma-rays and the isotropic luminosity \cite{norris00, band04,norris05}.
Attempts at modeling the spectral lags have relied on observer-angle 
dependences of the Doppler boost \cite{nakamura00,salmonson01b}.
In these correlations the isotropic equivalent luminosity was used, in 
the absence of jet signatures, and they must be considered tentative for now. 
However, if confirmed, they could be invaluable for independently estimating 
GRB redshifts.  A third one is a correlation between the spectral properties 
and the isotropic luminosity \cite{amati01,borryd01,bagoly02,lloydrr02}. These 
measures can be effectively calibrated only for light curves or spectra obtained 
with the same instrument and when redshifts are available. For the above methods 
the calibration set consists typically of a dozen bursts, which is 
insufficient for being considered reliable. The data set is now rapidly
increasing with new Swift redshift determinations, but the Swift spectra are
less well constrained at energies above 150 keV than with Beppo-SAX.

In recent years, attention has been drawn to a correlation between the 
photon spectral peak energy $E_{pk}$ and the apparent isotropic energy 
$E_{iso}$, or the isotropic luminosity $L_{iso}$ \cite{amati02}
of the form $E_{pk} \propto E_{iso}^{1/2}$. For GRB, this correlation
has been calibrated on a sample consisting of around ten bursts. While 
strongly suggestive, the sample is relatively small and the dispersion 
is large, so its usefulness as a distance measure is precarious.
Another suggestive result is that for some X-ray flashes whose redshift has 
been measured the correlation continues to hold \cite{lamb05a}, even 
though the peak energy is $\sim 1.5$ and the $E_{iso}$ is $\sim 3$ orders 
of magnitude lower than for GRB. 

An interesting development of the $E_{pk}-E_{iso}$ correlation is the 
proposal of a relationship between the collimation-corrected total 
energy ($E_j$ and the photon spectral peak. Modeling the jet break 
under the assumption that the jet expands into a uniform density external 
medium, this has the form $E_{pk}\propto E_j^{0.7}$ for a jet assumed to
propagate in a homogeneous medium \cite{ghirl04a,dailx04,ghirl04,ghirl05}, 
and the relation appears to be tighter than the previously discussed 
$E_{pk} \propto E_{iso}^{1/2}$ (Amati) relation. 
This requires for calibration both an observed redshift and a light curve 
break. If it holds up for a larger calibration sample than the current 
($\sim$ 18 bursts so far), it could be of promise as a cosmological tool 
\cite{lamb05b,ghirl05b}. There are however problems to resolve before this 
becomes competitive with SNIa as a cosmological tool \cite{fribloom05}.
The main ones are the presence of outliers \cite{nakar05}, the lack of a 
large low redshift sample for calibration (since using a high redshift 
sample requires assumptions about the cosmology which it is supposed to 
test), an evaluation of observational biases and 
selection effects, and the dependence of the results on model assumptions
about the external medium and jet properties. The latter may be circumvented
by relying only on observables, e.g. $E_{peak}$, fluence (or peak flux) and
the break time $t_{br}$ \cite{liangzh05,nava06}. The dispersion, however,
remains so far about 1.5-2 times larger than for SNIa.

\section{Recent Results from Swift and Follow-up Observations}
\label{sec:swift}

Compared to previous missions, the Swift results represent a significant
advance on two main accounts. First, the sensitivity of the Burst Alert 
Detector (BAT, in the range 20-150 keV) is somewhat higher than that of the 
corresponding instruments in CGRO-BATSE, BeppoSAX and HETE-2 \cite{band06}.
Second, Swift can slew in less then 100 seconds in the direction determined
by the BAT instrument, positioning its much higher angular resolution X-ray
(XRT) and UV-Optical (UVOT) detectors on the burst \cite{gehr05_md05}

As of December 2005, at an average rate of 2 bursts detected per week,
over 100 bursts had been detected by BAT in about a year (compared to 300/year
by BATSE; note the BAT field of view is 2 sr, versus $4\pi$ in BATSE).
Of these $\sim 100$ bursts, 90\% were detected and followed with the XRT 
within $350$ s from the trigger, and about half within 100 s \cite{burr05_md05}, 
while $\sim 30\%$ were detected also with the UVOT \cite{roming05_md05}.  
Of the total, over 23 resulted in redshift determinations. Included in this 
total sum are nine short GRB, of which five had detected X-ray afterglows, 
three had optical, and one had a radio afterglow, with five host galaxy 
detections and redshift determinations. These were the first ever
short GRB afterglows detected and followed.

The new observations bring the total redshift determinations to over
50 since 1997, when BeppoSAX enabled the first one. The median redshift
of the Swift bursts is $z\simg 2$, which is a factor $\sim 2$ higher
than the median of the BeppoSAX and HETE-2 redshifts \cite{berger05a}. 
This is a statistically significant difference between the Beppo-SAX
and Swift redshift samples \cite{jakob_zdist,bagoly06}. This may
in part be ascribed to the better sensitivity of the BAT detector, but 
mostly to the prompt and accurate positions from XRT and UVOT, making 
possible ground-based detection at a stage when the afterglow is much 
brighter, by a larger number of robotic and other telescopes.
As of January 2006 the highest Swift-enabled redshift is that of GRB 050904, 
obtained with the Subaru telescope, $z=6.29$ \cite{kawai05}, and the second 
highest is GRB 050814 at $z=5.3$, whereas the previous Beppo-SAX era record 
was $z=4.5$.  The relative paucity of UVOT detections versus XRT
detections may be ascribed in part to this higher median redshift, and
in part to the higher dust extinction at the implied  shorter rest-frame
wavelenghts for a given observed frequency \cite{roming05_md05}, although
additional effects may be at work too.

\begin{figure}[ht]
\begin{center}
\centerline{\epsfxsize=4.in \epsfbox{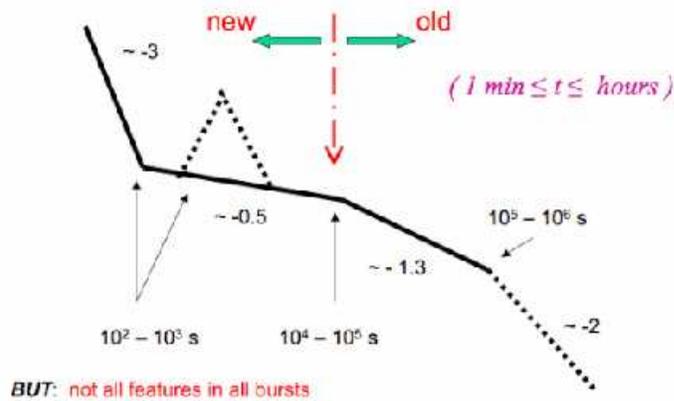}}
\caption{Schematic features seen in early X-ray afterglows detected with the 
Swift XRT instrument (e.g. \cite{zhang06_ag,nousek06} (see text).
\label{fig:xrt-lc} }
\end{center}
\end{figure}

In some of the bursts,  both of the ``long" ($t_\gamma \simg 2$ s) and 
``short" ($t_\gamma \leq 2$ s) categories as defined by BATSE, the Swift 
BAT results show faint soft gamma-ray extensions or tails, which extend the 
duration by a substantial factor beyond what BATSE would have detected 
\cite{gehr05_md05}.  A rich trove of information on the burst and afterglow 
physics has come from detailed XRT light curves, starting on average 100 seconds 
after the trigger, together with the corresponding BAT light curves and spectra.
This suggests a canonical X-ray afterglow picture \cite{nousek06,zhang06_ag,chinc05}
which includes one or more of the following:\hfill
\lbr
1) an initial steep decay $F_X \propto t^{-\alpha_1}$ with a temporal index
$3 \siml \alpha_1 \siml 5$, and an energy spectrum $F_\nu \propto \nu^{-\beta_1}$
with energy spectral index $1 \siml \beta_1 \siml 2$ (or photon number index
$2 \siml \alpha+1 \siml 3$), extending up to a time $300 \s \siml t_1
\siml  500 \s$;\hfill
\lbr
2) a flatter decay portion $F_X \propto t^{-\alpha_2}$ with temporal index
$0.2 \siml \alpha_2 \siml 0.8$ and energy index $0.7 \siml \beta_2 \siml 1.2$, 
at times $10^3 \s \siml t_2 \siml 10^4 \s$;\hfill
\lbr
3) a ``normal" decay $F_X \propto t^{-\alpha_3}$ with $1.1 \siml \alpha_3
\siml 1.7$ and $0.7 \siml \beta_3 \siml 1.2$ (generally unchanged
the previous stage), up to a time $t_3 \sim 10^5 \s$, or in some cases
longer;\hfill
\lbr
4) In some cases, a steeper decay $F_X \propto
t^{-\alpha_4}$ with $2\siml \alpha_4 \siml 3$, after $t_4\sim 10^5 \s$;\hfill
\lbr
5) In about half the afterglows, one or more X-ray flares are observed,
sometimes starting as early as 100 s after trigger, and sometimes as
late as $10^5 \s$. The energy in these flares ranges from a percent
up to a value comparable to the prompt emission (in GRB 050502b).
The rise and decay times of these flares is unusually steep, depending
on the reference time $t_0$, behaving as $(t-t_0)^{\pm \alpha_{fl}}$
with $3 \siml \alpha_{fl} \siml 6$, and energy indices which can be
also steeper (e.g. $\beta_{fl}\siml 1.5$) than during the smooth decay 
portions. The flux level after
the flare usually decays to the value extrapolated from the value before
the flare rise (see Figure \ref{fig:xrt-lc}). The above characteristics 
are derived mainly from long bursts, but interestingly, at least one of 
the short bursts shows similar features. However, the evidence for late
time activity is more sketchy in short bursts, so that the analogy must
be considered with caution.
  
%

An exciting result from Swift was the detection of the long burst
GRB 050904, which broke through the astrophysically and psychologically
significant redshift barrier of $z\sim 6$, which is thought to mark the
approximate end of the ``dark ages", when re-ionization of the intergalactic 
medium by the first generation of light sources approaches completion.
This burst was very bright, both in its prompt $\gamma$-ray emission 
($E_{\gamma , iso} \sim 10^{54}$ erg) and in its X-ray afterglow. Prompt
ground-based optical/IR upper limits and a J-band detection suggested a
photometric redshift $z>6$ \cite{haislip05}, and spectroscopic confirmation
soon followed with the 8.2 m Subaru telescope, giving $z=6.29$ \cite{kawai05}.
There are several striking features to this burst.
One is the enormous X-ray brightness, exceeding for a full day the X-ray
brightness of the most distant X-ray quasar know to-date, SDSS J0130+0524,
by up to a factor $10^5$ in the first minutes \cite{wat05}. The implications
as a tool for probing the intergalactic medium are thought-provoking. 
Another feature is the extremely variable X-ray light curve, showing many large
amplitude flares extending up to at least a day. A third exciting feature
is the report of a brief, very bright IR flash \cite{boer05}, comparable in
brightness to the famous $m_V\sim 9$ optical flash in GRB 990123.

A third major advance from Swift was the discovery and localization of short
GRB afterglows. As of December 2005 nine short bursts had been localized by
Swift, while in the same period HETE-2 discovered two, and one was identified
with the IPN network. In five of the short bursts, GRB 050509b, 050709, 050724 
and 051221a  an X-ray afterglow was measured and followed up, with GRB 050709, 
050724 and 051221a showing also an optical afterglow, and 050724 also a radio 
afterglow, while 040924 had an optical afterglow but not an X-ray one 
\cite{fox05_md05}.  These are the first afterglows detected for short bursts. 
Also, for the first time, host galaxies were identified for these short bursts, 
which in a number of cases are early type (ellipticals) and in other cases are 
irregular galaxies (e.g. \cite{proch06}. The redshifts of four of them are in 
the range $z\sim 0.15-0.5$, while another one was determined to be $z=0.8$ (and 
less securely, it has been argued that this latter may instead be $z \simeq 1.8$
\cite{berger05_md05}). The median $z$ is $\siml 1/3-1/2$ that of the long bursts.
There is no evidence for significant star formation in these host environments 
(except for GRB 050709, \cite{fox05_0709,proch06}, which is compatible with what 
one expects for an old population, such as neutron star mergers or neutron 
star-black hole mergers, the most often discussed progenitor candidates 
(although it would also be compatible with other progenitors involving old 
compact stars). While the evidence for a neutron star or black merger is
suggestive, the evidence is not unequivocal. E.g. the observations suggest 
a typical time delay of at least several Gyr between the start formation 
epoch and the explosion of short GRBs \cite{nakar06_sho,zhengrr06}. There 
are a number of unresolved issues related to this (see \S \ref{sec:short}).
The first short burst afterglow followed up by Swift, GRB 050509b, was a 
rather brief ($\sim 30$ ms), moderate luminosity ($L_{iso}\sim 10^{50}$ 
erg s$^{-1}$ but low fluence ($E_{iso}\sim 2\times 10^{48}$ erg) burst with 
a simple power-law X-ray afterglow  which could only be followed for 
several hundreds of seconds \cite{gehr05_0509}.
The third one, GRB 050724, was brighter, $E_{iso}\sim 3\times 10^{50}$ erg,
and could be followed in X-rays with Swift for at least $10^5$ s 
\cite{barth05_0724}, and with Chandra up to $2\times 10^6$ s \cite{gehr06_kitp}.
The remarkable thing about this burst's X-ray afterglow is that it shows some
similarities to the typical X-ray light curves described above for long GRB -- 
except for the lack of a slow-decay phase, and for the short prompt emission 
which places it the the category of short bursts, as well as the elliptical 
host galaxy candidate. It also has also bumps in the X-ray light curve at 100 s 
and at $3\times 10^4$ s, which resemble some of the long burst X-ray flares
and whose origin is unclear. The first bump or flare has the same fluence as 
the prompt emission, while the late one has $\sim 10\%$ of that. The 
interpretation of these pose interesting challenges, as discussed below
and in S \ref{sec:short},

A fourth exciting result from Swift was the detection of a long burst, 
GRB 060218, which was seen also with the XRT and UVOT instruments 
\cite{camp06_060218}, and which is associated with SN 2006aj 
\cite{modjaz06_060218,mirab06_2006aj,sol06_2006aj,cob06_2006aj}.
The redshift is $z=0.033$, and the contribution to the optical light curve
as well as the spectrum are similar to those of the Ic type SN1998bw.
The result this is so exciting is that it is the first time that a GRB/SN
has been observed minutes after the $\gamma$-ray trigger at X-ray and UV/optical
wavelengths. This is discussed further in \S \ref{sec:sn}.

\section{Theoretical Framework}
\label{sec:theory}

\subsection{The Relativistic Fireball Model}
\label{sec:fireball}

As discussed in the introduction, the ultimate energy source of GRB is
convincingly associated with a catastrophic energy release in stellar mass
objects. For long bursts, this is almost certainly associated with the late 
stages of the evolution of a massive star, namely the collapse of its core 
\cite{woo93,pac98}, which at least in some cases is associated with a detectable
supernova. For short bursts, it has been long assumed \cite{pac86,eic89} that 
they were associated with compact binary mergers (NS-NS or NS-BH), a view 
which is gaining observational support \cite{berger05a,fox05_0709}, although 
the issue cannot be considered settled yet. In both cases, the central 
compact object is likely to be a black hole of several solar masses 
(although it might, temporarily, be a fast rotating high-mass neutron
star, which eventually must collapse to black hole). In any case, the
gravitational energy liberated in the collapse or merger involves of order
a few solar masses, which is converted into free energy  on timescales of
milliseconds inside a volume of the order of tens of kilometers cubed. This 
prompt energy is then augmented by a comparable amount of energy release 
in a similar or slightly larger volume over a longer timescale of seconds to
hundreds of seconds, by the continued infall or accretion of gas onto the
central object, either from the central parts of the massive progenitor star
or from the debris of the disrupted compact stars which was temporarily held 
up by its rotation. 

The principal result of the sudden release of this large gravitational 
energy (of order a solar rest mass) in this compact volume is the conversion 
of a fraction of that energy into neutrinos, initially in thermal equilibrium, 
and gravitational waves (which are not in thermal equilibrium), while a
significantly smaller fraction ($10^{-2}-10^{-3}$) goes into a high temperature 
fireball ($kT \simg$ MeV) consisting of $e^\pm$, $\gamma$-rays and baryons.
The fireball is transparent to the gravitational waves, and beyond several
interaction lengths, also to the neutrinos. This leads to the prompt emission 
(on timescales of a few seconds) of roughly comparable energy amounts 
(several $\times 10^{53}$ ergs) of thermal $\nu_e {\bar \nu_e}$ with typical 
energies 10-30 MeV, and of gravitational waves mainly near $10^2-10^3$ Hz. 
These two, by far most dominant, energy forms are so far undetected, and are
discussed further in \S \ref{sec:vhe}. A smaller fraction of the liberated energy,
or order $10^{50}-10^{52}$ ergs remains trapped in a  $e^\pm$, $\gamma$-ray
and baryon fireball, which can also contain a comparable (or in some scenarios
a larger) amount of magnetic field energy. This amount of energy is observed,
mainly as non-thermal gamma-rays. While smaller than the predicted thermal 
neutrino and gravitational wave fluence, this is nonetheless a formidable 
electromagnetic energy output, much more intense than any other explosive
event in the universe. While the total energy is comparable to the electromagnetic 
and kinetic energy of supernovae, the difference is that in supernovae the energy
is doled out over months, mainly at optical wavelengths, while in GRB  most of the
electromagnetic energy is spilled out in a matter of seconds, and mainly at 
$\gamma$-ray wavelengths.

The leading model for the electromagnetic radiation observed from GRBs 
is based on the relativistic fireball created in the core collapse or merger.
The photon luminosity inferred from the energies and timescales discussed and
from the observations is many orders of magnitude larger than the Eddington 
luminosity $L_E=4\pi GM m_p c/\sigma_T = 1.25\times 10^{38}(M/M_\odot)$ 
erg s$^{-1}$, above which radiation pressure exceeds self-gravity, so the 
fireball will expand. The first (thermal) fireball models were assumed to
reach relativistic expansion velocities \cite{cavrees78,pac86,goo86,shepi90}.
However, the ultimate expansion velocity depends on the baryon load of the 
fireball \cite{pac90}. If the fireball energy involved all the baryons in the core
(solar masses) the expansion would be sub-relativistic. However, near the black 
hole the density is reduced due to accretion and centrifugal forces, it is 
likely that baryons are much depleted in the region where the fireball forms, 
with a tendency to form high entropy (high energy/mass ratio) radiation bubbles.
Dynamically dominant magnetic fields would also tend to involve fewer baryons.
A phenomenological argument shows that the expansion must, indeed, be highly 
relativistic. This is based on the fact that most of the GRB spectral energy
is observed above 0.5 MeV, so that the mean free path for the $\gamma\gamma 
\to e^\pm$ process in an isotropic plasma (an assumption appropriate for a
sub-relativistically expanding fireball) would be very short. This leads to a
contradiction, since many bursts show spectra extending above 1 GeV, so the 
flow must be able to avoid degrading these via photon-photon interactions to 
energies below the threshold $m_e c^2=$ 0.511 MeV\cite{hb94}. To avoid this, 
it seems inescapable that the flow must be expanding with a very high Lorentz factor
$\Gamma$, since then the relative angle at which the photons collide is less
than $\Gamma^{-1}$ and the threshold for the pair production is then diminished.
This condition is
\beq
\Gamma \simg 10^2 [(\eps_\gamma / {\rm 10 GeV}) (\eps_t /{\rm MeV}) ]^{1/2}~,
\label{eq:gg}
\enq
in order for photons $\eps_\gamma $ to escape annihilation against target photons 
of energy $\eps_t \sim 1$ Mev \cite{m95,hb94}.  I.e., a relativistically expanding 
fireball is expected, with bulk Lorentz factors $\Gamma = 100\Gamma_2 \simg 1$.

\subsection{Reference frames and timescales in relativistic flows}
\label{sec:frame}

The emitting gas is moving relativistically with velocity $\beta=v/c=
(1-1/\Gamma^2)^{1/2}$ relative to a laboratory frame $K_\ast$ which may
be taken to be the origin of the explosion or stellar frame (which, aside
from a cosmological Doppler or redshift factor is  the same as the Earth
frame $K$ of an observer). The lengths, times,  thermodynamic and radiation 
quantities of the gas are best evaluated in the gas rest frame (the 
comoving frame) $K'$, and are obtained in the stellar/lab frame through
Lorentz transformations. Thus, a proper length $dr'$ in the comoving
frame has a stellar/lab frame length $dr_\ast=dr'/\Gamma$ (if both 
ends $r_{\ast 1}$ and $r_{\ast 2}$ of the length $dr_\ast=
r_{\ast 1}- r_{\ast 2}$ in $K_\ast$ are measured at the same time so 
$dt_\ast=0$ in $K_\ast$; i.e. the usual Fitz-Gerald contraction).
Similarly, a proper time interval $dt'$ in the comoving frame has a 
duration $dt_\ast= dt'\Gamma$ in $K_\ast$ (provided the times $t_{\ast 1}$ 
and $t_{\ast 2}$ of $dt_\ast=t_{\ast 1}-t_{\ast 2}$ in $K_\ast$ are measured
at same positions $x_{\ast 1}$ and $x_{\ast 2}$ in $K_\ast$ so $dx_\ast =0$;
the usual time dilation effect). The time needed in the stellar/lab frame 
$K_\ast$ for the gas to move from $x_{\ast 1}$ to $x_{\ast 2}$ is the usual 
$dt_\ast=t_{\ast 1} -t_{\ast 2}= dr_\ast /\beta c \approx dr_\ast /c$.
%
\begin{figure}[ht]
\begin{center}
\centerline{\epsfxsize=5.in \epsfbox{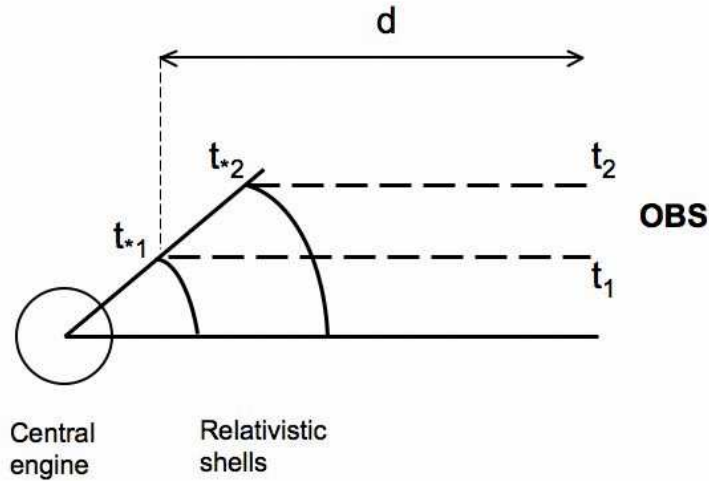}}
\end{center}
\caption{Illustration of the emission from spherical relativistic shells in the
source frame and the relativistic time delay leading to the relation between source 
frame time and observer time.
\label{fig:reltime} }
\end{figure}

When it comes to observations at Earth of radiation emitted from the 
relativistically moving gas, even though the Earth frame $K$ is essentially 
the same as the $K_\ast$ stellar/lab frame, in addition to the above 
Lorentz transformations one has to consider also the classical light 
travel time delay (Doppler) effect, e.g. \cite{rl79}. In the observer
frame $K$ one can use the same spatial coordinates $r\equiv r_\ast$ and
$dr\equiv dr_\ast$ as in $K_\ast$, but the actual time of arrival of 
signals as measured by an observer, which is for brevity denoted just $t$,
differs from $t_\ast$ by the above Doppler effect, $t\neq t_\ast$. 
Since this observed time  $t$ is the actual observable, it is customary
to describe GRB problems in terms of $t$ (remembering it is $\neq t_\ast$)
and $r$ (which is $=r_\ast$). Considering a gas which expands radially
in a direction at an angle $\cos \theta=\mu$ respect to the observer line 
of sight, if a first photon is emitted when the gas is at the radius 
$r_{\ast 1}=r_1$ (which is at a distance $d$ from the  observer) 
at $t_{\ast 1}$, this photon arrives at the observer at an observer time
$t_1= t_{\ast 1} - d/c$. A second photon emitted from a radius $r_{\ast 2}
=r_2$ at time $t_{\ast 2}$ will arrive at an observer time $t_2=t_{\ast 2} +
(d/c - \beta\mu dt_\ast)$, where $dt_\ast=t_{\ast 1} -t_{\ast 2}$. This is
illustrated in the source frame in  Figure \ref{fig:reltime}.
For an observer close to the line of sight the observed time difference 
between the arrival of the two photons is
\beq
dt = dt_\ast (1-\beta \mu) \simeq dt_\ast (1/2 \Gamma^2 +\theta^2/2) 
   \simeq dr/(2\Gamma^2 c)(1+\Gamma^2\theta^2) \simeq dr/(2\Gamma^2 c)~,
\label{eq:tobs}
\enq
where we assumed $\Gamma \gg 1$ for an approaching gas ($\mu=\cos\theta>0$) 
along a radial direction well inside the light cone $\theta \ll \Gamma^{-1}$. 

\begin{figure}[ht]
\begin{center}
\centerline{\epsfxsize=5.in \epsfbox{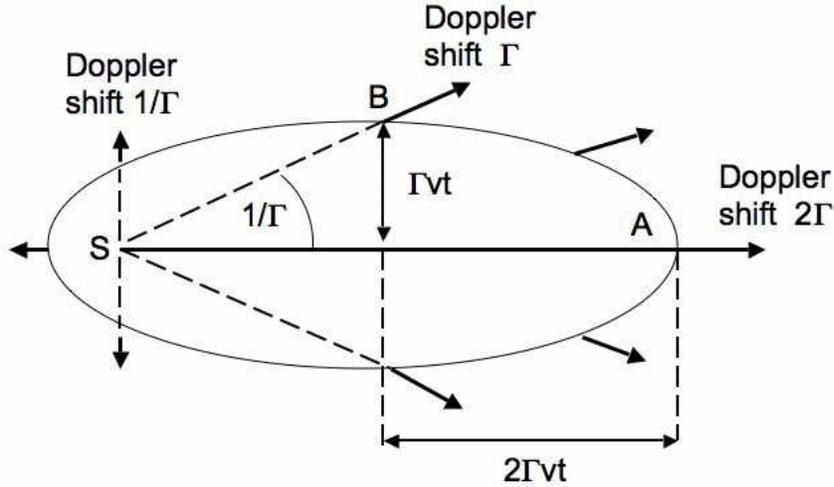}}
\label{fig:observertime}
\end{center}
\caption{For a distant observer (located to the right) viewing a shell 
which expands spherically from $S$ with $\Gamma=\sqrt{1-(v/c)^2} \gg 1$,
the locus of the points from which radiation reaches it at a later time $t$
appears as a spheroid (equal arrival time surface). Most of the radiation
arrives from the forward (right) hemisphere, which is strongly Doppler boosted
inside the light cone $1/\Gamma$ (after \cite{rees66}). The apparent transverse
radius of the ellipsoid is $r_{\perp} \simeq \Gamma ct$, and its semi-major
axis is $r_\parallel \simeq 2\Gamma ct$, where $t$ is observer time.}
\label{fig:relellipse}
\end{figure}

While both $dt$ and $dt_\ast$ are in the same reference frame, $K=K_\ast$,
the difference is that $dt_\ast$ is the time difference between emission of
the two photons, and $dt$ is the time difference between reception of the 
two photons. The general relation between the observer frame and comoving 
frame quantities is given through the Doppler factor $\cal D$,
\beq
{\cal D}= [\Gamma(1-\beta\mu)]^{-1} 
\label{eq:doppler}
\enq
where ${\cal D}\sim 2\Gamma$ for an approaching gas with $\Gamma \gg 1$,
$\mu \to 1$ and $\theta<\Gamma^{-1}$ (blueshift), or ${\cal D}\sim 1/2\Gamma$
for a receding gas with $\mu \to -1$ (redshift). Thus, the relation between 
the comoving frame $dt'$ and the observer frame time $dt$ is 
\beq
dt= {\cal D}^{-1} dt' =\Gamma(1-\beta\mu) dt' \simeq dt'/2\Gamma~,
\label{eq:tcom}
\enq
where an approaching gas is assumed with $\theta<\Gamma^{-1}$ (while
$dt_\ast= dt' \Gamma$). This is illustrated in terms of observer-frame 
quantities in Figure \ref{fig:relellipse}.  Note that in all the above
transformations we have neglected cosmological effects, which would 
result in multiplying any reception or observer-frame times by an additional 
factor $(1+z)$ for signals emanating from a source at redshift $z$.
%
%

The relation between the source frame and observer frame frequency, solid angle, 
specific intensity, temperature, volume, specific emissivity, specific absorption
coefficient and radial width are obtained in terms of the Doppler factor using 
relativistic invariants \cite{rl79}, 
$\nu={\cal D}\nu'$, $d\Omega={\cal D}^{-2}d\Omega'$, $I_\nu(\nu)={\cal D}^3 
I'_{\nu'}(\nu')$, $T(\nu)={\cal D}T'(\nu')$, $dV={\cal D}dV'$, $j_{\nu}(\nu)
={\cal D}^2 j'_{\nu'}(\nu')$, $\mu_\nu(\nu)={\cal D}^{-1}\mu'_{\nu'}(\nu')$,
$\delta r ={\cal D}^{-1}\delta r'$.  Here $\mu_\nu=n \sigma_\nu$ (in cm$^{-1}$, 
where $n$ is density and $\sigma_\nu$ is absorption cross section), and both 
$\nu\mu_\nu$ and the optical depth are invariants. 

\subsection{Relativistic dynamics}
\label{sec:dynamics}

From general considerations, an outflow arising from an initial energy $E_0$ 
imparted to a mass $M_0 \ll E_0/c^2$ within a radius $r_0$ will lead to an
expansion, which due to the initial high optical depth can be considered 
adiabatic. The pressure will be dominated by radiation, so the adiabatic
index is $\gamma_a=4/3$, and the comoving temperature $T'$ (or comoving
random Lorentz factor per particle $\gamma'$) evolves with comoving volume 
$V'$as $T\propto V'^{1-\gamma_a}$. With a comoving volume $V'\propto r^{3}$
(equation [\ref{eq:comvol}])  this means $T'\propto \gamma' \propto r^{-1}$.
By conservation of energy, this decrease in internal energy per particle
is balanced by an increase in its expansion-related energy, i.e. the bulk 
kinetic energy per particle or bulk Lorentz factor $\Gamma$, so that
$\gamma\Gamma=$ constant, so that $\Gamma\propto r$. This expansion occurs 
at the expense of the comoving frame internal energy. Since the bulk Lorentz
factor per particle cannot increase beyond the initial value of random
internal energy per particle, $\gamma_0=\eta=E_0/M_0c^2$, the bulk Lorentz
factor only grows until it reaches $\Gamma_{max} \sim \eta = E_o/M_o c^2$,
which is achieved at a radius $r/r_0 \sim \eta$. Beyond this radius the flow
begins to coast, with $\Gamma \sim \eta \sim $ constant 
\cite{pac86,goo86,shepi90,pac91}, 
\beq
\Gamma(r) \sim \cases{ (r/r_0) ~,& for $r/r_0 \siml \eta,~~r \siml r_s$;\cr
                     \eta   ~,& for $r/r_0 \simg \eta,~~ r \simg r_s$ \cr}~, 
\label{eq:Gamma}
\enq
which defines a saturation radius $r_s\sim r_0\eta$ beyond which the Lorentz
factor has saturated. Another way to understand the initial acceleration 
\cite{mlr93} is that initially, at $r=r_0$, the gas particles have a bulk Lorentz 
factor $\Gamma\sim 1$ and have an isotropical distribution of velocities with 
random Lorentz factors $\gamma \sim \eta=E_0/M_0 c^2$. As the particles expand 
outward, when they have reached a radius $r$ their velocity vectors will confined 
inside an angle $(r/r_0)^{-1}$ of the radial direction. A transformation to a 
comoving frame moving radially with a bulk Lorentz factor $\Gamma(r) \sim r/r_0$ is
needed for the velocity distribution to be isotropic in the comoving frame,
as it should be. 

As particles initially contained inside $r_0$ move outwards 
with velocity vectors which are increasingly radial, they form a radially 
expanding shell whose lab-frame width is initially $\delta r\sim \delta r_o \sim r_o$. 
The radial velocity spread is $(c-v)/c = 1-\beta \sim \Gamma^{-2}$, which
causes a gradual spread of the lab-frame radial width $\delta r/r \sim 
\delta v/v \sim \Gamma^{-2}$. For typical values of $r_0\sim 10^6-10^7$ cm
and $\eta \siml 10^3$ this is negligible until well beyond the saturation
radius, and a noticeable departure from the approximately constant width 
$\delta r \sim r_o$ starts to become appreciable only for radii in excess of
a spreading radius $r_\delta$ where $\delta r \sim r \delta v/c \sim 
r_\delta \eta^{-2} \simg r_o$. The laboratory frame width is therefore 
\cite{mlr93}
\beq
\delta r \sim ~\hbox{max}~ [\delta r_o~,~ r/\Gamma^2 ]~
\sim~ \cases{ \delta r_o~,& ~~~~for $r \siml r_\delta $;\cr
              r/\Gamma^2~,& ~~~~for $r \simg r_\delta$ ~.\cr}  
\label{eq:labwidth}
\enq
where the spreading radius $r_\delta \sim \delta r_0 \eta^2$ is a factor 
$\eta$ larger than the saturation radius $r_s\sim \delta r_0 \eta$.

The comoving radial width $\delta r'$ is related to the lab width $\delta r$ 
through $\delta r' =\delta r {\cal D}\sim \delta r \Gamma$. Hence
\beq
\delta r' \sim \cases{
\delta r_0 \Gamma \sim r~& ~~~~for $r\siml r_s$; \cr
\delta r_0\eta ~& ~~~~for $r_{s} \siml r \siml r_\delta$; \cr
r/\eta~& ~~~~for $r \simg r_\delta$~,\cr}    
\label{eq:comwidth}
\enq

Since the dimensions transverse to the motion are invariant, the comoving 
volume is $V'\propto r^2\delta r'$, which behaves as \cite{mlr93}
\beq
V'\sim 4\pi r^2\delta r' \sim \cases{
4\pi r^3 ~&~~~~for $r \siml r_s$;\cr
4\pi \eta \delta r_0~ r^2 ~&~~~~for $r_s \siml r \siml r_\delta$;\cr
4\pi\eta^{-1} r^3 ~&~~~~for $r \simg r_\delta$;\cr}
\label{eq:comvol}
\enq
and the comoving particle density $n'\propto V'^{-1}$.
For an adiabatic expansion (valid for the high initial optical depths) 
and a relativistic gas polytropic index $4/3$ (valid as long as the pressure 
is dominated by radiation),  one  has
\beq
{E' \over E_0}={T'\over T_0}= \left({V_0 \over V'}\right)^{1/3} 
\simeq \cases{ 
(\delta r_o/r) &~for $r <r_s$;\cr
(\delta r_o/r_s)(r_s/r)^{2/3}=\eta^{-1/3}(\delta r_o/r)^{2/3}
                                              &~for $r_s <r< r_\delta$;\cr
(\delta r_o/r_s)(r_s/r_\delta)^{2/3}(r_\delta /r)=\eta^{1/3}(\delta r_o/r)
                                              &~for $r > r_\delta$;\cr}
\label{eq:adiab}
\enq
where $E',~T',~\rho',~V'$ are comoving internal energy, temperature, density
and volume.

The above equations refer to the release of an energy $E_0$ and mass $M_0$ 
corresponding to $\eta= E_0/M_0 c^2$, originating inside a region of
dimension $\delta r_0\sim r_0$. This mass and energy leaves that original
region in a lab-frame (or observer frame) light-crossing time 
$\delta t_0 \sim \delta r_0/c$. For typical core collapse or compact merger 
stellar scenarios, the energy release volume is of the order of several 
Schwarzschild radii of the ensuing black hole (BH), few times $2GM_{BH}/c^2$
with $M_{BH}\simg 2 M_\odot$, say $r_0\sim 10^7 m_1$ cm, where 
$m_1=M_{BH}/10 M_\odot$, with a light-crossing timescale $t_0\sim r_0/c 
\sim 3\times 10^{-4}m_1$ s. This is of the order of the dynamical (Kepler) 
timescale near the last stable circular orbit in a temporary accretion disk 
feeding the newly formed black hole (or near the light cylinder of an initial 
fast-rotating magnetar or neutron star, before it collapses to a black hole).

\subsection{Optical Depth and Photosphere}
\label{sec:phot}

As shown by \cite{pac86,goo86,shepi90} a relativistically expanding fireball 
initially has $e^\pm$ pairs in equilibrium which dominate the scattering optical 
depth, but the pairs fall out of equilibrium and recombine below a comoving 
temperature $T'\sim 17$ keV, and thereafter only a residual freeze-out density 
of pairs remains, which for $\eta$ not too large (in practice $\eta \siml 10^5$ 
i.e. baryon loads not too small) is much less than the density of ``baryonic" 
electrons associated with the protons, $n_e=n_p$.  For a typical burst 
conditions the initial black-body temperature $T'_0$ at $r_0\sim 10^7$ cm 
is a few MeV, and pair recombination occurs at radii below the  saturation
radius. The scattering optical depth of a minishell (and of the whole outflow)
is still large at this radius, due to the baryonic electrons. For a minishell
of initial width $\delta r_0$ the optical depth varies as \cite{mlr93}
\beq
\tau_T=(M_0 \sigma_T /4 \pi r^2 m_p)=\tau_0 (\delta r_0/r)^2
\label{eq:taumini}
\enq
where $\tau_0=(E_0\sigma_T/4\pi r_0^2 m_p c^2\eta)$, for $\delta r_o \siml
r/\Gamma^2$ or $r\siml r_\delta$.  Assuming a burst 
with total energy $E_0=10^{52}E_{52}$ and total duration $t_{grb}$ divided 
into minishells of duration $\delta t_0=3\times 10^{-4}$ s, each of energy
$10^{47.5}E_{47.5}$ erg, these becomes optically thin at
\beq
r_t = \tau_0^{1/2}r_0 =3 \times 10^{11} m_1^{-1} (E_{47.5}/\eta_2)^{1/2}~{\rm cm},
\label{eq:miniphot}
\enq
where henceforth the notation $Q_x$ (where $x$ is a number) indicates the 
quantity $Q$ in units of $10^x$ times its c.g.s. units.

For bursts of some substantial duration, e.g. an outflow duration 
$t_{grb}=10$ s as above, at any instant different parts of the flow have 
different densities, and are above or below the saturation radius, 
so a continuous outflow picture is more appropriate \cite{pac90}. 
In this ``wind" regime one defines the dimensionless entropy as 
$\eta=L/{\dot M} c^2$, and instead of integral conservation laws one 
uses the relativistic fluid differential equations. The Lorentz factor 
again grows linearly and saturates at the same radius $r_s=r_0\eta$ 
(equation [\ref{eq:Gamma}]), where $r_0=\delta r_0$ is the minimum 
variability radius, and the adiabatic behavior of equation (\ref{eq:adiab}) 
is the same for the temperature, etc. The particle density follows from
the mass conservation equation,
\beq
n'_p=({\dot M}/4\pi r^2 m_p c \Gamma) =(L/4\pi r^2 m_p c^3 \eta \Gamma) 
\label{eq:np}
\enq
and the optical depth is $\tau_T(r)=\int_r^\infty n_e'\sigma_T 
[(1-\beta)/(1+\beta)]^{1/2} dr'$ $\sim n'_e \sigma_T (r/2\Gamma)$
which yields for the global photosphere \cite{pac90}
\beq
r_{ph}\simeq ({\dot M} \sigma_T /8\pi m_p c \Gamma^2) 
       =(L\sigma_T /8\pi m_p c^3 \eta \Gamma^2)
       \simeq 6\times 10^{11} L_{51}\eta_2^{-3}~{\rm cm}.
\label{eq:rphotgen}
\enq
The comoving temperature of the flow behaves as $T'\propto r^{-1},~r^{-2/3}$
and the observer-frame temperature $T=T'\Gamma$ is $T\sim T_0,~
T\sim T_0(r/r_s)^{-2/3}$ for $r<r_s,~r>r_s$ (equations [\ref{eq:adiab}]).

The radiation escaping from a radius $r$ (e.g. the photospheric radius $r_{ph}$)
which is released at the same stellar frame time $t_\ast$ would arrive at the 
observer only from within angles inside the light cone, $\theta <\Gamma^{-1}$. 
The observer-frame time delay between light coming from central line of sight and 
the edges of the light cone (the so-called angular time \cite{goo86}) is
\beq
t_{ang} \simeq t \simeq (r/c)(1-\beta)\sim r/2c\Gamma^2 ~.
\label{eq:tang}
\enq
This is because the `edge' of the light cone corresponds to an angle of $1/\Gamma$ 
from the line of sight, and therefore $ct_{ang} \sim r(1-\cos\theta) \sim r(1-\beta) 
\sim r/2\Gamma^2$, since at $\theta = 1/\Gamma$, $\cos\theta \approx \beta$. This time 
is the same as the observer-frame time of equation (\ref{eq:tobs}).  Note that 
if the outflow duration $t_{grb}$ is shorter than $t_{ang}$ of equation (\ref{eq:tang}), 
the latter is the observed duration of the photospheric radiation (due to the 
angular time delay). Otherwise, for $t_{grb}>t_{ang}$, the photospheric radiation is 
expected for a lab-frame duration $t_{grb}$.

\subsection{Thermal vs. Dissipative Fireballs and Shocks} 
\label{sec:shocks}

The spectrum of the photosphere would be expected to be a black-body
\cite{pac86,goo86,shepi90}, at most modified by comptonization at the
higher energy part of the spectrum.  However, the observed $\gamma$-ray 
spectrum observed is generally a broken power law, i.e., highly non-thermal. 
In addition, a greater problem is that the expansion would lead to a 
conversion of internal energy into kinetic energy of expansion, so even 
after the fireball becomes optically thin, it would be highly inefficient, 
most of the energy being in the kinetic energy of the associated protons, 
rather than in photons. For a photosphere occurring at $r<r_s$, which 
requires high values of $\eta$, the radiative luminosity in the observer 
frame is undiminished, since $E'_{rad}\propto r^{-1}$ but $\Gamma\propto r$ so
$E_{rad}\sim$ constant, or $L{ph}\propto r^2 \Gamma^2 {T'}^4 \propto$ constant, 
since $T'\propto r^{-1}$. However for the more moderate values of $\eta$
the photosphere occurs above the saturation radius, and whereas the kinetic
energy of the baryons is constant $E_{kin}\sim E_0\sim$ constant the radiation
energy drops as $E_{rad}\propto (r/r_s)^{-2/3}$, or $L_{ph}\sim L_0
(r_{ph}/r_s)^{-2/3}$ \cite{mr93a,mr00b}. 

A natural way to achieve a non-thermal spectrum in an energetically efficient 
manner is by having the kinetic energy of the flow re-converted into random 
energy via shocks, after the flow has become optically thin 
\cite{rm92,mr93a,mlr93,katz94a,rm94,sapi95}.
Such shocks will be collisionless, i.e. mediated by chaotic electric and magnetic
fields rather than by binary particle interactions, as known from interplanetary
experiments and as inferred in supernova remnants and in active galactic nuclei 
(AGN) jets . As in these well studied sources, these shocks can be expected to
accelerate particles via the Fermi process to ultra-relativistic energies 
\cite{be87,acht01,ell02,lempel03_accel,sokol06_accel}, and
the relativistic electron component can produce non-thermal radiation via the 
synchrotron and inverse Compton (IC) processes. A shock is essentially 
unavoidable as the fireball runs into the external medium, producing a blast 
wave. The external medium may be the interstellar medium (ISM), or the 
pre-ejected stellar wind from the progenitor before the collapse.
For an outflow of total energy $E_0$ and terminal coasting bulk Lorentz 
factor $\Gamma_0=\eta$ expanding in an external medium of average particle 
density $n_{ext}$, the external shock becomes important at a deceleration radius 
$r_{dec}$ for which $E_0= (4\pi/3) r_{dec}^3 n_0 m_p c^2 \eta^2$ \cite{rm92},
\beq
r_{dec} \sim (3E_0/4\pi n_{ext} m_p c^2 \eta^2)^{1/3} \sim
 5.5 \times 10^{16} E_{53}^{1/3} n_o^{-1/3} \eta_{2.5}^{-2/3} ~{\rm cm}~.
\label{eq:rdec}
\enq
At this radius the initial bulk Lorentz factor has decreased to approximately 
half its original value, as the fireball ejecta is decelerated by the swept-up 
external matter. The amount of external matter swept at this time is a
fraction $\eta^{-1}$ of the ejecta mass $M_0$, $M_{ext} \sim M_0/\eta$ (in
contrast to the sub-relativistic supernova expansion, where deceleration
occurs when this fraction is $\sim 1$).

The light travel time difference between a photon originating from $r=0$ and 
a photon originating from matter which has moved to a radius $r$ with a Lorentz 
factor $\Gamma$ is $\Delta t \sim (r/c)(1-\beta) \sim r/2c \Gamma^2$ \cite{rees66},
and the emission from a photosphere or from a shock emission region at radius 
$r$ moving at constant $\Gamma$ is also received from within the causal light
cone angle $\Gamma^{-1}$ on an observer angular timescale $t\sim r/2c\Gamma^2$
\cite{rm92,mr93a}. For an explosion which is impulsive (i.e. essentially 
instantaneous as far as observed relativistic time delays) a similarity solution 
of the relativistic flow equations shows that the bulk of the ejected matter at 
a radius $r$ is mainly concentrated inside a region of width $\Delta r\sim 
r/2\Gamma^2$ \cite{bm76,bm76b}. The time delay between radiation along the
central line of sight originating from the back and front edges of this shell 
also arrive with a similar time delay $t\sim r/2c\Gamma$. Thus, the timescale 
over which the deceleration is observed to occur is generally
\beq
t_{dec} \sim  r_{dec}/(2c\Gamma^2) \sim 10 (E_{53}/n_o)^{1/3}\eta_{2.5}^{-8/3}~{\rm s},
\label{eq:tdec}
\enq
and this is the observer timescale over which the external shock radiation 
is detected. This is provided that the explosion can be taken to be impulsive, 
which can be defined as the outflow having a source-frame (and observer frame) 
duration $t_{grb}< t_{dec}$ (see however \S \ref{sec:duration}).
Variability on timescales shorter than $t_{dec}$ may occur on the cooling 
timescale or on the dynamic timescale for inhomogeneities in the external 
medium, but is not ideal for reproducing highly variable profiles\cite{sapi97},
and may therefore be applicable to the class of long, smooth bursts.
However, it can reproduce bursts with several peaks\cite{panmes98a}, and if
the external medium is extremely lumpy ($\Delta n_o/n_o \simg 10^5-10^6$) it
might also describe spiky GRB light curves \cite{dermit99}.

\begin{figure}[ht]
\begin{center}
\centerline{\epsfxsize=5.in \epsfbox{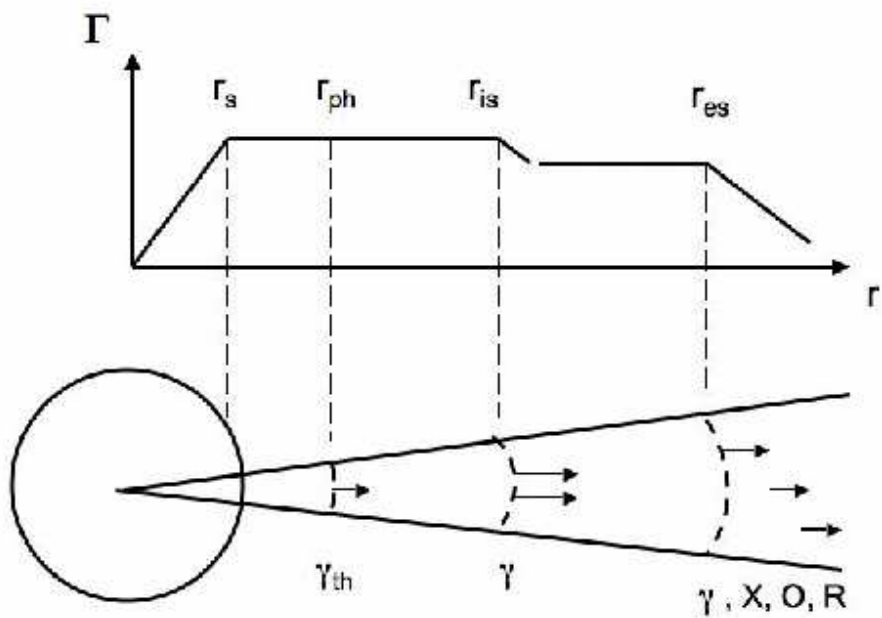}}
\end{center}
\caption{Jet Lorentz factor schematic behavior and examples of nominal locations 
of the saturation radius $r_s$, photospheric radius $r_{ph}$, internal shock 
(or magnetic dissipation) radius $r_{is}$ and external shock $r_{es}$. The
photosphere produces thermal $\gamma$-rays, the internal shock/dissipation region
produces the non-thermal $\gamma$-rays, the external shock region produces the
afterglow.  }
\label{fig:jetlorentz}
\end{figure}

Before the ejecta runs into the external medium, ``internal shocks" can
also occur as faster portions of the ejecta overtake  slower ones,
leading to $pp$ collisions and $\pi^0$ decay gamma-rays \cite{px94} 
and to fast time-varying MeV gamma-rays \cite{rm94}. The latter can be 
interpreted as the main burst itself. If the outflow is described by an
energy outflow rate $L_o$ and a mass loss rate ${\dot M_o}=dM_o/dt$ 
starting at a lower radius $r_l$, maintained over a time $T$, then the 
dimensionless entropy is $\eta= L_o/ {\dot M_o c^2}$, and the behavior is 
similar to that in the impulsive case, $\Gamma \propto r$ and comoving 
temperature $T' \propto r^{-1}$, followed by saturation $\Gamma_{max} 
\sim \eta$ at the radius $r/r_o \sim \eta$ \cite{pac90}. For variations
of the output energy or mass loss of order unity, the ejected shells of 
different Lorentz factors $\Delta \eta \sim \eta$ are initially separated 
by $c t_v $ (where $t_v \leq T$ are the typical variations in the energy 
at $r_l$), and they catch up with each other at an internal shock (or
dissipation) radius 
\beq
r_{dis} \sim  c t_v \eta^2 \sim 3\times 10^{14} t_{v,0} \eta_2^2 ~ {\rm cm},
\label{eq:rdis}
\enq
The time variability should reflect the variability of the central engine, 
which might be expected e.g. from accretion disk intermittency, flares, etc. 
\cite{napapi92}. The radiation from the disk or flares, however, cannot be 
observed directly, since it occurs well below the scattering photosphere of the 
outflow and the variability of the photons below it is washed out \cite{rm94}.
The comoving Thomson optical depth is $\tau_T=n_e' \sigma_T r/\Gamma$, and 
above the saturation radius $r_s=r_o\eta$ where $\Gamma=\eta$, the radius of 
the photosphere ($\tau_T=1$), is given from equation (\ref{eq:rphotgen}) as
\beq
r_{ph} \sim \cases{ 
   1.2\times 10^{12} L_{51} \eta_2^{-3}~{\rm cm}, & for $r>r_s$;\cr
   1.2\times 10^{10} L_{51}^{1/3}r_{07}^{2/3} \eta_2^{-1/3}~{\rm cm} & for $r<r_s$.}
\label{eq:rphot}
\enq
The location of this baryonic photosphere defines a critical dimensionless 
entropy $\eta_\ast=562 (L_{51}/r_{07})^{1/4}$ above (below) which the 
photosphere occurs below (above) the saturation radius \cite{mr00b}. 
In order for internal shocks to occur above the wind photosphere 
and above the saturation radius (so that most of the energy does not come out 
in the photospheric quasi-thermal radiation component) one needs to have
$3.3 \times 10^1 (L_{51} r_{0,7}/t_{v,0})^{1/5} \siml \eta \siml
5.62 \times 10^2 (L_{51}/r_{0,7})^{1/4} $.
The radial variation of the bulk Lorentz factor and the location of the various 
characteristic radii discussed above is shown in Figure \ref{fig:jetlorentz}.

Such internal shock models have the advantage\cite{rm94} that they allow an
arbitrarily complicated light curve, the shortest variation timescale 
$t_{v,min} \simg 10^{-4}$ s being limited only by the dynamic timescale 
at $r_0 \sim c t_{v,min} \sim 10^7 r_{0,7}$ cm, where the energy input may 
be expected to vary chaotically, while the total duration is $t_{grb}\gg t_v$.  
Such internal shocks have been shown explicitly  to reproduce (and be required 
by) some of the more complicated light curves\cite{sapi97,kps97,panmes99} (see 
however \cite{dermit99,rydpet02}). The gamma-ray emission of GRB from internal 
shocks radiating via a synchrotron and/or inverse Compton mechanism reproduces 
the general features of the gamma-ray observations \cite{frewax01,spada00}. 
There remain, however, questions concerning the low energy ($20-50$ keV) 
spectral slopes for some bursts (see \S {sec:spectrum}).
Alternatively, the main $\gamma$-ray bursts could be (at least in part) 
due to the early part of the external shock \cite{rm92,dermit99}.
Issues arise with the radiation efficiency, which for internal shocks,
is estimated to be moderate in the bolometric sense (5-20\%), higher
values ($\siml 30-50\%$) being obtained if the shells have widely differing 
Lorentz factors \cite{spada00,belob00b,kobasar01}, although in this case
one might expect large variations in the spectral peak energy $E_{peak}$
between spikes in the same burst, which is problematic. The total efficiency 
is substantially affected by inverse Compton losses \cite{pm96,pilla98}.  The 
efficiency for emitting in the BATSE range is typically low $\sim 1-5\%$, both 
when the MeV break is due to synchrotron \cite{kumar99,spada00,guetta01} and 
when it is due to inverse Compton \cite{panmes00}.

\subsection{Duration, reverse shocks, thin and thick shells}
\label{sec:duration}

In the following discussion we assume for simplicity a uniform external 
medium.  For a baryonic outflow such as we have been considering, the timescale 
$t_0\sim r_0/c\sim$ ms represents a minimum variability timescale in the 
energy-mass outflow. (Note, however, if the gamma-ray emission arises from 
local dissipation events, such as e.g. magnetic reconnection in a Poynting 
flux dominated outflow, the minimum timescales could be smaller than the 
timescales of the central source variations). On the other hand, the total 
duration $t_{grb}$ of the outflow, during which the central engine keeps pouring 
out energy and matter, is likely to be substantially longer than the minimum
variability timescale $t_0$. The temporary accretion disk must have an 
outer radius larger than  $r_0$, and a total accretion (or jet energization) 
time $t_{grb} \gg t_0$ (or the magnetar has a spin-down time $t_{grb} \gg t_0$).  
Thus, in general the total lab-frame width of the outflow ejecta will be 
$\Delta \approx ct_{grb}$, which may be viewed as composed of many radial 
minishells whose individual widths are $\delta r \sim c\delta t_0$ or larger. 
While the saturation radius is still $r_s\sim \delta r_0\eta$ where $\delta r_0
\sim r_0$ corresponds to the shortest variability time (and the smallest
minishells coast after this $r_s$), the entirety of the ejecta reaches
coasting speed only after its leading edge has moved to a larger radius 
$r_{s'}\sim \Delta  \eta$, and the ejecta as whole starts to spread at a 
larger radius $r_\Delta \sim \Delta  \eta^2$ (even though individual minishells 
of initial width $\delta r \ll \Delta$ start to spread individually at the 
smaller radius $r_\delta \sim \delta r ~ \eta^2$).

In general, whatever the duration $t_{grb}$ of the outflow, one expects the
external shock to have both a forward shock (blast wave) component propagating
into the external medium, and a reverse shock propagating back into the ejecta
\cite{mr93a}. The forward shock and the reverse shock start forming as
soon as the outflow starts, although their radiation is initially weak and
increases progressively. The forward shock is highly relativistic, $\Gamma
\sim \eta$ from the very beginning, but the reverse shock starts initially as 
a sub-relativistic sound  wave (relative to the contact discontinuity or shock 
frame) and becomes progressively stronger as more external matter is swept up. 
(This describes the more frequently encountered ``thin shell" case, see below;
the reverse shock becomes stronger with time only if the external density profile
is shallower than $r^{-2}$, whereas the reverse shock strength is constant
for an $r^{-2}$ profile at $r<r_\delta$).

For an impulsive regime outflow, where $t_{grb} < t_{dec}$, i.e. when the 
outflow time is shorter than the time-delayed duration of the external shock 
when it starts to decelerate, equation [\ref{eq:tdec}], this deceleration time
can be taken to be the observable duration of the peak emission from the
external shock. 
Thereafter the expansion goes into a self-similar expansion
with $\Gamma\propto r^{-3/2}$ \cite{bm76,rm92}. In this case, $t_{dec}$ is 
also the observer time at which 
the reverse shock finishes crossing the ejecta, and at that time the reverse
shock Lorentz factor ${\bar \Gamma_r}$ relative to the contact discontinuity
frame has become marginally relativistic, ${\bar \Gamma_r} \sim 1$, while
relative to the external gas or the observer, the reverse shocked gas is still 
moving at almost the same speed as the forward shocked gas \cite{mr93a,mrp94}. 
One consequence of this is that while the forward shocked protons have 
highly relativistic random Lorentz factors, those in the reverse shock are 
marginally relativistic, and consequently the electrons in the forward shock are 
much more relativistic than those in the reverse shock, leading to a much softer 
(optical) spectrum of the reverse shocks \cite{mr93b,mr97a} (see \S \ref{sec:reverse})

However, when the outflow time $t_{grb}$ exceeds the deceleration time 
$t_{dec}$ of equation (\ref{eq:tdec}), the eternal shock dynamics is different 
\cite{sapi95}. In this case there is an initial intermediate regime 
$\Gamma \propto r^{-1/2}$ (obtained, for a constant external density $\rho$ and a 
constant kinetic luminosity $L$ at $t<t_{grb}$ from momentum balance in
the shock frame, $L/(r^2\Gamma^2 \propto \rho\Gamma^2$), and the transition to 
a self-similar expansion $\Gamma\propto r^{-3/2}$ \cite{bm76,sari97} occurs at 
the observer time $t_{grb}$, instead of at $t_{dec}$. Thus, the observer time 
for the transition to the self-similar expansion is
\beq
T=\max [ t_{grb}, 10 (E_{53}/n_0)^{1/3}\eta_{2.5}^{-8/3}~{\rm s} ]
\label{eq:duration}
\enq
This defines a critical initial Lorentz factor $\Gamma_0\sim \eta$ of the 
burst by setting $T=t_{grb}$  in place of $t_{dec}$ in equation (\ref{eq:tdec}), 
\beq
\Gamma_{BM} \simeq 300 (E_{53}/n_0)^{1/8} (T/10~{\rm s})^{-3/8} ~.
\label{eq:GammaBM}
\enq
For $\eta < \Gamma_{BM}$, $T=t_{dec}$ we have the usual ``thin shell" case, 
where deceleration and transition to the self-similar expansion occurs at the 
usual $r_{dec}$, $t_{dec}$, and at this time the reverse shock has crossed the
ejecta and is marginally relativistic. For $\eta>\Gamma_{BM}$ we have a ``thick 
shell" case, where deceleration and transition to the self-similar regime occurs 
at $T=t_{grb}$ and $r_{BM}\sim 2cT \Gamma_{BM}^2$, when $\Gamma\sim \Gamma_{BM}$.  
In this $\eta > \Gamma_{BM}$ case, the reverse shock becomes relativistic, 
and by the time it has crossed the ejecta (at time $T=t_{grb}$) the reverse 
shock Lorentz factor in the contact discontinuity frame is ${\bar \Gamma_r} \sim 
\eta/2\Gamma_{BM} \gg 1$, and the forward shock Lorentz factor at this time is 
$\Gamma\simeq \Gamma_{BM}$.

\subsection{Spectrum of the Prompt GRB Emission}
\label{sec:spectrum}

The prompt emission observed from classical GRB (as opposed to XRFs or SGRs) 
has most of its energy concentrated in the  gamma-ray energy range 0.1-2 MeV. 
The generic phenomenological photon spectrum is a broken power law \cite{band93} 
with a break energy in the above range, and power law extensions down into 
the X-ray, and up into the 100 MeV to GeV ranges (although a substantial
fraction of GRB have soft X-ray excesses above this, and some are classified 
as X-ray rich (XRR) \cite{preece00}, a classification intermediate between
XRF and GRB). For classical GRB the photon energy flux $F_E\propto E^{-\beta}$ 
has typical indices below and above the typical observed break energy $E_{br}\sim 
0.2$ MeV of $\beta_1 \sim 0$ and $\beta_2\sim 1$ \cite{band93}. (Pre-BATSE 
analyses sometimes approximated this as a bremsstrahlung-like spectrum with an
exponential cutoff at $E_{br}$, but BATSE showed that generally the extension
above the break is a power law). A synchrotron interpretation is thus natural,
as has been argued e.g. since the earliest external shock synchrotron models
were formulated. 

The simplest synchrotron shock model starts from the conditions 
behind the relativistic forward shock or blast wave \cite{rm92,mr93a}. 
The post-shock particle and internal energy density follow from the
relativistic strong shock transition relations \cite{bm76},
\bea
n_2 &= &(4\Gamma_{21}+3) n_1 \simeq 4\Gamma_{21} n_1, \nonumber \\
e_2 &= &(\Gamma_{21}-1) n_2 m_p c^2 \simeq \Gamma_{21} n_2 m_p c^2
\simeq 4\Gamma_{21}^2 n_1 m_p c^2,
\label{eq:shock}
\ena
where it is assumed that the upstream material is cold.  Here $n$ is number 
density and $e$ is internal energy density, both measured in the 
comoving frames of the fluids, $\Gamma_{21} \simeq \Gamma$ is the relative 
Lorentz factor between the fluids 2 (shocked, downstream) and 1 (unshocked, 
upstream), and the Lorentz factor of the shock front itself is 
$\Gamma_{sh}=\sqrt{2} \Gamma_{21}$, valid for $\Gamma_{21} \gg 1$. For 
internal shocks the jump conditions can be taken approximately the same, but 
replacing $\Gamma_{21}$ by a lower relative Lorentz factor $\Gamma_r\sim 1$.

The typical proton crossing a strong shock front with a relative bulk Lorentz
factor $\Gamma_{21}$ acquires (in the comoving frame) an internal energy 
characterized by a random (comoving) Lorentz factor $\gamma_{p,m} \sim 
\Gamma$ \cite{mr93a}. The comoving magnetic field behind the shock can build up 
due to turbulent dynamo effects behind the shocks \cite{mr93a,mr93b} (as  also 
inferred in supernova remnant shocks). More recently, the Weibel instability has 
been studied in this context \cite{medloeb99,nord05,medvedev06_sim,spit06}. While 
the efficiency of this process remains under debate, one can parametrize the
resulting magnetic field as having an energy density behind the shock which is
a fraction $\eps_B$ of the equipartition value relative to the proton random energy
density behind the shock, $B'\sim [32 \pi \eps_B n_{ex} (\gamma'_p -1)m_p c^2]^{1/2}
\Gamma$, where the post-shock proton comoving internal energy is $(\gamma'_p-1)
m_p c^2 \sim ~1$ (or $\sim \Gamma $) for internal (external) shocks \cite{mr93a,rm94}.
Scattering of electrons (and protons) by magnetic irregularities upstream and
downstream can lead to a Fermi acceleration process resulting in a relativistic
power law distribution of energies $N(\gamma)\propto \gamma^{-p}$ with $p\geq 2$. 
It should be stressed that although the essential features of this process are 
thought to be largely correct, and it is widely used for explaining supernova
remnant, AGN and other non-thermal source radiation spectra, the details are only 
sketchily understood, \cite{be87,acht01,ell02,lempel03_accel,keswax05,sokol06_accel}. 
(Possible difficulties with the simplest version of Fermi acceleration and
alternative possibilities were discussed, e.g. in \cite{barbrab05,nord04,duffy05}).
The starting minimum (comoving) Lorentz factor of the thermal electrons injected 
into the acceleration process, $\gamma_{e,m}$ would in principle be the same as 
for the protons, $\Gamma$, (they experience the same velocity difference), 
hence both before and after acceleration they would have $\sim (m_e/m_p)$ less 
energy than the protons. However, the shocks being collisionless, i.e. mediated 
by chaotic electric and magnetic fields, can  redistribute the proton energy 
between the electrons and protons, up to some fraction $\eps_e$ of the thermal 
energy equipartition value with the protons, so $\gamma_{e,m} \sim \eps_e (m_p/m_e) 
\Gamma$ \cite{mr93b,mrp94}. If only a fraction $\zeta_e\leq 1$ of all the 
shocked thermal electrons is able to achieve this $\eps_e$ initial equipartition 
value to be injected into the acceleration process, then the initial minimum electron
random comoving Lorentz factor is $\gamma_m \sim (\eps_e/\zeta_e)(m_p/m_e) 
\Gamma$ \cite{byk96}, where henceforth we ignore the subscript $e$ in $\gamma_{e,m}$.
More accurately, integrating over the power law distribution, one has
$\gamma_m = g(p) (m_p/m_e) (\epsilon_e/\zeta_e) \Gamma \sim 310
[g(p)/(1/6)] (\epsilon_e/\zeta_e) \Gamma$, where $g(p)=(p-2)/(p-1)$.
The observer frame synchrotron spectral peak is 
\beq
\nu_m \sim \Gamma (3/8\pi)(eB'/m_e c)\gamma_m^2\sim 
   2\times 10^6 B'\gamma_m^2 \Gamma ~{\rm  Hz}~,
\label{eq:num}
\enq
and the optically thin  synchrotron spectrum is \cite{rl79}
\beq
F_\nu \propto \cases{ \nu^{1/3}    & for $\nu< \nu_m$ ; \cr
                    \nu^{-(p-1)/2} & for $\nu>\nu_m$ },
\label{eq:syncslow}
\enq
assuming that the radiative losses are small (adiabatic regime).
For the forward external shock at deceleration, typical values are, e.g. 
$B'\sim 30 (\eps_{B,-1} n_{ex})^{1/2} \eta_{2.5}$ G, $\Gamma\sim \eta \sim 
3\times 10^2$, $\gamma_m\sim 10^5 (\eps_{e,-1}/\zeta_{e,-1})\eta_{2.5}$ and 
$\nu_m\sim 2\times 10^{20} (\eps_{e,-1}/\zeta_{e,-1})^2 (\eps_{B,-1}n_{ex})^{1/2}
\eta_{2.5}^4$ Hz, while for internal shocks typical values are, e.g.
$B'\sim 3\times 10^5 (\eps_{B,-1} n'_{13})^{1/2} \Gamma_{rel,0}$ G, $\Gamma_{rel} 
\sim 1\Gamma_{rel,0}$, $\gamma_m\sim 3\times 10^3 (\eps_{e,-1}/\zeta_{e,-1})
\Gamma_{rel,0}$ and $\nu_m\sim 2\times 10^{19} (\eps_{e,-1}/zeta_{e,-1})^2 
(\eps_{B,-1}n'{13})^{1/2} \Gamma_{r,0}^3\eta_{2.5}$ Hz. 
For the prompt emission, the high energy slope $\beta_2=(p-1)/2$ is close to 
the mean high energy slope of the Band fit, while the low energy slope can 
easily approach $\beta_1\sim 0$ considering observations from, e.g., 
a range of $B'$ values (a similar explanation as for the flattening of the 
low energy synchrotron slope in flat spectrum radio-quasars). 
The basic synchrotron spectrum is modified at low energies by synchrotron
self-absorption \cite{mr93b,mrp94,katz94a,grapisa99b}, where it makes the 
spectrum steeper ($F_\nu \sim \nu^2$ for an absorption frequency $\nu_a <\nu_m$).
It is also modified at high energies, due to inverse Compton effects 
\cite{mr93b,mrp94,rm94,dercm00,sariesin01,zhames01b}, extending into the GeV range.

The synchrotron interpretation of the GRB radiation is the most straightforward.
However, a number of effects can modify the simple synchrotron spectrum.
One is that the cooling could be rapid, i.e.  when the comoving synchrotron 
cooling time $t'_{sy}= 9m_e^3 c^5/ 4e^4 B'^2 \gamma_e) \sim 7\times 
10^8/B'^2\gamma_e ~{\rm s}$ is less than the comoving dynamic time 
$t'_{dyn}\sim r/2c\Gamma$, the electrons cool down to 
$\gamma_c= 6\pi m_e c /\sigma_T B'^2 t'_{dyn}$ and the spectrum above 
$\nu_c\sim \Gamma (3/8\pi)(eB'/m_e c)\gamma_c^2$ is $F_\nu \propto \nu^{-1/2}$ 
\cite{sapina98,ghiscel99}. Also, the distribution of observed low energy spectral 
indices $\beta_1$ (where $F_\nu\propto \nu^{\beta_1}$ below the spectral peak) 
has a mean value $\beta_1\sim 0$, but for a fraction of bursts this slope reaches 
positive values $\beta_1>1/3$ which are incompatible with a simple synchrotron 
interpretation \cite{preece00}. Possible explanations include synchrotron 
self-absorption in the X-ray \cite{grapisa00} or in the optical range 
up-scattered to X-rays \cite{panmes00}, low-pitch angle scattering  or jitter
radiation \cite{medvedev00,medvedev06_jitter}, observational selection biases 
\cite{lloyd01} and/or time-dependent acceleration and radiation \cite{lloyd02}, 
where low-pitch angle diffusion can also explain high energy indices steeper 
than predicted by isotropic scattering. Other models invoke a photospheric
component and pair formation \cite{mr00b}, see below. 

There has been extensive work indicating that the apparent clustering of the 
break energy of prompt GRB spectra in the 50-500 keV range may be real 
\cite{preece00}, rather than due to observational selection effects \cite{pina96}.
I.e. the question is, if this is a real clustering, what is the physical reason
for it. (Note, however, that if X-ray flashes, or XRF, discussed  below, form a 
continuum with GRB, then this clustering stretches out to much lower energies; 
at the moment, however, the number of XRFs with known break energies is small).  
Since the synchrotron peak frequency observed is directly dependent on the bulk Lorentz 
factor, which may be random, the question arises whether this peak is indeed
due to synchrotron, or to some other effect. An alternative is to attribute a
preferred peak to a black-body at the comoving pair recombination temperature 
in the fireball photosphere \cite{eiclev00}. In this case a steep low energy 
spectral slope is due to the Rayleigh-Jeans part of the photosphere, and the 
high energy power law spectra and GeV emission require a separate explanation. 
For such photospheres to occur at the pair recombination temperature in the 
accelerating regime requires an extremely low baryon load. For very large 
baryon loads, a related explanation has been invoked \cite{tho94},
considering scattering of photospheric photons off MHD turbulence in the coasting
portion of the outflow, which up-scatters the adiabatically cooled photons
up to the observed break energy. 

Pair formation can become important \cite{rm94,pm96,pilla98} in internal shocks
or dissipation regions occurring at small radii, since a high comoving 
luminosity implies a large comoving compactness parameter
\beq
\ell'=n'_\gamma \sigma_T r_{dis}/\Gamma \sim 
(\alpha L \sigma_T /4 \pi r_{dis} m_e c^3 \Gamma^3) \simg 1 ~,
\label{eq:compac}
\enq
where $\alpha\siml 1$ is the luminosity fraction above the electron rest mass.
Pair-breakdown may cause a continuous rather then an abrupt heating and lead 
to a self-regulating moderate optical thickness pair plasma at sub-relativistic 
temperature, suggesting a comptonized spectrum \cite{ghiscel99}. Copious pair 
formation in internal shocks may in fact extend the photosphere beyond the 
baryonic photosphere value (\ref{eq:rphot}). A generic model has been proposed 
\cite{mr00b,mrrz02,raml02,ryde04,ryde05,ram05} which includes the emission of a thermal 
photosphere as well as a non-thermal component from internal shocks outside of it, 
subject to pair breakdown, which can produce both steep low energy spectra, preferred 
breaks and a power law at high energies. A moderate to high scattering depth can 
lead to a Compton equilibrium which gives spectral peaks in the right energy range 
\cite{pw04a,pw04b}. An important aspect is that Compton equilibrium of internal 
shock electrons or pairs with photospheric photons lead to a high radiative 
efficiency, as well as to spectra with a break at the right preferred energy 
and steep low energy slopes \cite{rm05,peermr05a,peermr06a}. It also leads to possible
physical explanations for the Amati \cite{amati02} or Ghirlanda \cite{ghirl04}
relations between spectral peak energy and burst fluence \cite{rm05,tho06}.

\subsection{Alternative Prompt Emission Models}
\label{sec:alternative}

There are several alternative models for the prompt GRB emission, which so
far have not found wide use for explaining the observations. The most plausible
of these, despite the technical difficulties which impair its applicability,
considers the main $\gamma$-ray burst emission to arise from  magnetic 
reconnection or dissipation processes, if the ejecta is highly magnetized or 
Poynting dominated \cite{usov94,tho94,mrp94,mr97b,drenk02,lyubland03,tho06}.
The central engine could also in principle be a temporary highly magnetized 
neutron star or magnetar \cite{whee00}. These scenarios would lead to alternative
dissipation radii, instead of equation (\ref{eq:rdis}), where reconnection leads 
to particle acceleration, and a high radiative efficiency is in principle 
conceivable due to the very high magnetic field. An external shock would follow 
after this, whose radius in the ``thin shell" limit would be again given by 
equation (\ref{eq:rdec}), with a standard forward blast wave but no (or a weaker) 
reverse shock \cite{mrp94,mr97a}, due to the very high Alfv\'en speed in the ejecta. 
For a long duration outflow, however, the dynamics and the deceleration radius 
would  be similar to the ``thick shell" case of \S \ref{sec:duration}, i.e. the
case with a relativistic reverse shock \cite{lyubland03}. Following the claim of 
an observed high gamma-ray polarization in the burst GRB 021206 \cite{coburn03}, 
there was increased attention on such models for some time (e.g. \cite{lyubland03}), 
and on whether the usual baryonic (i.e. sub-dominant magnetization) jets might 
also be able to produce such high polarization
\cite{waxman03,grankon03,gran03,nakpirwax03,lyut03,laz04,eiclev03,darder04}.
The issue may remain unresolved, as the observational analysis appears to be
inconclusive \cite{rutfox04,coburn03b,wigger04}.

Other alternative models include different central energy sources such as strange 
stars (\cite{cd96,bbdfl03,drago05,pachaens05}) and charged black hole electric 
discharges \cite{ruffini01}, while retaining essentially similar fireball shock scenarios.  
A model unifying SGR, XRF and GRB \cite{far99,far04} postulates a very thin 
($10^{-4}$ rad) precessing, long-lived  magnetized jet.
This requires a separate explanation for the light-curve (``jet") breaks, and the 
interaction during precession with the massive stellar progenitor is unclear.
Another speculative radiation scenario considers non-fluid ejecta in the form 
of discrete "bullets" \cite{heibeg99}, or ``cannon-balls" ejected at relativistic 
velocities, which assume no collective interactions (i.e. no collisionless shocks)
and instead rely on particle-particle interactions, and produce prompt emission by 
blue-shifted bremsstrahlung and produce afterglows by IC scattering progenitor or 
ambient photons \cite{dado02,darder04}. The predictions are similar to those of the 
standard fluid jet with shocks or dissipation. However, the basic ansatz of coherent 
bullet formation, acceleration to relativistic velocities and their survival against 
plasma instabilities is an unanswered issue in this model. It is also farther from 
astrophysical experience, whereas other well-observed systems such as AGN jets, 
which are known to be fluid (as is almost everything else in astrophysics at high 
energy per particle values) involve dynamical and radiation physics concepts which 
are quite plausibly extended to the GRB context. Fluid or plasma GRB outflow and 
jet models are better supported by theoretical work and simulations, and are so far 
not only compatible with observations  but have produced predictions borne out by 
observations. Nonetheless, even in this standard scenario, the models remain largely 
phenomenological. The detailed nature of the underlying central engine and progenitor 
are poorly known, and the micro-physics of particle acceleration, magnetic field 
amplification in shocks  and/or reconnection or dissipation is not well understood, 
and the radiation mechanisms are, at least for the prompt emission, subject of 
discussion.

\section{Afterglow Radiation Models}
\label{sec:ag}

\subsection{The standard model}
\label{sec:standardag}

The external shock starts to develop as soon as the ejecta expands into the
external medium.  As the ejecta plows ahead, it sweeps up an increasing amount of 
external matter, and the bolometric luminosity of the shock increases as $L\propto 
t^2$  (equating in the contact discontinuity frame the kinetic flux 
$L/4\pi r^2 \Gamma^2$ to the external ram pressure $\rho_{ext} \Gamma^2$ while 
$\Gamma\sim \Gamma_0= \eta\sim$ constant, $r\propto 2 \Gamma^2 ct \propto t$ 
\cite{rm92}). The luminosity peaks after $\Gamma$ has dropped to about half its 
initial value, at a radius $r_{dec}$ at an observer time $t_{dec}$ given by 
equations (\ref{eq:rdec},\ref{eq:tdec}). Thereafter, as more matter is swept up,
the bulk Lorentz factor and the radius vary as \cite{rm92,panmes98b} as
\bea
& \Gamma \propto r^{-3/2} \propto t^{-3/8}, & ~~~r \propto t^{1/4}~~{\rm (adiabatic)},
\nonumber \\
& \Gamma \propto r^{-3} \propto t^{-3/7}, & ~~~r \propto t^{1/7}~~{\rm (radiative)},
\label{eq:dynamics}
\ena
or in general $\Gamma \propto r^{-g}\propto t^{-g/(1+2g)}~,~r\propto t^{1/(1+2g)}$
with $g=(3,3/2)$ for the radiative (adiabatic) regime. In the adiabatic case the 
radiative cooling  time, e.g. synchrotron, is longer than the observer-frame 
dynamical time $t\sim r/ 2 c \Gamma^2$, so the energy is approximately 
conserved $E = (4\pi/3) r^3 n_0 m_p c^2 \Gamma^2 \sim$ constant (c.f. equation 
[\ref{eq:rdec}]), while in the radiative case the cooling time is shorter than 
the dynamic time and momentum is conserved (as in the snow-plow phase of 
supernova remnants), $n_o r^3 \Gamma\sim$ constant. Thus, after the external 
shock luminosity peaks, one expects the bolometric luminosity to decay as 
$L\propto t^{-1}$ in the adiabatic regime \cite{rm92} or steeper in the radiative 
regime, in a gradual fading. The observed time-radius relation is more generally
$t\sim r/ K c \Gamma^2$, where $K=2$ in the constant $\Gamma$ regime, and
$K=4$ in the self-similar (BM) regime \cite{wax97c,sari97}.

The spectrum of radiation is likely to be due to synchrotron radiation,  whose 
peak frequency in the observer frame is $\nu_m \propto \gamma B' \gamma_e^2$,
and both the comoving field $B'$ and electron Lorentz factor $\gamma_e$ are
likely to be proportional to $\gamma$ \cite{mr93a}. This implies that as
$\gamma$ decreases, so will $\nu_m$, and the radiation will move to longer
wavelengths. Consequences of this are the expectation that the burst would
leave a radio remnant \cite{pacro93} after some weeks, and before that an
optical \cite{katz94b} transient. The observation of linear polarization at 
the few percent level observed in a number of optical or IR afterglows (e.g.
\cite{jvp00}) supports the paradigm of synchrotron emission as the dominant 
emission mechanism in the afterglow.

The first self-consistent afterglow calculations \cite{mr97a} took into account
both the dynamical evolution and its interplay with the relativistic particle
acceleration and a specific relativistically beamed radiation mechanism
resulted in quantitative predictions for the entire spectral evolution,
going through the X-ray, optical and radio range. For a spherical fireball
advancing into an approximately smooth external environment, the bulk
Lorentz factor decreases as in inverse power of the time (asymptotically
$t^{-3/8}$ in the adiabatic limit), and the accelerated electron minimum random
Lorentz factor and the turbulent magnetic field also decrease as inverse power
laws in time. The synchrotron peak energy corresponding to the time-dependent
minimum Lorentz factor and magnetic field then moves to softer energies as
$t^{-3/2}$. These can be generalized in a straightforward manner when in the
radiative regime, or in presence of density gradients, etc.. The radio  spectrum
is initially expected to be self-absorbed, and becomes optically thin after 
$\sim$ hours. For times beyond $\sim 10$ minutes, the dominant radiation  is from 
the forward shock, for which the flux at a given frequency and the synchrotron peak
frequency decay as \cite{mr97a}
\beq
F_\nu \propto t^{-(3/2)\beta}~~,~~\nu_m\propto t^{-3/2}~,
\label{eq:Fnu}
\enq
as long as the expansion is relativistic. This is referred to as the ``standard"
(adiabatic) model, where $g=3/2$ in $\Gamma\propto r^{-g}$ and $\beta=d\log F_\nu
/ d\log\nu$ is the photon spectral energy flux slope. More generally \cite{mr99} 
the relativistic forward shock flux and frequency peak are given by
\beq
F_\nu\propto  t^{[3-2g(1-2\beta)]/(1+2g)}$ and $\nu_m\propto t^{-4g/(1+2g)}~.
\label{eq:Fnugen}
\enq
where $g=(3/2,3)$ for the adiabatic (radiative) regime. The transition to the 
non-relativistic expansion regime has been discussed, e.g. by 
\cite{wrm97,dailu99,livwax00}.  
A reverse shock component is also expected \cite{mr93b,mr97a,sapi99,mr99}, with 
high initial optical brightness but much faster decay rate than the forward 
shock, see \S \ref{sec:reverse}).  Remarkably, the simple ``standard" model
where reverse shock effects are ignored is a good approximation for modeling
observations starting a few hours after the trigger, as during 1997-1998.

\begin{figure}[ht]
\begin{center}
\centerline{\epsfxsize=4.in \epsfbox{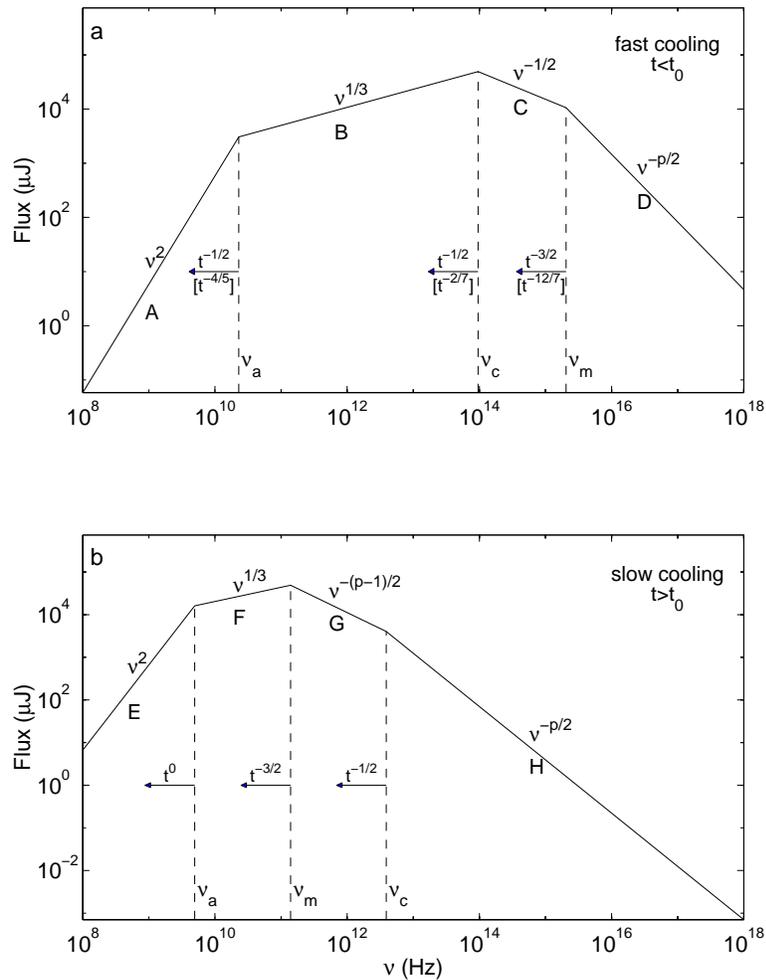}}
\end{center}
\caption{Fast cooling and slow cooling synchrotron spectra \cite{sapina98}}
\label{fig:sync-spectra}
\end{figure}

The afterglow spectrum at a given instant of time depends on the flux 
observed at different frequencies from electrons with (comoving) energy 
$\gamma_e m_c c^2$ and bulk Lorentz factor $\Gamma$, whose observed peak 
frequency is $\nu=\Gamma \gamma_e^2 (eB'/2 \pi m_ec)$. Three critical frequencies
are defined by the three characteristic electron energies. These are
$\nu_m$ (the ``peak" or injection frequency corresponding to $\gamma_m$), 
$\nu_c$ (the cooling frequency), and $\nu_M$ (the maximum synchrotron 
frequency). There is one more frequency, $\nu_a$, corresponding to the
synchrotron self-absorption at lower frequencies. 
For a given behavior of $\Gamma$ with $r$ or $t$ (e.g. adiabatic,
$\Gamma \propto r^{-3/2}$) and values of the isotropic equivalent kinetic 
energy of the explosion, of the electron index (e.g. $p=2.2$) and the
efficiency factors $\eps_e,~\zeta_e,~\eps_B$, one can obtain the
time dependence of the characteristic observer-frame frequencies,
including also a cosmological redshift factor $z$ \cite{zhames04}
\bea
\nu_m &=& (6\times 10^{15} ~\mbox{Hz})~ (1+z)^{1/2} g(p)^2
(\epsilon_e/\zeta_e)^2 \epsilon_B^{1/2} E_{52}^{1/2} t_d^{-3/2} \label{eq:numt}\\
\nu_c &=& (9\times 10^{12} ~\mbox{Hz})~ (1+z)^{-1/2} \epsilon_B^{-3/2}
n^{-1} E_{52}^{-1/2} t_d^{-1/2} \label{eq:nuct}\\
\nu_a &=& (2\times 10^9 ~\mbox{Hz})~ (1+z)^{-1} (\epsilon_e/\zeta_e)^{-1}
\epsilon_B^{1/5} n^{3/5} E_{52}^{1/5} \label{eq:nuat}\\
F_{\nu,max} &=& (20~\mbox{mJy})~ (1+z) \epsilon_B^{1/2} n^{1/2} E_{52}
d_{L,28}^{-2},~ \label{eq:Fnumaxt}
\ena
where $t_d=(t/{\rm day})$ and $g(p)=(p-2)/(p-1)$.
The final GRB afterglow synchrotron spectrum is a four-segment broken power 
law \cite{sapina98,mrw98,grasa02,zhames04} separated by the typical frequencies 
$\nu_a$, $\nu_m$, and $\nu_c$ (Figure \ref{fig:sync-spectra}). Depending on the 
order between $\nu_m$ and $\nu_c$, there are two types of spectra \cite{sapina98}. 
For $\nu_m < \nu_c$, called the ``slow cooling case'', the spectrum is
\begin{equation}
 F_\nu= F_{\nu,max}\left\{ \begin{array}{l@{\quad \quad}l}
              (\nu_a/\nu_m)^{1/3}(\nu/\nu_a)^2  & \nu < \nu_a \cr
              (\nu/\nu_m)^{1/3} & \nu_a \le \nu<\nu_m \cr
              (\nu/\nu_m)^{-(p-1)/2}  &  \nu_m \le \nu < \nu_c \cr
              (\nu_c/\nu_m)^{-(p-1)/2}(\nu/\nu_c)^{-p/2} & \nu_c \le
                \nu \le \nu_M
          \end{array} \right.
\label{eq:slowc}
\end{equation}
For $\nu_m > \nu_c$, called the ``fast cooling case'', the
spectrum is
\begin{equation}
 F_\nu= F_{\nu,max}\left\{ \begin{array}{l@{\quad \quad}l}
              (\nu_a/\nu_c)^{1/3}(\nu/\nu_a)^2  & \nu < \nu_a \cr
              (\nu/\nu_c)^{1/3} & \nu_a \le \nu < \nu_c \cr
              (\nu/\nu_c)^{-1/2}  &  \nu_c \le \nu < \nu_m \cr
              (\nu_m/\nu_c)^{-1/2}(\nu/\nu_m)^{-p/2} & \nu_m \le \nu
                \le \nu_M
          \end{array} \right.
\label{eq:fastc}
\end{equation}

A useful tabulation of the temporal indices $\alpha$ and spectral indices
$\beta$ is given in Table 1 of \cite{zhames04}, corresponding to the various 
forward shock spectral regimes of equations (\ref{eq:slowc}),(\ref{eq:fastc}),
for a homogeneous or a wind external medium. In the above, the normalization 
$F_{\nu,max}$ is obtained by multiplying the total number of radiating 
electrons $4\pi r^3 n_1/3$ by the peak flux from a single electron\cite{sapina98}, 
which is only a function of $B$ and is independent of the energy ($\gamma_e$) of 
the electron\cite{sapina98,wiga99}.
There are more complicated regimes for various cases of self-absorption
\cite{grapisa00}, e.g. there can also be an intermediate fast cooling optically
thick power law segment of the synchrotron spectrum where $F_\nu \propto 
\nu^{11/8}$.

The predictions of the fireball shock afterglow model \cite{mr97a} were made
in advance of the first X-ray detections by Beppo-SAX \cite{cos97} allowing
subsequent follow-ups \cite{jvp97,metz97,fra98} over different wavelengths,
which showed a good agreement with the standard model, e.g.
\cite{vie97a,wrm97,tav97,wax97a,wax97b,rei97}.
The comparison of increasingly sophisticated versions of this theoretical model
(e.g. \cite{sapina98,wiga99,piran99,derbc00,dercm00,grasa02}) against an increasingly
detailed array of observations (e.g. as summarized in \cite{jvp00}) has provided
confirmation of this generic fireball shock model of GRB afterglows.

A snapshot spectrum of the standard model at any given time consists generally 
of three or four segment power law with two or three breaks, such as those 
shown in Figure \ref{fig:sync-spectra}. (More rarely, a five segment power law
spectrum may also be expected \cite{grapisa00}). The observations (e.g. 
\cite{jvp00}) are compatible with an electron spectral index $p\sim 2.2-2.5$, 
which is typical of shock acceleration, e.g. \cite{wax97a,sapina98,wiga99}, etc.  
As the remnant expands the photon spectrum moves to lower frequencies, and the 
flux in a given band decays as a power law in time, whose index can change as 
the characteristic frequencies move through it.
Snapshot spectra have been deduced by extrapolating measurements at different
wavelengths and times, and assuming spherical symmetry and using the model time
dependences \cite{wax97b,wiga99}, fits were obtained for the different physical
parameters of the burst and environment, e.g. the total energy $E$, the magnetic
and electron-proton coupling parameters ${\eps}_B$ and ${\eps}_e$ and the external
density $n_o$.  
These lead to typical values $n_o\sim 10^{-2}-10$ cm$^{-3}$, $\eps_B\sim 10^{-2}$,
$\eps_e\sim 0.1-0.5$ and $E\sim 10^{52}-10^{54}$ ergs (if spherical; but see
\S \ref{sec:jet}).

\subsection{Prompt Flashes and Reverse Shocks}
\label{sec:reverse}

An interesting development was the observation \cite{ak99} of a prompt and 
extremely bright ($m_v\sim 9$) optical flash in the burst GRB 990123, 
the first data point for which was at 15 seconds after the GRB started 
(while the gamma-rays were still going on). This observation was followed 
by a small number of other prompt optical flashes, generally not as bright.  
A prompt multi-wavelength flash, contemporaneous 
with the $\gamma$-ray emission and reaching such optical magnitude levels 
is an expected consequence of the reverse component of external shocks
\cite{mr93b,mr97a,sapi99,mr99}. Generally the reverse shock can expected to 
be mildly relativistic (thin shell case; see, however, below). In this case the 
thermal Lorentz factor of the reverse electrons is roughly $\gamma _{e}^{r}\sim 
\eps_e m_{p}/m_{e}$ (whereas in the forward shock, the thermal Lorentz factor 
of the electrons is $\gamma _e^{f}\sim \epsilon _{e}\Gamma m_{p}/m_{e}$.
In this case the reverse electrons radiate much softer radiation than 
the forward shock electrons. This follows also from the fact that the
reverse shock has a similar total energy as the forward shock, but consists
of $\Gamma$ times more electrons, hence the energy per electron is $1/\Gamma$
times smaller \cite{mr97a}.  In general, since the pressure (and hence the
magnetic energy density) is the same in the forward and reverse shocked
regions, one has  the following relations between forward and reverse shock
radiation properties \cite{sames00}:
1) The peak flux of the reverse shock, at any time, is larger by a factor of
$\Gamma$ than that of the forward shock, $F_{\nu ,max }^{r}=\Gamma F_{\nu,max }^{f}$;
2) The typical frequency of the minimal electron in the reverse shock is
smaller by a factor of $\Gamma^{2}$, $\nu _{m}^{r}=\nu _{m}^{f}/\Gamma^{2}$;
3) The cooling frequency of the reverse and forward shock are equal,
$\nu_{c}^{r}=\nu _{c}^{f}=\nu_c$ (under the assumption that $\eps_B$ is the 
same in the forward and reverse shocked gas; this might not be true if the
ejecta carries a significant magnetic field from the source);
4) Generally (also in refreshed shocks) $\nu _{a}^{r,f}<\nu _{m}^{r,f}$ and 
$\nu _{a}^{r,f}<\nu _{c}$. The self-absorption frequency of the reverse shock 
is larger than that of the forward shock. The characteristic frequencies
and flux temporal slopes for a standard afterglow are given by the case (r)
with $s=0$ in Table 1 below.

The prompt optical flashes, starting with GRB 990123, have been generally interpreted
\cite{sapi99,mr99,nakarpi05_rev} as the radiation from a reverse (external) shock, 
although a prompt optical flash could be expected from either an internal shock or 
the reverse part of the external shock, or both \cite{mr97a}. The decay rate of the 
optical flux from reverse shocks is much faster (and that of internal shocks is faster 
still) than that of forward shocks, so the emission of the latter dominate after 
tens of minutes \cite{gou03}.  Such bright prompt flashes, however, appear to be 
relatively rare. Other early optical flashes, e.g. in GRB 021004, GRB 021211, 
GRB 041219a, GRB 050904 are also consistent with the reverse shock interpretation
\cite{kobazh03a,zhakm03,fox03a,fox03b,wei03,fanzw05,weiyf06}.
After the launch of Swift, new prompt optical observations with robotic
telescopes have greatly added to the phenomenology of prompt flashes
(see \S \ref{sec:swift}

\subsection{Dependence on external density, injection variability and anisotropy} 
\label{sec:refresh}

If the external medium is inhomogeneous, e.g. $n_{ext} \propto r^{-k}$, the energy
conservation condition is $\Gamma^2 r^{3-k} \sim$ constant, so $\Gamma \propto 
t^{1/(4-k)}$, $r\propto t^{-(3-k)/(8-2k)}$, which changes the temporal decay 
rates \cite{mrw98}.  This might occur if the external medium is a stellar wind 
from the evolved progenitor star of a long burst, e.g. $n_{ext}\propto r^{-2}$,
such light curves fitting some bursts better with this hypothesis 
\cite{chevli00,lichev01}.

Another departure from a simple injection approximation is one where $E_0$ (or
$L_0$) and $\Gamma_0$ are not a simple a delta function or top hat functions.
An example is if the mass and energy injected during the burst duration
$t_{grb}$ (say tens of seconds) obeys $M(>\gamma) \propto \gamma^{-s}$,
$E(>\gamma)\propto \gamma^{1-s}$, i.e. more energy emitted with lower Lorentz
factors at later times, but still shorter than the gamma-ray pulse duration 
\cite{rm98,sames00}. The ejecta dynamics becomes
\beq
\Gamma(r) \propto r^{-(3-k)/(1+s)} \propto t^{-(3-k)/(7+s-2k)}~~,~~
r \propto t^{(1+s)/(7+s-2k)}.
\label{eq:Gammarefr}
\enq 
This can drastically change the temporal decay rate, extending the afterglow
lifetime in the relativistic regime. If can provide an explanation for 
shallower decay rates, if the progressively slower ejecta arrives continuously, 
re-energizing the external shocks (``refreshed" shocks) on timescales comparable 
to the afterglow time scale \cite{rm98,kupi00,dailu00,sames00}. While observational 
motivations for this were present already in the Beppo-SAX era, as discussed in 
the above references, this mechanism has been invoked more recently in order to 
explain the Swift prompt X-ray afterglow shallow decays (see \S \ref{sec:shallow}).  
When the distribution of $\Gamma$ is discontinuous, it can also explain a sudden 
increase in the flux, leading to bumps in the light curve. After the onset, 
the non-standard decay rates for the forward and reverse shock are tabulated
for different cases \cite{sames00} in Table \ref{tab:refr} 
\begin{table*}
\begin{center}
\def\frh{}
\def\dn{2(7+s-2k)}
\begin{tabular}{cccccccccc}
\hl\hl
 & $\nu_m$ & $ F_{\nu_m}$ & $\nu_c$ & $F_\nu$:~$\nu_m<\nu<\nu_c$  & $F_\nu$:~$\nu>\max(\nu_c,\nu_m)$ \cr

\hl

f & -$\frh\frac{24-7k+sk}{\dn}$   & $\frh\frac{6s-6+k-3sk}{\dn}$ &
          -$\frh\frac{4+4s-3k-3sk}{\dn}$& -$\frh\frac{6-6s-k+3sk+\beta(24-7k+sk)}{\dn}$ & 
          -$\frh\frac{-4- 4s+k+sk+\beta(24-7k+sk)}{\dn}$ \cr
\hl
r &  -$\frh\frac{12-3k+sk}{\dn}$   & $\frh\frac{6s-12+3k-3sk}{\dn}$ &
        -$\frh\frac{4+4s-3k-3sk}{\dn}$& -$\frh\frac{12-6s-3k+3sk+\beta(12-3k+sk)}{\dn}$ & 
        -$\frh\frac{8-4 s-3k+sk+\beta(12-3k+sk)}{\dn}$ \cr
\hl
\hl
\end{tabular}
\caption{ Temporal exponents of the peak frequency $\nu_m$, the maximum flux
$F_{\nu_m}$, the cooling frequency $\nu_c$ and the flux in a given bandwidth
$F_\nu$, for the forward (f) and reverse (r) shocks, calculated both in the 
adiabatic regime $\nu_m < \nu <\nu_c$ ($F_\nu\propto F_{\nu_m}(\nu_m/\nu)^\beta 
\propto t^{-\alpha}\nu^{-\beta}$, where $\beta=(p-1)/2$), and in the cooling 
regime $\nu_c<\nu_m<\nu$ ($F_\nu\propto(\nu_c/\nu_m)^{1/2}(\nu_m/\nu)^\beta \propto
t^{-\alpha}\nu^{-\beta}$ where $\beta=p/2$). For $s=1$ this gives the usual (i.e. 
without ``refreshment") forward and reverse shock behavior \cite{sames00,rm98}. }
\label{tab:refr}
\end{center}
\end{table*}

Other types of non-standard decay can occur if the outflow has a transverse 
$\theta$ dependent gradient in the energy or Lorentz factor, e.g. as some power 
law $E\propto \theta^{-a},~\Gamma\propto \theta^{-b}$ \cite{mrw98}. Expressions 
for the temporal decay index $\alpha (\beta,s,d,a,b,..)$ in $F_\nu\propto t^\alpha$
are given by \cite{mrw98,sames00}, which now depend also on $s$, $d$, $a$, $b$, 
etc.  (and not just on $\beta$ as in the standard relation of equ.(\ref{eq:Fnu}).
The result is that the decay can be flatter (or steeper, depending on $s$, $d$, 
etc)) than the simple standard $\alpha= (3/2)\beta$. Such non-uniform outflows 
have been considered more recently in the context of jet breaks based on 
structured jets (\S \ref{sec:jet}).

Evidence for departures from the simple standard model was present even before
the new Swift observations, by e.g. sharp rises or humps in the light curves 
followed by a renewed decay, as in GRB 970508 \cite{ped98,pir98a}, or shallower 
than usual light curve decays. Time-dependent model fits \cite{panmr98} to the 
X-ray, optical and radio light curves of GRB 970228 and GRB 970508 indicated 
that in order to explain the humps, a non-uniform injection or an anisotropic 
outflow is required. Another example is the well-studied wiggly optical light
curve of GRB 030329, for which refreshed shocks provide the likeliest explanation
\cite{granapi03}. Other ways to get light curve bumps which are not too steep
after $\sim$ hours to days is with micro-lensing \cite{garnavich00,graloeb01}, 
late injection \cite{zhames01a,iokakz05}, or inverse Compton effects 
\cite{sariesin01,zhames01b,harrison01}. The changes in the shock physics and 
dynamics in highly magnetized or Poynting dominated outflows were discussed, 
e.g. in \cite{usov94,tho94,mr97b,grankon01,granapi03,lyubland03,zhako05}.
More examples and references to departures from the standard model are 
discussed, e.g. in \cite{jvp00,zhames04}. Departures from spherical symmetry
and jet effects are discussed in the next two subsections.

\subsection{Equal arrival time surface and limb brightening effect}
\label{sec:limb}

As illustrated in Figure \ref{fig:relellipse}, for a distant observer the photons
from a spherically expanding shell are received from an equal-arrival time surface
which is an ellipsoid (if $\Gamma=$ constant). The photons arriving from the line 
of sight originated at larger radii than photons arriving from the light-cone at
$\theta \sim \Gamma$. At smaller radii the outflow had a higher magnetic field
and higher density, so the radiation from the $1/\Gamma$ edge is harder and more 
intense. Thus an interesting effect, which arises even in spherical outflows, 
is that the effective emitting region seen by the observer resembles a ring
\cite{wax97b,goo97,panmes98c,sari98,grapisa99a}. This limb brightening effect
is different in the different power law segments of the spectrum. When one considers 
the change in $\Gamma$ due to deceleration, the ellipsoid is changed into an egg 
shape, which is similarly limb-brightened. This effect is thought to be implicated
in giving rise to the radio diffractive scintillation pattern seen in several
afterglows, since this requires the emitting source to be of small dimensions
(the ring width), e.g. in GRB 970508 \cite{waxfk98}. This provided an  important 
observational check, giving a direct confirmation of the relativistic source 
expansion and a direct determination of the (expected) source size 
\cite{waxfk98,katzps98}. The above treatments were based on the simple 
asymptotic scaling behavior for the Lorentz factor $\Gamma\sim$ constant
at $r\leq r_{dec}$ and $\Gamma \propto r^{-3/2}$ ($\Gamma \propto r^{-3}$)
at $r\simg r_{dec}$ for the adiabatic (fully radiative) cases (\S \ref{sec:shocks}). 
More exact treatments are possible \cite{biancoruf05a,biancoruf05b} based on 
following analytically and numerically the detailed dynamical evolution equations 
for the Lorentz factor through and beyond the transition between pre-deceleration
and post-deceleration. The shape of the equi-temporal surfaces is modified,
and the expected light curves will be correspondingly changed. 
The exact afterglow behavior will depend on the unknown external medium 
density and on whether and what kind of continued of continued energy 
injection into the shock occurs, which introduces an additional layer of 
parameters to be fitted.

\subsection{Jets}
\label{sec:jet}

The spherical assumption is valid even when considering a relativistic outflow
collimated within some jet of solid angle $\Omega_j < 4\pi$, provided the
observer line of sight is inside this angle, and $\Gamma \simg \Omega_j^{-1/2}$
\cite{mlr93},  so the light-cone is inside the jet boundary (causally disconnected)
and the observer is unaware of what is outside the jet. However, as the ejecta is 
decelerated, the Lorentz factor eventually drops below this value, and a change 
is expected in the dynamics and the light curves \cite{rho97,rho99}. It is
thought that this is what gives rise to the achromatic optical light curve
breaks seen in many afterglows \cite{kul99b,fra01}.

The jet opening angle can be obtained form the observer time $t_j$ at which
the flux $F_\nu$ decay rate achromatically changes to a steeper value, assuming 
that this corresponds to the causal (light-cone) angle $\Gamma(t)^{-1}$ having 
become comparable to (and later larger than) the jet half-angle $\theta_j$ \cite{rho97}.
Assuming a standard adiabatic dynamics and a uniform external medium, the jet 
opening half-angle is
\beq
\theta_j \sim 5 {\rm deg}~t_{j,d}^{3/8}E_{53}^{-1/8}n_{ex}^{1/8}
                (\eta_\gamma/0.2)^{1/8}([1+z]/2)^{-3/8}
\label{eq:thetaj}
\enq
where $E_{53}$ is the isotropic equivalent gamma-ray energy in ergs,
$t_{j,d}=t_j/{\rm day}$ and $\eta_\gamma$ is radiative efficiency \cite{fra01}.
The degree of steepening of the observed flux light curve can be estimated
by considering that while the causal angle is smaller than the jet opening 
angle, the effective transverse area from which radiation is received is 
$A\sim r_\perp^2 \sim (r/\Gamma)^2 \propto t^2\Gamma^2$, whereas after the 
causal angle becomes larger than the jet angle, the area is $A\sim r^2\theta_j^2$. 
Thus the flux after the break, for an adiabatic behavior $\Gamma \propto t^{-3/8}$
(valid if there is no sideways expansion) is steeper by a factor $\propto \Gamma^2 
\propto t^{-3/4}$ \cite{mr99}, a value in broad agreement with observed breaks. 
After this time, if the jet collimation is simply ballistic (i.e. not due to 
magnetic or other dynamical effects) the jet can start expanding sideways at 
the comoving (relativistic) speed of sound, leading to a different decay
$\Gamma\propto t^{-1/2}$ and $F_\nu \propto t^{-p} \propto t^{-2}$ \cite{rho99}. 

A collimated outflow greatly alleviates the energy requirements of GRB.
If the burst energy  were emitted isotropically, the energy required spreads
over many orders of magnitude, $E_{\gamma,iso} \sim 10^{51}- 10^{54}$ erg 
\cite{kul99b}.  However, taking into account the jet interpretation of light
curve breaks in optical afterglows \cite{pankum01,fra01,pankum02} the 
spread in the total $\gamma$-ray energy is reduced to one order of magnitude, 
around a less demanding mean value of $E_{\gamma,tot}\sim 1.3 \times 10^{51}$ 
erg \cite{bloofraku03}.
This is not significantly larger than the kinetic energies in core-collapse 
supernovae, but the photons are concentrated in the gamma-ray range, and the
outflow is substantially more collimated than in the SN case. Radiative 
inefficiencies and the additional energy which must be associated with
the proton and magnetic field components increase this value (e.g. the
$\eta_\gamma$ factor in equation [\ref{eq:thetaj}]), but this energy is 
still well within the theoretical energetics $\siml 10^{53.5}-10^{54}$ erg
achievable in {\it either} NS-NS, NS-BH mergers \cite{mr97b} or in collapsar 
models \cite{woo93,pac98,pop99} using MHD extraction of the spin energy of a 
disrupted torus and/or a central fast spinning BH. It is worth noting that 
jets do not invalidate the usefulness of spherical snapshot spectral fits, 
since the latter constrain only the {\it energy per solid angle} \cite{mrw99}.

Equation (\ref{eq:thetaj}) assumes a uniform external medium, which fits
most afterglows, but in some cases a wind-like external medium ($n_{ext}
\propto r^{-2}$) is preferred \cite{pankum00,chevli00,lichev01}. For an
external medium varying as $n_{ext}=Ar^{-k}$ one can show that the the Lorentz
factor initially evolves as $\Gamma \propto (E/A)^{1/2} r^{-(3-k)/2} \propto
(E/A)^{1/(8-2k)} t^{-(3-k)/(8-2k)}$, and the causality  (or jet break) 
condition $\Gamma \sim \theta_j^{-1}$ leads to a relation between the observed 
light curve break time $t_j$ and the inferred collimation angle $\theta_j$
which is different from equation (\ref{eq:thetaj}), namely 
$\theta_j \propto (E/A)^{-1/(8-2k)} (t_j/[1+z])^{(3-k)/(8-2k)} 
\propto (E/A)^{-1/4} (t_j/[1+z])^{1/4}$, where the last part is for $k=2$.
Another argument indicating that the medium in the vicinity of at least some 
long-GRB afterglows is not stratified, e.g. as $r^{-2}$, is the observation
of a sharp jet-break in the optical afterglow lightcurves (as, e.g. in 
GRB 990510, 000301c, 990123). As pointed out by \cite{kupa00}, relativistic 
jets propagating in a wind-like external medium are expected to give rise 
to a very gradual and shallow break in the afterglow lightcurve. 

The discussion above also makes the simplifying assumption of a uniform jet 
(uniform energy and Lorentz factor inside the jet opening angle, or top-hat jet 
model).  In this case the correlation between the inverse beaming factor 
$f_b^{-1}= (\theta_j^2 /2)^{-1}$ (or observationally, the jet break time from 
which $\theta_j$ is derived) and the isotropic equivalent energy or fluence 
$E_{\gamma,iso}$ is interpreted as due to a distribution of jet angles, larger 
angles leading to lower $E_{\gamma,iso}$, according to $E_{\gamma,iso}
\propto \theta_j^{-2}$. There is, however, an equally plausible interpretation 
for this correlation, namely that one could have a universal jet profile such
that the energy per unit solid angle $dE_\gamma/d\Omega \propto \theta^{-2}$,
where $\theta$ is the angle measured from the axis of symmetry 
\cite{rossi02,zhames02b}. (To avoid a singularity, one can assume this law to
be valid outside some small core solid angle). This model also explains the 
\cite{fra01,pankum01} correlation, the different $E_{iso}$ would be due to the 
observer being at different angles relative to the jet axis. This hypothesis 
has been tested in a variety of ways \cite{guetta05,nakar04,kugra03,graku03}.  
Attempts to extend the universal $\theta^{-2}$ jet structure to include X-ray 
flashes (\S \ref{sec:xrfobs}), together with use of the Amati relation between 
the spectral peak energy $E_{peak}$ and $E_{\gamma,iso}$ (\S \ref{sec:correl}), 
leads to the conclusion that a uniform top-hat model is preferred over a 
universal $\theta^{-2}$ jet model \cite{lamb05a}. 
Uniform jets seen off-axis have also been considered as models for XRF in
a unified scheme, e.g. \cite{yamin04,grarrp05}.
On the other hand, another type of universal jet profile with a Gaussian 
shape \cite{zhadai04,daizha05} appears to satisfy both the 
jet break-$E_{\gamma,iso}$ and $E_{peak}-E_{\gamma,iso}$ correlations for 
both GRB and XRFs. More extensive discussion of this is in \cite{zhames04}. 

The uniform and structured jets are expected to produce achromatic breaks
in the light curves, at least for wavebands not too widely separated. However, 
in some bursts there have been indications of different light curve break 
times for widely separated wavebands, e.g. GRB 030329, suggesting different 
beam opening angles for the optical/X-ray and the radio components 
\cite{berger03_twocomp}. Such two-component jets could arise naturally in
the collapsar model, e.g. with a narrow, high Lorentz factor central jet 
producing $\gamma$, X-ray and optical radiation, and a wider slower outflow, 
e.g. involving more baryon-rich portions of the envelope producing 
radio radiation \cite{ramcr02}. A wider component may also be connected to a 
neutron-rich part of the outflow \cite{peng05}. More recent discussions 
of possible chromatic breaks are in \cite{fanpir06,panmes06}. 

\section{Current Theoretical Issues in Afterglow Models}
\label{sec:newag}

The afterglow is generally assumed to become important after the time
when the self-similar $\Gamma\propto r^{-3/2}$ behavior starts. From
equation (\ref{eq:duration} for the deceleration time $t_{dec}\sim 
(r_{dec}/2c \Gamma^2)$ and taking into account the gradual transition to 
the self-similar regime \cite{koba06}, this is approximately
\beq
t_{ag} \sim T \sim {\rm Max}[ t_{grb}(1+z)~,~ 
15 (E_{53}/n_0)^{1/3}\eta_{2.5}^{-8/3}[(1+z)/2]~{\rm s}]~,
\label{eq:tag}
\enq
where $t_{grb}$ is the duration of the outflow, i.e. an upper limit for 
the duration of the prompt $\gamma$-ray emission, and a cosmological time 
dilation factor is included. (Note that in some 
bursts the $\gamma$-rays could continue in the self-similar phase). 
The afterglow emission from the forward and the reverse shock emission 
starts immediately after the ejecta starts to sweeps up matter, but it 
does not peak (and become dominant over the prompt emission or its decaying
tail) until the time $\sim t_{ag}$, marking the beginning of the self-similar 
blast wave regime.

Denoting the frequency and time dependence  of the afterglow spectral energy 
flux as $F_\nu(t)\propto \nu^{-\beta}t^{-\alpha}$, the late X-ray afterglow 
phases (3) and (4) of \S \ref{sec:swift} seen by Swift are similar to those 
known previously from Beppo-SAX (the theoretical understanding of which is 
discussed in \S \ref{sec:ag} and in \cite{zhames04}). The ``normal" decay phase 
(3), with temporal decay indices $\alpha \sim 1.1-1.5$ and spectral energy 
indices $\beta\sim 0.7-1.0$, is what is generally expected from the evolution 
of the forward shock in the Blandford-McKee self-similar late time regime, 
under the assumption of synchrotron emission.

\subsection{Early steep decay}
\label{sec:steep}

Among the new afterglow features  detected by Swift (see Figure \ref{fig:xrt-lc}), 
the steep initial decay phase $F_\nu \propto t^{-3}- t^{-5}$ in X-rays of the long 
GRB afterglows is one of the most striking. There are several possible mechanisms 
which could cause this. The most obvious first guess would be to attribute 
it to the cooling following cessation of the prompt emission (internal shocks
or dissipation). If the comoving magnetic field in the emission region is
random [or transverse], the flux per unit frequency along the line of sight
in a given energy band, as a function of the electron energy index $p$,
decays as $F_\nu \propto t^{-\alpha}$ with $\alpha={-2p}~[(1-3p)/2]$ in the
slow cooling regime, where $\beta=(p-1)/2$, and it decays as
$\alpha=-2(1+p),~[-(2-3p)/2]$ in the fast cooling regime where $\beta=p/2$,
i.e. for the standard $p=2.5$ this would be $\alpha=-5,~[-3.25]$ in the
slow cooling or $\alpha=-7,~[-2.75]$ in the fast cooling regime, for
random [transverse] fields \cite{mr99}. In some bursts this may be the 
explanation, but in others the time and spectral indices do not correspond well.
In addition, if the flux along the line of sight decays as steeply as above,
the observed flux would be dominated by the so-called high latitude emission,
which is discussed next.

At present, the most widely considered explanation for the fast decay, 
both of the initial phase (1) and of the steep flares, attributes it to the 
off-axis emission from regions at $\theta >\Gamma^{-1}$ (the curvature 
effect, or high latitude  emission \cite{km00}. In this case, after
the line of sight gamma-rays have ceased, the off-axis emission observed
from $\theta>\Gamma^{-1}$ is $(\Gamma\theta)^{-6}$ smaller than that from
the line of sight. Integrating over the equal arrival time region, this
flux ratio becomes $\propto (\Gamma\theta)^{-4}$. Since the emission from
$\theta$ arrives $(\Gamma\theta)^2$ later than from $\theta=0$, the observer
sees the flux falling as $F_\nu\propto t^{-2}$, if the flux were frequency 
independent. For a source-frame flux $\propto \nu'^{-\beta}$, the observed 
flux per unit frequency varies then as 
\beq 
F_\nu\propto (t-t_0)^{-2-\beta}
\label{eq:hilat}
\enq
i.e. $\alpha=2+\beta$. This ``high latitude" radiation, which for observers 
outside the line cone at $\theta > \Gamma^{-1}$ would appear as prompt 
$\gamma$-ray emission from dissipation at radius $r$, appears to observers 
along the line of sight (inside the light cone) to arrive delayed by $t\sim 
(r\theta^2/2c)$ relative to the trigger time, and its spectrum is softened 
by the Doppler factor $\propto t^{-1}$ into the X-ray observer band. 
For the initial prompt decay, the onset of the afterglow (e.g. phases
2 or 3), which also come from the line of sight, may overlap in time
with the delayed high latitude emission.  In equation (\ref{eq:hilat}) 
$t_0$ can be taken as the trigger time, or some value comparable or
less than equation (\ref{eq:tag}). This can be used to constrain the
prompt emission radius \cite{lazbeg05b}. When $t_{dec}<T$, the emission can 
have an admixture of high latitude and afterglow, and this can lead to
decay rates intermediate between the two \cite{obrien06}.
Values of $t_0$ closer to the onset of the decay also lead to steeper slopes. 
It is possible to identify for various bursts values of $t_0$ near the
rising part of the last spike in the prompt emission which satisfy the
subsequent steep decay slope \cite{liazha06}.  Structured jets, when
viewed on-beam produce essentially the same slopes as homogeneous jets,
while off-beam observing can lead to shallower slopes \cite{dyks05}.
For the flares, if their origin is assumed to be internal (e.g. some form of
late internal shock or dissipation) the value of $t_0$ is just before the 
flare, e.g the observer time at which the internal dissipation starts to be 
observable \cite{zh05_md05}.  This interpretation appears, so far, compatible 
with most of the Swift afterglows \cite{zhang06_ag,nousek06,pan06_ag}. 

Alternatively, the initial fast decay may be due to the emission of 
a cocoon of exhaust gas \cite{peer06}, where the temporal and spectral 
index are explained through an approximately power-law behavior of escape 
times and spectral modification of multiply scattered photons.
The fast decay may also be due to the reverse shock emission, if
inverse Compton up-scatters primarily synchrotron optical photons into
the X-ray range. The decay starts after the reverse shock has crossed
the ejecta and electrons are no longer accelerated, and may have both a
line of sight and an off-axis component \cite{koba05}.
This poses strong constraints on the Compton-y parameter, and cannot
explain decays much steeper than $\alpha=-2$, or $-2-\beta$ if the 
off-axis contribution dominates. Models involving bullets, whose origin, 
acceleration and survivability is unexplained, could give a prompt decay 
index $\alpha \sim -3$ to $-5$ \cite{dado05}, with a bremsstrahlung energy 
index $\beta \sim 0$ which is not observed in the fast decay; switching to 
a synchrotron or IC mechanisms requires additional parameters. 
Finally, a patchy shell model, where the Lorentz factor is highly 
variable in angle, would produce emission with $\alpha \sim -2.5$. Thus, such 
mechanisms may explain the more gradual decays, but not the more extreme 
$\alpha=-5,-7$ values encountered in some cases. It should be noted, however,
that the Swift X-ray observations suggest that the steep decay is a direct
continuation of the prompt emission \cite{obrien06}, which in turn suggests
that the prompt and the fast decaying emission arise from the same physical 
region, posing a problem for the models in this paragraph (but not for the
high latitude emission model).

\subsection{Shallow decay}
\label{sec:shallow}

The slow decay portion of the X-ray light curves ($\alpha\sim -0.3-0.7$), 
ubiquitously detected by Swift, is not entirely new, having been detected 
in a few cases by BeppoSAX. This, as well as the appearance of wiggles and 
flares in the X-ray light curves after several hours were the motivation for 
the ``refreshed shock" scenario \cite{rm98,sames00} (\S \ref{sec:refresh}). Refreshed 
shocks can flatten the afterglow light curve for hours or days, even if the 
ejecta is all emitted promptly at $t=T \siml t_\gamma$, but with a range of 
Lorentz factors, say $M(\Gamma) \propto \Gamma^{-s}$, where the lower 
$\Gamma$ shells arrive much later to the foremost fast shells which have already
been decelerated.  Thus, for an external medium of density $\rho\propto r^{-k}$ 
and a prompt injection where the Lorentz factor spread relative to ejecta mass and 
energy  is $M(\Gamma)\propto \Gamma^{-s}$, $E(\Gamma)\propto \Gamma^{-s+1}$, the 
forward shock flux  temporal decay is given by \cite{sames00}
\beq
\alpha=[(k-4)(1+s)+\beta(24 -7k +sk)]/[2(7+s-2k)]~,
\label{eq:shallow}
\enq
(for more details, see Table \ref{tab:refr}).
It needs to be emphasized that in this model all the ejection can be prompt
(e.g. over the duration $\sim T$ of the gamma ray emission) but the low $\Gamma$
portions arrive at (and refresh) the forward shock at late times, which can
range from hours to days. I.e., it is not the central engine which is active
late, but its effects are seen late.  Fits of such refreshed shocks to observed 
shallow decay phases in Swift bursts \cite{grankum05} lead to a $\Gamma$ 
distribution which is a broken power law, extending above and below a peak 
around $\sim 45$.

There is a different version of refreshed shocks, which does envisage central 
engine activity extending for long periods of time, e.g. $\siml$ day (in contrast 
to the $\siml$ minutes engine activity in the model above). Such long-lived 
activity may be due to continued fall-back into the central black hole 
\cite{woo05_md05} or a magnetar wind \cite{zhames01a,dai04,nousek06}. 
One characteristic of both types of refreshed models is that after the 
refreshed shocks stop and the usual decay resumes, the flux level shows a 
step-up relative to the previous level, since new energy has been injected.

From current analyses, the refreshed shock model is generally able to explain 
the flatter temporal X-ray slopes seen by Swift, both when it is seen to join 
smoothly on the prompt emission (i.e.  without an initial steep decay phase) 
or when seen after an initial steep decay. Questions remain concerning the
interpretation of the fluence ratio in the shallow X-ray afterglow and the 
prompt gamma-ray emission, which can reach $\siml 1$ \cite{obrien06}. This 
requires a higher radiative efficiency in the prompt gamma-ray emission than 
in the X-ray afterglow. One might speculate that this might be achieved if 
the prompt outflow is Poynting-dominated, or if a more efficient afterglow 
hides more of its energy in other bands, e.g. in GeV, or IR. Alternatively
\cite{ioka05,grankopi06}, a previous mass ejection might have emptied a cavity 
into which the ejecta moves, leading to greater efficiency at later times
(although this would not work above the cooling frequency, which from the
spectrum appears to be required in about half the cases), or otherwise the 
energy fraction going into the electrons increases $\propto t^{1/2}$. Other
possible ways of addressing this include the afterglow coming from off-axis 
directions \cite{eicgra06}, and exploring plausible reasons for having
underestimated in previous studies the energy of the ejecta \cite{grankopi06}. 

\subsection{X-ray flares}
\label{sec:flares}

Refreshed shocks can also explain some of the X-ray flares whose rise and decay
slopes are not too steep. However, this model encounters difficulties with
the very steep flares with rise or decay indices  $\alpha\sim \pm 5,~\pm 7$, 
such as inferred from the giant flare of GRB 0500502b \cite{burr05a} around 300 s 
after the trigger. Also, the flux level increase in this flare is a factor 
$\sim 500$ above the smooth afterglow before and after it, implying a comparable 
energy excess in the low versus high $\Gamma$ material. An explanation based
on inverse Compton scattering in the reverse shock \cite{koba05} can explain a
single flare at the beginning of the afterglow, with not too steep decay.
For multiple flares, models invoking encountering a lumpy external medium 
have generic difficulties  explaining steep rises and decays 
\cite{nakpir03,zhang06_ag}, although extremely dense, sharp-edged  lumps, 
if they exist, might satisfy the steepness \cite{dermer05_md05}. 

Currently the more widely considered model for the flares ascribes them to 
late central engine activity \cite{zhang06_ag,nousek06,pan06_ag}. 
The strongest argument in favor of this is that the energy budget is 
more easily satisfied, and the fast rise/decay is straightforward to explain.
In such a model the flare energy can be  comparable to the prompt emission, 
the fast rise comes naturally from the short time variability leading to
internal shocks (or to rapid reconnection), while the rapid decay may be
due to the high latitude emission following the flare, with $t_0$ reset to 
the beginning of each flare (see further discussion in \cite{zh05_md05}).
However, some flares are well modeled by refreshed forward shocks, while
in others  this is clearly ruled out and a central engine origin is
better suited \cite{wu05}. Aside from the phenomenological desirability 
based on energetics and timescales, a central engine origin is conceivable, 
within certain time ranges, based on numerical models of the burst origin in 
long bursts. These are interpreted as being due to core collapse of a massive 
stellar progenitor, where continued infall into fast rotating cores can continue
for a long time \cite{woo05_md05}. However, large flares with a fluence which 
is a sizable fraction of the prompt emission occurring hours later remain 
difficult to understand. It has been argued that gravitational instabilities 
in the infalling debris torus can lead to lumpy accretion \cite{perna06}.
Alternatively, if the accreting debris torus is dominated by MHD effects,
magnetic instabilities can lead to extended, highly time variable accretion
\cite{proga03}, which may give rise to GRB X-ray flares \cite{progazh06}.

\subsection{Late steep decay and jet breaks}
\label{sec:jetbreak}

The late steep decay decay phase (4) of \S \ref{sec:swift} is seen in a 
modest fraction of the Swift bursts, mainly in X-rays, and mainly but 
not exclusively in long bursts. The natural interpretation
is that these are caused by the fact that the outflow is collimated into 
a jet break: when the decrease of the ejecta Lorentz factor leads to the 
light-cone angle becoming larger than the jet angular extent, $\Gamma_j(t) 
\simg 1/\theta_j$ (e.g. \S \ref{sec:jet}), the light curve steepens 
achromatically. For the Swift bursts, it is noteworthy that this final 
steepening has been seen in less than $\sim 10\%$ of the afterglows followed, 
and then with reasonable confidence mainly in X-rays. The corresponding 
optical light curve breaks have been few, and not well constrained.  
The UVOT finds afterglows in only $\sim 30\%$ of the bursts, and ground-based 
optical/IR telescopes have yielded few continued late time light curves
monitored. This is unlike the case with the $\sim 20$ Beppo-SAX bursts, for 
which an achromatic break was reported in the optical \cite{fra01}, while 
in rarer cases there was an X-ray or radio break reported, which in a few 
cases appeared to occur at a different time than the optical break 
(e.g. \cite{berger03}). 

The relative paucity of optical breaks in Swift afterglows may be an 
observational selection effect due to the larger median redshift, and 
hence fainter and redder optical afterglow at the same  observer epoch. 
At higher redshifts the break occurs later in the observer frame, which 
compounds a possible reluctance to commit large telescope time on more 
frequently reported bursts (roughly 2/week from Swift versus an earlier 
2/month with Beppo-SAX).  One can speculate that the apparent scarcity
of detected light curve breaks might indicate that at higher redshifts 
the opening angle is intrinsically larger. However, continued monitoring
of the X-ray light curves with both Swift and Chandra is resulting in a
growing number of bursts with  high quality late X-ray light curves 
showing in some cases a clear break, and others the absence of a break
up to weeks (also in short bursts, e.g. \cite{burr06_051221a,grupe06_050724}).
This is an evolving topic, with some indications that light curve breaks 
may not (or not always) appear achromatic \cite{fanpir06,panmes06}.

\subsection{Prompt optical flashes and high redshift afterglows}
\label{sec:promptopt}

Optical/UV afterglows have been detected with the Swift UVOT telescope in
roughly half the bursts for which an X-ray afterglow was seen. For a more
detailed discussion of the UVOT afterglow observations see \cite{roming05_md05}.
Of particular interest is the ongoing discussion on whether ``dark GRB" are 
really optically deficient, or the result of observational bias \cite{berger05a}.
Another puzzle is the report of a bimodal intrinsic brightness distribution in 
the rest-frame R-band \cite{liangzh06,nard05}.  This suggests possibly the 
existence of two different classes of long bursts, or at least two different 
types of environments. 

Compared to a few years ago, a much larger role is being played by 
ground-based robotic optical follow-ups, due to the increased rate of 
several arc-second X-ray alerts from XRT, and the larger number of robotic 
telescopes brought on-line in the last years. For the most part, these 
detections have yielded optical decays in the $\simg$ few 100 s range, 
initial brightness $m_V\sim 14-17$ and temporal decay slopes $\alpha\sim 
1.1-1.7$ previously associated with the evolution of a forward shock 
\cite{fox05_md05,berger05_md05}. In a few cases, a prompt optical detection 
was achieved in the first 12-25 s \cite{rykoff05a,rykoff05b,vestrand05}.

\begin{figure}[ht]
\begin{center}
\centerline{\epsfxsize=5.in \epsfbox{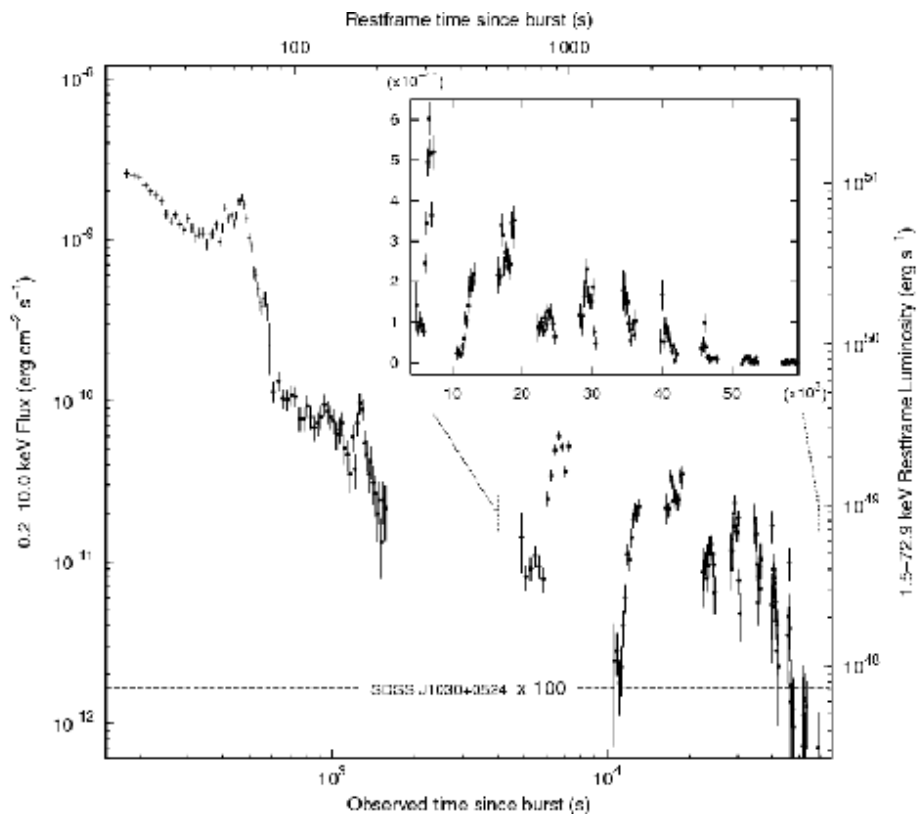}}
\end{center}
\caption{The X-ray afterglow of the GRB 050094 at $z=6.29$ \cite{wat05}, 
showing for comparison the flux level of one of the most lumnious X-ray 
quasars at a comparable redshift, SDSS J1030+524 (multiplied by 100).
The inset shows the GRB variability in the 10-70 ks timeframe.}
\label{fig:050904_xrlc}
\end{figure}

The most exciting prompt robotic IR detection (and optical non-detection)
is that of GRB 050904 \cite{boer05,haislip05}. This object, at the
unprecedented high redshift of $z=6.29$ \cite{kawai05}, has an X-ray brightness 
exceeding for a day that of the brightest X-ray quasars (see Figure 
\ref{fig:050904_xrlc}) \cite{wat05}. 
Its O/IR brightness in the first 500 s (observer time) was comparable to that
of the extremely bright ($m_V\sim 9$) optical flash in GRB 990123, with a 
similarly steep time-decay slope $\alpha\sim 3$ \cite{boer05}. Such prompt, 
bright and steeply decaying optical emission is expected from the reverse shock 
as it crosses the ejecta, marking the start of the afterglow \cite{mr97a,sapi99,mr99}.

However, aside from the two glaring examples of 990123 and 050904, in the last
six years there have been less than a score of other prompt optical flashes, 
typically with more modest initial brightnesses $m_v\simg 13$. There are a  
number of possible reasons for this paucity of optically bright flashes, if
ascribed to reverse shock emission. One is the absence or weakness of a reverse 
shock, e.g. if the ejecta is highly magnetized \cite{mr97a}. A moderately 
magnetized ejecta is in fact favored for some  prompt flashes \cite{zhakm03}. 
Alternatively, the deceleration might occur in the thick-shell regime 
($T \gg t_{dec}$. see eq. (\ref{eq:tag}), which can result in the reverse 
shock being relativistic, boosting the optical reverse shock spectrum into 
the UV \cite{kob00} (in this case a detection by UVOT might be expected, unless
the decay is faster than the typical 100-200 s for UVOT slewing and integration).
Another possibility, for a high comoving luminosity, is copious pair formation 
in the ejecta, causing the reverse shock spectrum to peak in the IR \cite{mrrz02}. 
Since both GRB 990123 and GRB 050904 had 
$E_{iso}\sim 10^{54}$ erg, among the top few percent of all bursts, the latter 
is a distinct possibility, compatible with the fact that the prompt flash in 
GRB 050904 was bright in the IR I-band but not in the optical. On the other 
hand, the redshift $z=6.29$ of this burst, and a Ly-$\alpha$ cutoff at 
$\sim 800$ nm would also ensure this (and GRB 990123, at $z=1.6$, was detected 
in the V-band). However, the observations of optical flashes in these two objects 
but not in lower $E_{iso}$ objects appears compatible with having a relativistic 
(thick shell) reverse shock with pair formation. Even in the absence of pairs,
more accurate calculations of the reverse shock \cite{nakarpi04,mcmahon06} 
find the emission to be significantly weaker than was estimated earlier.
Another possibility is that the cooling frequency in reverse shock is 
typically not much larger than the optical band frequency. In this case the 
optical emission from the reverse shock drops to zero very rapidly soon after
the reverse shock has crossed the ejecta and the cooling frequency drops below 
the optical and there are no electrons left to radiate in the optical band 
\cite{mcmahon06}.

\section{Short GRB in the Swift Era}
\label{sec:short}

\subsection{Short GRB observations}
\label{sec:shortobs}

\begin{figure}[ht]
\begin{center}
\centerline{\epsfxsize=5.in \epsfbox{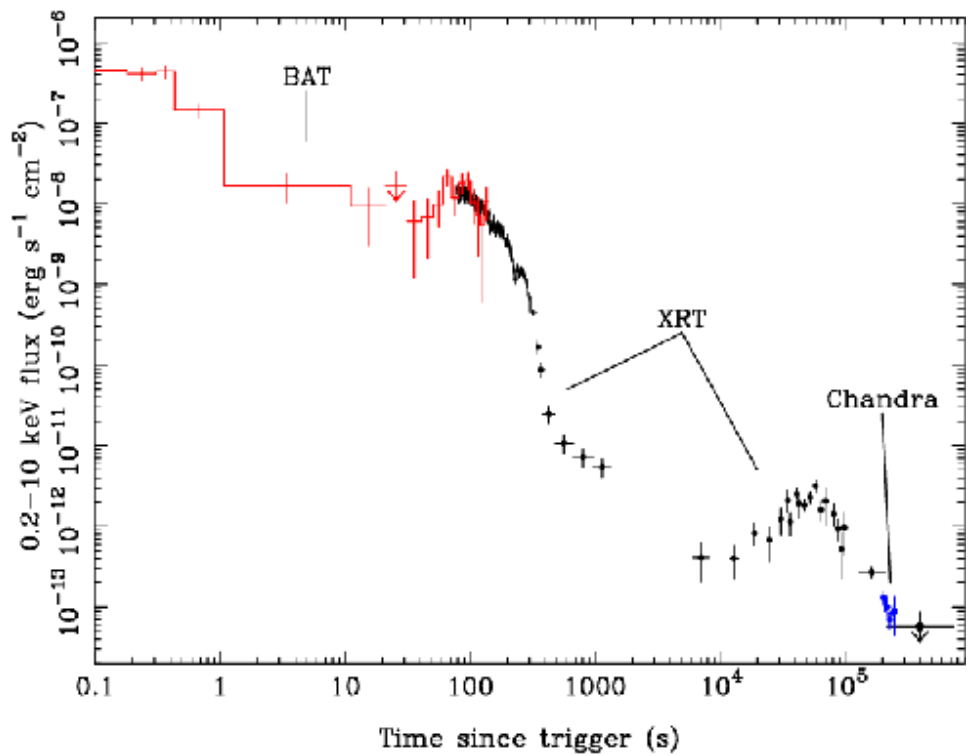}}
\end{center}
\caption{The afterglow of GRB 050724 \cite{barth05_0724}, showing the 
Swift results on the prompt BAT emission extrapolated to the X-ray range 
and the subsequent XRT emission, as well as the late Chandra follow-up.} 
\label{fig:050724_xrlc}
\end{figure}

Swift, and in smaller numbers HETE-2, have provided the first bona fide 
short burst X-ray afterglows followed up starting $\sim 100$ s after the 
trigger, leading to localizations and  redshifts. In the first of these, 
GRB 050509b \cite{gehr05_0509} the extrapolation of the prompt BAT emission 
into the X-ray range, and the XRT light curve from 100 s to about 1000 s 
(after which only upper limits exist, even with Chandra, due to the 
faintness of the burst) can be fitted with a single power law of 
$\alpha \sim$ 1.2 (1.12 to 1.29 90\% conf), or separately as $\alpha_{BAT}=$
1.34 (0.27 to 2.87 90\% conf) and $\alpha_{XRT}=$1.1 (0.57 to 2.36 90\% conf).
The X-ray coverage was sparse due to orbital constraints, the number of X-ray 
photons being small, and no optical transient was identified, probably due to 
the faintness of the source.  An optical host was however identified, an
elliptical galaxy \cite{bloo06}. The next one, discovered by HETE-2, was 
GRB 050709 \cite{vil06_050709}. Its host \cite{fox05_0709} is an irregular 
galaxy at $z=0.16$ (and the observations ruled out any supernova association).
Even earlier, HETE-2 reported the short GRB 040924 \cite{van04_040924}, with 
a soft gamma-ray prompt emission and a faint broken power law optical 
afterglow \cite{huang05_040924}.  A proposed host galaxy at $z=0.86$ shows 
star formation, and evidence for an associated 1998bw-like SN contribution 
to the light curve \cite{sod06_040924}, which suggests this is perhaps the 
short end of the long burst or XRF distribution. The next Swift short burst, 
GRB 050724, was relatively bright, and besides X-rays, it also yielded both 
a decaying optical and a radio afterglow \cite{berger05_4sho}. This burst, 
together with  a significant part of other short bursts, is associated 
with an elliptical host galaxy. It also had a low-luminosity soft 
gamma-ray extension of the short hard gamma-ray component (which would 
have been missed by BATSE), and it had an interesting  X-ray afterglow 
extending beyond $10^5$ s \cite{barth05_0724} (Figure \ref{fig:050724_xrlc}). 
The soft gamma-ray extension, 
lasting up to 200 s, when extrapolated to the X-ray range overlaps well with 
the beginning of the XRT afterglow, which between 100 and 300 s has 
$\alpha\sim -2$, followed by a much steeper drop $\alpha\sim -5-7$ out to 
$\sim 600 s$, then a more moderate decay $\alpha \sim -1$. An unexpected 
feature is a strong flare peaking at $5\times 10^4$ s, whose energy is 
10\% of the prompt emission, while its amplitude is a 10 times increase 
over the preceding slow decay.  Among more recent Swift short bursts, such 
as GRB 050813 \cite{retter05_050813,sato05_050813} had an X-ray afterglow, 
a possible elliptical host, and was reported to be near a galaxy cluster at 
$z=1.7-1.9$ \cite{berger05_md05}. GRB 051210 \cite{lap06_05120} had an X-ray 
power law afterglow, with bumps or flares, and optical identifications still 
under consideration. GRB 051221a had X-ray and optical afterglows , and the 
host is a star forming galaxy at $z=0.55$ \cite{sod06_051221a}.

\subsection{Short GRB prompt and afterglow emission}
\label{sec:shortemission}

The main challenges for an understanding of the mechanism of short bursts 
are the relatively long, soft tail of the prompt emission, and the strength 
and late occurrence of the X-ray bumps or flares. A possible explanation 
for the extended long soft tails ($\sim 100 s$)  may be that the 
compact binary progenitor is a black hole - neutron star system 
\cite{barth05_0724}, for which analytical and numerical arguments
(\cite{davies04}, and references therein)  suggest that the disruption and
swallowing by the black hole may lead to a complex and more extended accretion 
rate than for double neutron stars (c.f. \cite{mil05,fab06}). The flares, for 
which the simplest interpretation might be to ascribe them to refreshed shocks 
(compatible with a short engine duration $T\siml t_\gamma \sim 2$ s and a
distribution of Lorentz factors), requires the energy in the slow material to 
be at least ten times as energetic as the fast material responsible for the 
prompt emission, for the GRB 050724 flare at $10^4$ s. The rise and decay 
times are moderate enough for this interpretation within the errors. On the 
other hand, if the decay slope is -2.8, this is steeper than expected for
refreshed shocks, but consistent with the high-latitude $-2-\beta$ model; 
a time origin $t_0$ can be determined at the beginning of the flare, and
late Chandra observations indicate that the decay after the resumes where
it had left off before the flare, which is more consistent with a late
engine activity interpretation \cite{liazha06}, requiring a factor 10 less
energy budget than the refreshed shock interpretation.  Another interpretation for 
such flares might be an accretion-induced collapse of a white dwarf in a binary, 
leading to a flare when the fireball created by the collapse hits the companion 
\cite{macfad05}, which might explain moderate energy one-time flares
of duration $\siml 10^2$ s.
However, for repeated, energetic flares, as also in the long bursts, the 
total energetics are easier to satisfy if one postulates late central 
engine activity (lasting at least half a day), containing $\sim 10\%$ of 
the prompt fluence \cite{barth05_0724}. A possible way to produce this 
might be temporary choking up of an MHD outflow \cite{proga03} (c.f. 
\cite{vanost01}), which might also imply a linear polarization of the 
X-ray flare \cite{fan05}. Such MHD effects could plausibly also explain 
the initial $\sim 100$ s soft tail. Another magnetic mechanism proposed for 
late X-ray flares in short bursts invokes a temporary post-merger massive
neutron star \cite{daiwwz06}.  However, a justification for substantial
$\simg 10^5$ s features remains so far on rather tentative grounds.

The similarity of the X-ray afterglow light curve with those of long bursts 
is, in itself, an argument in favor of the prevalent view that the afterglows 
of both long and short bursts can be described by the same paradigm, 
independently of any difference in the progenitors. This impression is 
reinforced by the fact that the X-ray light curve temporal slope is, on 
average, that expected from the usual forward shock afterglow model, and that 
in GRB 050724 the X-ray afterglow shows what appears like an initial steep
decay, a normal decay and a significant bump or flare. The identification of 
jet breaks in short bursts is  still preliminary, and the subject of debate.
In two short bursts (so far) evidence evidence has been reported for a jet 
break \cite{berger05_4sho,pan05_sho,burr06_051221a}. (However, in GRB 050724 
a late Chandra observation indicates no X-ray break \cite{grupe06_050724}). 
Taking these breaks as jet breaks, the average isotropic energy of these SHBs 
is a factor $\sim 100$ smaller, while the average jet opening angle (based
on the two breaks) is a factor $\sim 2$ larger than those of typical long GRBs
\cite{fox05_md05,pan05_sho}. Using standard afterglow theory, the bulk Lorentz 
factor decay can be expressed through $\Gamma(t_d) =6.5(n_o/E_{50})^{1/8} 
t_d^{-3/8}$, where $t_d=(t/{\rm day})$, $n_o$ is the external density in units 
of cm$^{-3}$, and $E_{50}$ is the isotropic equivalent energy in units of 
$10^{50}$ ergs.  If the jet break occurs at $\Gamma(t_j)=\theta_j^{-1}$, for a
single-sided jet the jet opening angle and the total jet energy $E_j$ are
\beq
\theta_j=  9^o (n_o/E_{50})^{1/8} t_{j,d}^{3/8}~~,
E_j =  \pi \theta_j^2 E \sim 10^{49} n_o^{1/4} (E_{50} t_{j,d})^{3/4}~{\rm erg}~.
\label{eq:short}
\enq
For the afterglows of GRB 050709 and GRB 050724, the standard afterglow 
expressions for the flux level as a function of time before and after the break 
lead to fits \cite{pan05_sho} which are not completely determined, allowing for 
GRB050709 either a very low or a moderately low external density, and for 
GRB050724 a moderately low to large external density. The main uncertainty is 
in the jet break time, which is poorly sampled, and so far mainly in X-rays.
A better determined case of an X-ray light curve break is that of GRB 051221a,
where combined Swift XRT and Chandra observations indicate a late break at
$t_j\sim 5$ days, leading to an estimated $\theta_j\sim 15$ degrees 
\cite{burr06_051221a}. This is similar to jet angles calculated numerically
for compact merger scenarios by \cite{janka05,aloy05}. It is worth noting,
however, that there are some indications that light curve breaks may not (or 
not always) be achromatic \cite{fanpir06,panmes06}. We note that chromatic breaks 
have been argued for in some long bursts, e.g. GRB 030329, suggesting different
beam opening angles for the optical/X-ray and the radio components 
(\cite{berger03_twocomp}; see also \cite{peng05}), and independently of whether 
this is the explanation, a similar phenomenon may be present in short bursts.

\subsection{Short burst hosts and progenitors}
\label{sec:shortprogenitor}

The most dramatic impact of Swift concerning short GRB, after the 
discovery and characterization of the afterglows, has been in providing
the first significant identifications of host galaxies, with the 
implications and constraints that this puts on the progenitor issue.
Out of ten short bursts detected until the 
end 2005, four of the hosts (GRB 040924, 050509b, 050724 and 050813; 
\cite{gehr05_0509,barth05_0724,berger05_4sho}) are elliptical galaxies, 
one (GRB 050709, \cite{fox05_0709}) is a nearby irregular galaxy, and one 
(GRB 050906, \cite{jakob_050906_host}) is a star-forming galaxy.
The number of elliptical hosts is of significant interest for the most 
frequently discussed progenitor of short GRB, the merger of neutron star 
binaries \cite{pac86,eic89,barth05_0724,leerrg05}, which would be relatively 
more abundant in old stellar population galaxies such as ellipticals. 
The argument partly depends on the expected long binary 
merger times, which in early population synthesis and merger simulations 
\cite{bloom99} was taken to be in excess of $10^8$ years. More
recent populations synthesis calculations \cite{belcz02} have
reduced this to the point where compact mergers could be expected in
substantial numbers also in young, e.g. star-forming galaxies, although 
statistically most mergers would be expected in old galaxies. 
The preponderance of claimed elliptical hosts, where star formation 
is absent, argues against alternative short burst origins such as
short-lived outflows from massive stellar collapses \cite{van03}.
The lack of any observed supernova emission weeks after the burst
\cite{hjorth05_050509b,fox05_0709} also argues against a massive stellar 
collapse (where a Ib/c supernova could be expected), and also against a
gravitational collapse of C/O white dwarfs to neutron stars leading to
a supernova Ia \cite{dado05_snIa}.  

An alternative interpretation of short bursts is that they may be the 
initial brief, hard spike seen in giant flares of soft gamma 
repeaters, or SGR \cite{hurl05_sgr,palm05_sgr}.  SGRs must be young 
objects, due to the fast field decay rate, and the total energy in giant 
SGR flares detected so far is at least two orders of magnitude too 
small to explain the short burst fluxes detected at $z\simg 0.2$.
The lack of recent star formation activity in the four mentioned 
elliptical hosts also indicates that at least some short bursts cannot
be ascribed to SGRs. A statistical analysis indicates that the fraction
of short bursts which could be due to SGRs is less than 15\% (or less than 
40\% at 95\% confidence level) \cite{nakar06_sgr}.
It is interesting that a correlation analysis of short bursts with X-ray 
selected galaxy clusters \cite{ghirl06_shb} gives a better than $2\sigma$ 
angular cross-correlation with clusters up to $z=0.1$, which compared
to model predictions would indicate that most short bursts originate
within $\sim 270$ Mpc. Any connection between alternative candidates
and a possible third category of bursts, intermediate between short and long 
\cite{horv98_third,muk98_third,mesz00_third_sky,horv06_third} remains so 
far unexplored.

\subsection{Short burst redshifts and progenitor lifetimes}
\label{sec:shortdistr}

With over a half dozen reasonably well studied short bursts (as of end of 
2005), their distribution in redshift space and among host galaxy types, 
including both ellipticals and spiral/irregulars \cite{proch06}, is similar 
to that of other old population objects, and thus is compatible with neutron 
star binaries or black hole-neutron star binaries \cite{nakar06_sho}. This 
progenitor identification, however, cannot be considered secure, so far. 
Nonetheless, the most striking thing about short hard burst (or SHBs) hosts 
is that it includes a number of ellipticals, with low star formation rate (SFR), 
e.g. 050909b, 050724; and even for those SHBs with star forming hosts, e.g. 
050709, 051221a, the SFR is lower than the median SFR for the long GRB hosts 
\cite{berger05_md05}. This confirms that they are a distinct population, as 
indicated also by their intrinsic spectral-temporal properties versus those of 
long bursts \cite{kou93,norris00,balazs03}. Using the BATSE flux distribution 
and the observed redshifts, the SHB local rates  are inferred to be at least
$\sim 10$ Gpc$^{-3}$ yr$^{-1}$ \cite{nakar06_sho,guepir06_sho} without beaming
corrections, and larger including beaming. The progenitor lifetimes lead to 
interesting constraints, e.g. the simple time delay distribution $P(t)\propto 
t^{-1}$ expected from galactic double neutron star systems appears in conflict 
with the low average redshift of SHBs \cite{galyam06_sho,nakar06_sho}, although
it is not ruled out \cite{guepir06_sho}. This has led to inferring a typical
lifetime for the progenitors of $\sim 6$ Gyr, and the suggestion that they
might be neutron star-black hole binaries, rather than double neutron stars.
However if the redshift $z\approx 1.8$ for GRB 050813 is correct, the lifetime
of the progenitor would be constrained to $\siml 10^3$ Gyr \cite{berger05_md05}.
On the other hand, consideration of the star formation history of both early
and late type galaxies suggests that at least half of the SHB progenitors have
lifetimes in excess of $\sim 10$ Gyr \cite{zhengrr06}.
Population synthesis models of double compact binaries \cite{belcz06_sho}
indicate two populations, with short ($10^{-2}-0.2$ Myr and long ($10^2-10^4$ Myr) 
merger times, with NS-NS and BH-NS binaries distributed roughly 1:1 and 4:1 
between these two merger time ranges, in apparent agreement with current
SHB redshift and host distributions between ellipticals and SFR galaxies.
The origin of a fraction of double neutron stars in globular clusters
\cite{grind06_sho} would help to explain short bursts which are offset from 
their host galaxy.

\section{Long GRB Progenitors in light of Swift}
\label{sec:progenitor}

\subsection{Long GRB hosts and progenitors}
\label{sec:longprog}

Out of the $\sim 90$ long bursts ($t_\gamma \simg 2$ s) detected by Swift up 
to the end of 2005, in all cases where a host galaxy was identified this was
of an early type, usually a blue star-forming galaxy
\cite{bloo05_md05,savaglio05_md05,sta06_host}. This was also the case for the 
thirty-some cases measured by Beppo-SAX (e.g. \cite{jvp00}) in the previous seven 
years. More recent observational studies have indicated also that the  long GRB 
host galaxy metallicity is generally lower than that of the average massive star 
forming galaxies \cite{lefloch03_host,lefloch05_md05,sta06_host}. This has 
implications for the expected redshift distribution of GRB \cite{nata05_grbdist}
(c.f. \cite{wi98}), indicating that $\sim 40\%$ of long GRB may be at $z\simg 4$. 
Long GRB may, in principle, be detectable up to $z\siml 25-30$ 
\cite{lamrei00,cialoe00,gou03}.

The preponderance of short-lived massive star formation in such young galaxies, 
as well as the identification of a SN Ib/c light curve peaking weeks after the 
burst in few cases, has provided strong 
support for a massive stellar collapse origin  of long GRBs, as argued by 
\cite{woo93,pac98,woo05_md05}.  The relatively long duration of the gamma-ray 
emission stage in these bursts ($2\s \siml t_\gamma \siml 10^3\s$) is  generally 
ascribed to a correspondingly long duration for the accretion of the debris 
\cite{macwoo99,pop99} falling into the central black hole which must form as the 
core of a massive star collapses. (For initial stellar masses in excess of about 
$28-30\msun$, the core is expected to collapse to a BH, e.g. \cite{fryer01}, while 
for smaller initial masses $10\msun \siml M_\ast \siml 28 \msun$ the collapsing 
core mass is below the Chandrasekhar mass and is expected to lead to a neutron star).
This accretion onto the black hole feeds a relativistic jet, which breaks through
the infalling core and the stellar envelope along the direction of the rotation 
axis. 

A related massive core collapse mechanism has been considered by \cite{van02,van03}
taking into account MHD effects in the disk and BH, in which 
the basic accretion time is short enough to be identified with short bursts, 
but magnetic tension can result in suspended accretion leading to long bursts.  
A mechanism based on the shorting of a charge separation built up around newly 
formed black holes has been discussed by \cite{ruffini01,ruffini03}.
Other mechanism invoked include collapse of a neutron star to a strange star
(e.g. \cite{cd96,bbdfl03,drago05}. The most widely adopted scenario,
from this list, is the first one, in which the long GRB derive their energy 
from either the gravitational energy liberated by the torus of debris 
accreting onto the central BH formed by the massive core collapse, or by 
the extraction of the rotational energy of the BH, mediated by the presence 
of the debris torus, whose accretion lifetime in both cases is identified 
with the duration of the "prompt" gamma-ray emitting phase of the burst.


\subsection{Supernova connection} 
\label{sec:sn}

In the year following the launch of Swift no supernovae were identified
in association with GRB. In fact, there are some upper limits on possible
supernovae, the most notable ones being on the short bursts 
GRB 050509b \cite{hjorth05_050509b} and GRB 050709 \cite{fox05_0709}. 
However, from the previous eight year period
there are two well documented cases of supernovae  associated with long 
bursts, and several more weaker cases, were the evidence suggests a 
long GRB-SN connection. The first evidence for a long GRB-supernova 
association was discovered in GRB 980425, which appeared associated 
with SN 1998bw \cite{gal98_sn,kul98b_sn}. 
This was a peculiar, more energetic than usual Type Ib/c supernova, where
the apparently associated GRBs properties seemed the same as usual, except 
for the redshift being extremely small ($z\sim 0.0085$). This implied the 
lowest ever long GRB isotropic equivalent energy $E_\gamma\sim 10^{48}$ erg, 
which resulted in the association being treated cautiously. However, using 
SN 1998bw as a template, other possible associations were soon claimed 
through detection of reddened bumps in the optical afterglow light-curves after
a time delay compatible with a supernova brightness rise-time, e.g.
in GRB 980326\cite{bloom99}, GRB 970228\cite{rei99,galama00}, 
GRB 000911\cite{laz01}, GRB 991208\cite{castro01}, GRB 990712\cite{sahu00}, 
GRB 011121\cite{bloom02a}, and GRB 020405\cite{price03}. 

The first unambiguous GRB-SN association was identified in GRB 030329,
at a redshift $z=0.169$, through both a supernova light-curve reddened 
bump and, more convincingly, by measuring in it a supernova spectrum of
type Ib/c (i.e. the same type as in 1998bw) \cite{sta03,hjorth03}. 
As a corollary, this observation rules out the ``supra-nova" model\cite{vieste98},
in which a core collapse to neutron star and a supernova was assumed to 
occur months before a second collapse of the NS into a BH and a GRB;
the delay between GRB 030329 and SN 2003dh is less than two days,
and is compatible with both events being simultaneous\cite{hjorth03}.
For pre-Swift GRB-SN associations, see, e.g. \cite{zehkk06,zehkh04}.

More recently, Swift observed with all three instruments, BAT, XRT and UVOT, an 
unusually long ($\sim 2000$ s), soft burst, GRB 060218 \cite{camp06_060218}, which 
was found to be associated with SN2006aj, a very nearby ($z=0.033$) type Ic supernova 
\cite{mas06,pia06,modjaz06_060218,mirab06_2006aj,sol06_2006aj,cob06_2006aj}.
This supernova light curve peaked earlier than most known supernovae, and its
time origin can also be constrained to be within less than a day from the GRB 
trigger. This is the first time that a connected GRB and supernova event has been
observed starting in the first $\sim 100$ s in X-rays and UV/Optical light, and
the results are of great interest. The early X-ray light curve shows a slow rise
and plateau followed by a drop  after $\sim 10^3$ s, with a power law spectrum 
and an increasing black-body like component which dominates at the end. The 
most interesting interpretation involves shock break-out of a semi-relativistic
component in a WR progenitor wind \cite{camp06_060218} 
(c.f. \cite{fanpir_060218,tanmm01}). 
After this a more conventional X-ray power law decay follows, and a UV component 
peak at a later time can be interpreted as due to the slower supernova envelope 
shock. Another GRB/SN detection based on Swift afterglow observations is that
associated with GRB 050525A \cite{dellavalle06}.

\subsection{Jet dynamics, cocoons and progenitors}
\label{sec:jetdyn}

For both long and short bursts, the most widely discussed central engine 
invokes a central black hole and a surrounding torus, either produced
by a massive stellar core collapse  (long bursts) or the merger of NS-NS 
or NS-BH binaries (short bursts). The latter mechanism is observationally
on a less firm footing than the first, and in both collapse and merger cases 
the black hole could be preceded by a temporary massive, highly magnetized
neutron star. There are two ultimate energy sources: the gravitational binding 
energy of the torus and the spin energy of the black hole. A possible third
is the magnetic energy stored during the collapse, which derives its energy 
from the other two. Two main ways have been discussed for extracting the 
accretion energy and black hole spin energy, namely a neutrino-driven wind 
\cite{eic89,ruffert97,ruffert98,macwoo99,fm03,leerrp05},
and the Blandford-Znajek\cite{bz77} mechanism. Both mechanisms lead to 
an optically thick $e^\pm$ jet or fireball, but the second is dynamically 
Poynting-dominated, i.e. dominated by strong magnetic fields threading 
the black hole\cite{mr97b,lwb00,van01a,li00}. Needless to say, identification 
of the content of the fireball and the mechanism of GRB 
prompt emission  would shed light on the mechanism that
powers the central engine. Hence the excitement following claims of a
very large gamma-ray polarization in GRB 021206 \cite{coburn03} suggesting a 
strongly magnetized central engine. This observation has been challenged
\cite{rutfox04,wigger04}. A strong gamma-ray polarization could in principle 
be expected from a pure Poynting-flux dominated jet \cite{lyut03}, or in a 
baryonic hydrodynamic jet with a globally organized magnetic field 
configuration\cite{waxman03,gran03,grankon03}.  A strong but less extreme 
magnetization of the jet is inferred from a combined reverse-forward shock 
emission analysis of GRB 990123 \cite{zhakm03,fan02}. 

In all models, an $e^\pm ,\gamma$ fireball is expected as a result of the 
dissipation associated with the transient core collapse or merger event.
The initial chaotic motions and shears also are expected to lead to build
up significant magnetic stresses \cite{tho94}. A combination of the
relativistic lepton (e$^\pm$) and MHD fields up to $\sim 10^{15}$ Gauss
can provide the driving stresses leading to a highly relativistic expansion
with $\Gamma_j \gg 1$. The fireball is very likely also to involve some 
fraction of baryons, and uncertainties in this ``baryon pollution" \cite{pac90} 
remain difficult to quantify until 3D MHD calculations capable of addressing 
baryon entrainment become available.  If the progenitor is a massive star, 
the expectation is that the fireball will likely be substantially collimated,
due to the transverse containing pressure of the stellar envelope, which, 
if fast-rotating, provides a natural fireball escape route along the 
centrifugally lightened rotation axis. 

The development of a jet and its Lorentz factor in a collapsar has been
discussed analytically in \cite{mr01,waxmes03,mat03,lazbeg05a}.
The essence of the dynamics of the jet in a burst from a massive star is that 
as long as the central BH accretes,  it injects along the rotation axis a 
relativistic jet, whose dimensionless entropy must be comparable to or larger 
than the final bulk Lorentz factor of the jet once it has emerged from the star, 
$\eta= (L/{\dot M}c^2)\simg \Gamma_j \simg 100$.  Even though such a jet is 
highly relativistic as it is injected, the overburden of the stellar core and 
envelope slow the jet head down to a sub-relativistic speed of advance, which 
gradually increases as the jet moves down the density gradient of the star. 
The difference between the injection and advance speed causes gas and energy 
spill-over into a transrelativistic cocoon of waste heat \cite{mr01,mat03,lazbeg05a}
surrounding the jet, which may be detectable \cite{ramcr02,peer06}.
By the time it reaches the boundary of the He core ($R_{He}\sim 10^{11}$ cm)
the jet head has reached a speed $v_j\sim c$. This takes, in the star's 
frame, $\sim 10$ s, hence the central engine must continue injecting 
energy and momentum into the  jet for at least this long. A very sharp drop 
in density is predicted by stellar models at this radius, beyond which a 
tenuous hydrogen envelope extends as a power law. In going down this
sharp gradient, the jet head Lorentz factor shoots up to a value comparable
to its final value, $\Gamma_j\simg 100$ (\cite{mr01,waxmes03}).Once the jet head
is relativistic, it becomes ballistic, and it is no longer affected by
whether the central engine energy injection continues or not. A constraint
on the mass of the envelope is that the mass overburden within the jet
solid angle must be less than the jet total energy divided by $\Gamma_j c^2$
(\cite{mat03}). If the star has lost is H envelope, this condition is guaranteed,
e.g. as in Wolf-Rayet type stars, where a stellar wind phase leads to envelope
loss previous to the core collapse phase. WR stars are, in fact, thought to be 
the progenitors of type Ib/c supernovae, which is the only type so far
seen in a few cases associated with GRB. A modest envelope, however, should
still be compatible with a high Lorentz factor, which could be tested through
detection of weak H lines in a GRB associated supernova (and may also be 
tested through TeV neutrino observations, \cite{razmw03_sn}).

The 2D development of a relativistic jet making its way out through a star 
have been calculated numerically by, e.g. \cite{aloy00,zhawm03}, while 
magnetically dominated jets are discussed by \cite{whee00,drenk02,proga05}.
Jets in compact mergers have calculated numerically by \cite{janka05,aloy05}.
The relativistic numerical calculations of GRB jets are, so far, mainly 
hydrodynamic, and involve approximations about the energy and momentum
injection at the lower boundary, the numerical difficulties in covering 
the entire dynamical range being extreme. The results \cite{zhawm03}
show that a jet of $\Gamma_j\sim 100$ can escape a star of radius
comparable to a WR ($R_\ast \sim 10^{11}$ cm). The angular structure
of the jet is, as expected, one where the Lorentz factor and energy per
solid angle tapers off towards the edges, where instabilities cause
mixing with and drag by the stellar envelope walls. An analytical argument
\cite{lazbeg05a} shows that this tapering off can result in an energy
profile $E_j(\theta)\propto \theta^{-2}$. Such a jet profile is a possible
interpretation \cite{rossi02,zhames02b} of the observational correlation 
between the isotropic equivalent jet energy and the jet break time derived 
from a sample of burst afterglows \cite{fra01,pankum01}.

\section{Very High Energy Photons and Non-Electromagnetic Emission}
\label{sec:vhe}

The highly relativistic nature of the outflows is inferred from and constrained 
by the observations of GeV photons, which indicate the need for bulk Lorentz 
factors of $\Gamma\simg 10^2$ \cite{fen93a,hb94,bh97}.  Such Lorentz factors result 
in synchrotron spectra which in the observer frame extend beyond 100 MeV, and 
inverse Compton (IC) scattering of such synchrotron photons leads to the 
expectation of GeV and TeV spectral components \cite{mrp94}. While $\siml 18$ GeV
photons have been observed (e.g. \cite{hur94}), TeV photons are likely to be 
degraded to lower energies by $\gamma\gamma$ pair production, either in the 
source itself, or (unless the GRB is at very low redshifts) in the intervening 
intergalactic medium \cite{coppi97,dejste02}.

Besides emitting in the currently studied sub-GeV electromagnetic channels, 
GRB are likely to be even more luminous in other channels, such as
neutrinos, gravitational waves and cosmic rays. For instance, nucleons
entrained in the fireball will have $\simg 100$ GeV bulk kinetic energies 
in the observer frame, which can lead to inelastic collisions resulting 
in pions, muons, neutrinos and electrons as well as their anti-particles. 
The main targets for the relativistic baryons are other particles in the 
relativistic outflow and particles in the external, slower moving environment. 
The expected flux and spectrum of 1--30~GeV neutrinos and $\gamma$-rays
resulting from pion decay due to interactions within the expanding plasma 
depends, e.g., on the neutron/proton ratio and on fireball inhomogeneities, 
while that due to interactions with the surrounding medium depends on the
external gas density and its distribution; and both depend on the Lorentz 
factor. Massive progenitors offer denser targets for nuclear collisions and a 
larger photon density for $p\gamma$ and $\gamma\gamma$ interactions, leading 
to modification of the photon spectra. On the other hand GRB from NS-NS mergers
would be characterized by neutron-rich outflows, leading to stronger 
5-10 GeV neutrinos and photons from $np$ collisions 
\cite{bahmes00,belob02,rossi05}. Photo-pion signatures of $\simg 100$  GeV 
photons and $10^{14}-10^{18}$ eV neutrinos may be expected to be relatively 
stronger in massive (high soft photon density) progenitors. Knowing what 
fraction of GRB, if any, arise from NS mergers is vital for facilitating 
interferometric gravitational wave detections, e.g. with LIGO. And,
conversely, detection with LIGO would provide important clues as to whether
short bursts are NS-NS (or NS-BH) mergers, or whether massive stellar collapses 
are asymmetric enough to produce substantial gravitational wave emission
and serve as a test of the relationship between long GRB and supernovae. 

The Fermi mechanism in shocks developing in the GRB outflow can also
accelerate protons to observer-frame energies up to $\sim 10^{20}$ eV 
\cite{wax95,wax04}. Internal shocks leading to the observed $\gamma$-rays
have a high comoving photon density and lead to $p\gamma$ photopion 
production and to $\simg 100$ TeV neutrinos \cite{waxbah97}. In external 
shocks due to deceleration by the external medium, the reverse shock moving 
into the ejecta can produce optical photons (\S \ref{sec:reverse}) which result 
in photopion production and $\simg 10^{19}$ eV neutrinos \cite{waxbah00}. 
Neutrinos in the TeV to EeV range may be easier to detect than those at 
$\sim10$~GeV energies, due to their higher interaction cross section, with
instruments currently under construction. Such neutrinos would serve as 
diagnostics of the presence of relativistic shocks, and as probes of the 
acceleration mechanism and the magnetic field strength. The flux and spectrum 
of $\simg 10^{19}$ eV neutrinos depends on the density of the surrounding gas, 
while the $\simg 10^{14}$ eV neutrinos depend on the fireball Lorentz factor. 
Hence, the detection of very high energy neutrinos would provide crucial 
constraints on the fireball parameters and GRB environment.

\subsection{UHE photons from GRB}
\label{sec:uhegam}

Ultra-high energy emission, in the range of GeV and harder, is
expected from electron inverse Compton in external shocks \cite{mrp94}
as well as from internal shocks \cite{pm96} in the prompt phase. The
combination of prompt MeV radiation from internal shocks and a more
prolonged GeV IC component for external shocks \cite{mr94} is a likely
explanation for the delayed GeV emission seen in some GRB \cite{hur94}.  
(An alternative invoking photomeson processes from ejecta protons impacting 
a nearby binary stellar companion is \cite{katz94b}).  The GeV photon 
emission from the long-term IC component in external afterglow shocks has been 
considered by \cite{dercm00,zhames01b,derishev01,wangdl01a,wangdl01b}. The IC 
GeV photon component is likely to be significantly more important \cite{zhames01b} 
than a possible proton synchrotron or electron synchrotron component at these
energies. Another possible contributor at these energies may be
$\pi^0$ decay from $p\gamma$ interactions between shock-accelerated
protons and MeV or other photons in the GRAB shock region
\cite{boetder98,tot99,fragile04}.  However, under the conservative
assumption that the relativistic proton energy does not exceed the
energy in relativistic electrons or in $\gamma$-rays, and that the
proton spectral index is -2.2 instead of -2, both the proton
synchrotron and the $p\gamma$ components can be shown to be
substantially less important at GeV-TeV than the IC component
\cite{zhames01b}.  Another GeV photon component is expected from the
fact that in a baryonic GRB outflow neutrons are likely to be present,
and when these decouple from the protons, before any shocks occur,
$pn$ inelastic collisions will lead to pions, including $\pi^0$,
resulting in UHE photons which cascade down to the GeV range
\cite{derishev99,bahmes00,rossi05}.  The final GeV spectrum results from a
complex cascade, but a rough estimate indicates that 1-10 GeV flux
should be detectable \cite{bahmes00} with GLAST \cite{glast_site} for 
bursts at $z\siml 0.1$.

In these models, due to the high photon densities implied by GRB
models, $\gamma\gamma$ absorption within the GRB emission region must
be taken into account \cite{baring00,lithwick01,razmz04_gev,pw04a,pw04b}.
One interesting result is that the observation of photons up to a certain energy, 
say 10-20 GeV with EGRET, puts a lower limit on the bulk Lorentz factor of
the outflow, from the fact that the compactness parameter (optical
depth to $\gamma\gamma$) is directly proportional to the comoving
photon density, and both this as well as the energy of the photons
depend on the bulk Lorentz factor. This has been used by \cite{lithwick01} 
to estimate lower limits on $\Gamma \siml 300-600$ for a number of specific 
bursts observed with EGRET.  On the other hand, for GRB with $\Gamma \simg 850$, 
TeV photons can escape the source \cite{razmz04_gev}.

Long GRB have recently been shown to be associated with supernovae
(\S \ref{sec:sn}). If GRB also accelerate cosmic rays, as suspected, then 
these could leave long-lasting UHE photon signatures in supernova remnants 
which were associated with GRB at the time of their explosion. One example 
may be the SN remnant W49B, which may be a GRB remnant. A signature of a 
neutron admixture in the relativistic cosmic ray outflow would be a TeV 
gamma-ray signature due to inverse Compton interactions following neutron 
decay \cite{iokkm04} (see also \cite{atoy05}). 
Continued magnetic outflows upscattering companion photons may also
signal GRB remnants \cite{ram04_uhe}.
The imaging of the surrounding emission could provide new constraints 
on the jet structure of the GRB.

The recent detection of delayed X-ray flares during the afterglow
phase of gamma-ray bursts (GRBs) with the Swift satellite (e.g. 
\cite{zhang06_ag,nousek06,pan06_ag}) suggests an inner-engine origin of 
these flares, at radii inside the deceleration radius characterizing the
beginning of the forward shock afterglow emission. Given the
observed temporal overlapping  between the flares and afterglows,
one expects an inverse Compton (IC) emission arising from such
flare photons scattered by forward shock afterglow electrons \cite{wanlm06}.
The jet may also IC upscatter shock break-out X-ray photons \cite{ramml02}.
This IC emission would produce  GeV-TeV flares, which may be detected by
GLAST and ground-based TeV telescopes. The detection of GeV-TeV flares
combined with low energy observations may help to constrain the poorly
known magnetic field in afterglow shocks.

At higher energies, a tentative $\simg 0.1$ TeV detection at the
$3\sigma$ level of GRB970417a has been reported with the water
Cherenkov detector Milagrito \cite{atk00}.  Another possible TeV
detection \cite{poirier03} of GRB971110 has been reported with the
GRAND array, at the $2.7\sigma$ level.  Stacking of data from the
TIBET array for a large number of GRB time windows has led to an
estimate of a $\sim 7\sigma$ composite detection significance
\cite{amenomori01}.  Better sensitivity is expected from the upgraded
larger version of MILAGRO, as well as from atmospheric Cherenkov telescopes 
under construction such as VERITAS, HESS, MAGIC and CANGAROO-III 
\cite{weekes00,ong05,horn06_hess,hold05_veritas,stam05_tev_rev,fer06_magic,kif04_cangaroo}.  
However, GRB detections in the TeV
range are expected only for rare nearby events, since at this energy
the mean free path against $\gamma\gamma$ absorption on the diffuse IR
photon background is $\sim$ few hundred Mpc \cite{coppi97,dejste02}.
The mean free path is much larger at GeV energies, and based on the
handful of GRB reported in this range with EGRET, several hundred
should be detectable with large area space-based detectors such as
GLAST \cite{glast04,zhames01b}.

\subsection{Cosmic rays from GRB}
\label{sec:crnu}

In the standard fireball shock model of the prompt $\gamma$-ray emission, 
say from internal shocks or magnetic dissipation, and also in the external 
afterglow shocks, the same acceleration mechanisms which lead to the 
non-thermal electron power laws implied by the observed photon spectra
must also lead to proton acceleration. Using the shock parameters
inferred from broad-band photon spectral fits, one infers that protons
can be accelerated to Lorentz factors up to $\siml 10^{11}$ in the
observer frame \cite{wax95,vie95}, i.e. to so-called GZK energy of
$E_p\sim 10^{20}$ eV. This is interesting mainly for ``baryonic" jets, 
where the bulk of the energy is carried by baryons, whereas in 
Poynting-dominated jets there would be much fewer protons to accelerate.  
Well below the GZK energy, protons interacting with the MeV photons 
present in GRB or with thermal nucleons are above the pion production 
threshold and can produce ultra-high energy neutrinos, as discussed below.
 
Discussions of GRB as cosmic ray sources are mainly oriented at exploring 
their contribution to the energy range above EeV ($10^{18}$ eV; e.g. 
\cite{wax04}), referred to as ultra-high energy cosmic rays, or UHECRs.
(A model where GRB are responsible for CRs ranging from PeV to GZK is 
\cite{wicder04}).  At EeV and higher energies the observed UHECR isotropy and the 
small expected magnetic deflection suggests an extra-galactic origin. The 
requirement that they are not attenuated by the cosmic microwave background through 
photomeson interactions constrains that they are originated within a volume inside 
a radius of 50-100 Mpc, the so-called ``GZK'' volume (e.g. \cite{cronin05}). 
Two broad classes of models suggested are the ``top-down'' scenarios, which 
attribute UHECR to decay of fossil Grand Unification defects, and the
``bottom-up'' scenarios, which assume UHECRs are accelerated in astrophysical 
sources. One of the most prominent candidate sources for bottom-up scenarios 
is GRBs \cite{wax95,vie95,mu95} (two others are AGNs, e.g. \cite{berezinsky05}
and cluster shocks, e.g. \cite{inoue05}). The most commonly discussed version of 
this scenario considers the UHECR to be protons accelerated in GRB internal 
shocks \cite{wax95,wax95b,wax04}, while another version attributes them to 
acceleration in external shocks \cite{vie95,vdg03,dermer05_md05}. 
(For UHECR acceleration in alternative GRB models, see, e.g. \cite{darder01,fargro05}).

The persuasiveness of this scenario is largely based on two coincidences, 
namely, the required condition to accelerate protons
to GZK energies is similar to the requirement for generating the prompt 
observed gamma-rays in GRB, and the observed UHECR energy injection rate
into the universe ($\sim 3\times 10^{44} ~{\rm erg~Mpc^{-3}~yr^{-1}}$) is 
similar to the local GRB $\gamma$-ray energy injection rate \cite{wax95,vie95}. 
These coincidences have been questioned, e.g. \cite{ste00,scuste00}, 
but these objections have been resolved using new data and further 
considerations \cite{wax04,vdg03}, and GRBs remain a promising candidate 
for UHECRs. However, there are some caveats of principle. The
internal shock scenario relies on the assumption that GRB prompt
gamma-ray emission is due to internal shocks. Although this is the leading
scenario, there is no strong proof so far, as is the case for the external 
shock (e.g., there are efficiency and spectrum issues, etc.). On the
other hand, a Poynting flux dominated GRB model would have to rely on
magnetic dissipation and reconnection, accelerating electrons and hence
also accelerating protons- but details remain to be investigated. The 
external shock model would have to rely on a magnetized medium \cite{vdg03} 
to reach the desired cosmic ray energy (as expected in pulsar wind bubbles 
\cite{kongran02} in the supranova scenario \cite{vieste98}, which however has become 
less likely since the almost simultaneous GRB 030329/SN 2003dh and the
more recent GRB 060218/SN2006aj association). 

Direct confirmation of a GRB orgin of UHECRs will be difficult. The next 
generation cosmic ray detectors such as the {\em Pierre Auger Observatory} 
\cite{auger_site} will have a substantially enhanced effective target area, 
which will greatly improve the cosmic ray count statistics. This will help 
to disentangle the two scenarios (top-down or bottom-up) and will reveal 
whether a GZK feature indeed exists. Within the bottom-up scenario, the 
directional information may either prove or significantly constrain the 
alternative AGN scenario, and may eventually shed light on whether GRBs are 
indeed the sources of UHECRs.

\subsection{UHE neutrinos contemporary with gamma-rays}
\label{sec:nuprompt}

Internal shocks in the GRB jet take place at a radius $r_i \sim 2
\Gamma_i^2 c \delta t \sim 5 \times 10^{12} \delta t_{-3}
\Gamma_{300}^2$ cm.  Here $\Gamma_i = 300 ~\Gamma_{300}$ is the bulk
Lorentz factor of the GRB fireball ejecta and $\delta t = 10^{-3}
\delta t_{-3}$ s is the variability time scale.  Observed
$\gamma$-rays are emitted from the GRB fireball when it becomes
optically thin at a radius $\simg r_i$.  Shock accelerated protons
interact dominantly with observed synchrotron photons with $\sim$MeV
peak energy in the fireball to produce a Delta resonance, $p\gamma
\rightarrow \Delta^+$ \cite{waxbah97}.  The
threshold condition to produce a $\Delta^+$ is $E_p E_{\gamma} = 0.2
\Gamma_i^2$ GeV$^2$ in the observer frame, which corresponds to a
proton energy of $E_p = 1.8 \times 10^{7} E_{\gamma, {\rm MeV}}^{-1}
\Gamma_{300}^{2}$ GeV. The subsequent decays $\Delta^+ \rightarrow n
\pi^+ \rightarrow n \mu^+ \nu_{\mu}
\rightarrow n e^+ \nu_e {\bar \nu}_{\mu} \nu_{\mu}$ produce high
energy neutrinos in the GRB fireball contemporaneous with
$\gamma$-rays \cite{waxbah97,rachmes98}.  Assuming that the secondary pions 
receive $20\%$ of the proton energy per interaction and each secondary lepton
shares 1/4 of the pion energy, each flavor of neutrino is emitted with
$5\%$ of the proton energy, dominantly in the PeV range.

\begin{figure}[ht]
\begin{center}
\centerline{\epsfxsize=4.in \epsfbox{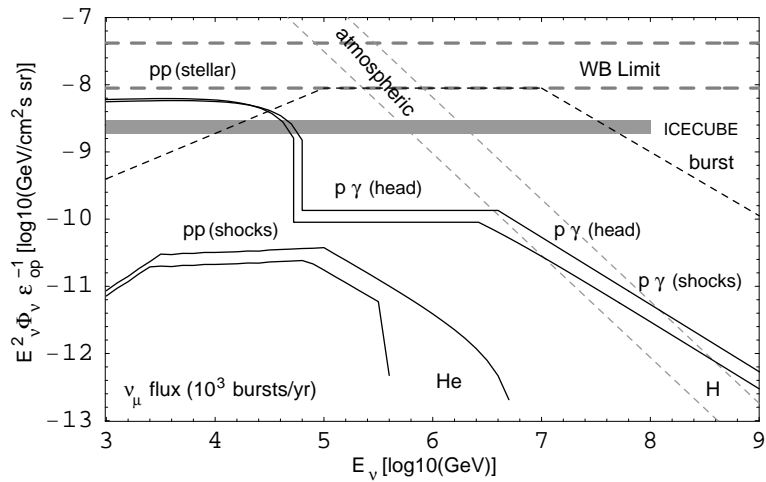}}
\end{center}
\caption{Diffuse muon neutrino flux arriving simultaneously with the 
$\gamma$-rays from shocks outside the stellar surface in observed GRB 
(dark short-dashed curve), compared to the Waxman-Bahcall (WB) diffuse 
cosmic ray bound (light long-dashed curves) and the atmospheric neutrino 
flux (light short-dashed curves). Also shown is the diffuse muon neutrino
precursor flux (solid lines) from sub-stellar jet shocks in two GRB progenitor 
models, with stellar radii $r_{12.5}$ (H) and $r_{11}$ (He). These neutrinos 
arrive 10-100 s before the $\gamma$-rays from electromagnetically detected 
bursts (with similar curves for $\nu_{\mu}$, $\nu_e$ and $\nu_{\tau}$) 
\cite{razmw03_tomo}.
\label{fig:nujet} }
\end{figure}

The diffuse muon neutrino flux from GRB internal shocks due to proton 
acceleration and subsequent photopion losses is shown as the short dashed 
line in Fig. \ref{fig:nujet}. The flux is compared to the Waxman-Bahcall 
limit of cosmic neutrinos, which is  derived from the observed cosmic ray 
flux \cite{waxbah00}.  The fluxes of all neutrino flavors are expected to 
be equal after oscillation in vacuum over astrophysical distances.

%

The GRB afterglow arises as the jet fireball ejecta runs into the
ambient inter-stellar medium (ISM), driving a blast wave ahead into it
and a reverse shock back into the GRB jet ejecta. This (external)
reverse shock takes place well beyond the internal shocks, at a radius
$r_e \sim 4\Gamma_e^2 c \Delta t \sim 2 \times 10^{17} \Gamma_{250}^2
\Delta t_{30}$ cm \cite{waxbah00}.  Here $\Gamma_{e} \approx 250
\Gamma_{250}$ is the bulk Lorentz factor of the ejecta after the
partial energy loss incurred in the internal shocks and $\Delta t = 30
\Delta t_{30}$ s is the duration of the GRB jet.  Neutrinos are
produced in the external reverse shock due to $p\gamma$ interactions
of internal shock accelerated protons predominantly with synchrotron
soft x-ray photons produced by the reverse shock.  The efficiency of
pion conversion from $p\gamma$ interactions in this afterglow scenario
is much smaller than in the internal shocks \cite{waxbah00}.

In the case of a massive star progenitor the jet may be expanding into
a wind, emitted by the progenitor prior to its collapse. In this case,
the density of the surrounding medium, at the external shock radius,
may be much higher than that typical ISM density of $n\simeq 1~{\rm
cm}^{-3}$. For a wind with mass loss rate of $10^{-5}M_\odot~{\rm
yr}^{-1}$ and velocity of $v_w=10^3~{\rm km/s}$, the wind density at
the typical external shock radius would be $\simeq 10^4~{\rm
cm}^{-3}$. The higher density implies a lower Lorenz factor of the
expanding plasma during the reverse shocks stage, and hence a larger
fraction of proton energy lost to pion production. Protons of energy
$E_p\simg 10^{18}$~eV lose all their energy to pion production in this
scenario \cite{waxbah00,vie98,dai00} producing EeV neutrinos.

\subsection{Precursor neutrinos}
\label{sec:nuprecursor}

As discussed before, the core collapse of massive stars are the most likely
candidates for long duration GRBs, which should lead to the formation of
a relativistic jet initially buried inside the star.  The jet burrows through
the stellar material, and may or may not break through the stellar
envelope\cite{meswax01}. Internal shocks in the jet, while it
is burrowing through the stellar interior, may produce high energy
neutrinos through proton-proton ($pp$) and photomeson ($p\gamma$)
interactions \cite{razmw03_tomo}.  High energy neutrinos are
produced through pion decays which are created both in $pp$ and
$p\gamma$ interactions.  The jets which successfully penetrate through
the stellar envelope result in GRBs ($\gamma$-ray bright bursts) and
the jets which choke inside the stars do not produce GRBs
($\gamma$-ray dark bursts).  However, in both cases high energy
neutrinos produced in the internal shocks are emitted through the
stellar envelope since they interact very weakly with matter.

High energy neutrinos from the relativistic buried jets are emitted as
precursors ($\sim$ 10-100 s prior) to the neutrinos emitted from the
GRB fireball in case of an electromagnetically observed burst. 
In the the case of a choked burst (electromagnetically undetectable) no 
direct detection of neutrinos from individual sources is possible.  
However the diffuse neutrino signal is boosted up in both scenarios.  
The diffuse neutrino flux from two progenitor star models are shown in
Fig. \ref{fig:nujet}, one for a blue super-giant (labeled H) of radius 
$R_\ast=3 \times 10^{12}$ cm and the other a Wof-Rayet type (labeled He)
of radius $R_\ast =10^{11}$ cm. The Waxman-Bahcall diffuse cosmic ray bound 
\cite{waxbah99}, the atmospheric flux and the IceCube sensitivity to 
diffuse flux are also plotted for comparison.  The neutrino component 
which is contemporaneous with the gamma-ray emission (i.e. which arrives
after the precursor) is shown as the dark dashed curve, and is plotted 
assuming that protons lose all their energy to pions in $p\gamma$ 
interactions in internal shocks.  

Most GRBs are located at cosmological distances (with redshift $z\sim
1$) and individual detection of them by km scale neutrino telescopes
may not be possible.  The diffuse $\nu$ flux is then dominated by a
few nearby bursts.  The likeliest prospect for UHE $\nu$ detection is
from these nearby GRBs in correlation with electromagnetic
detection. Detection of ultrahigh energy neutrinos which point back to
their sources may establish GRBs as the sources of GZK cosmic rays.

The detection of ultrahigh energy neutrinos by future experiments such as
ICECUBE \cite{icecube_site}, ANITA \cite{anita_site}, KM3NeT \cite{km3net_site},
and Auger \cite{auger_site} can provide useful information, such as particle 
acceleration, radiation mechanism and magnetic field, about the sources and 
their progenitors. High energy neutrino astrophysics is an imminent prospect, 
with Amanda already providing useful limits on the diffuse flux from GRB 
\cite{sta04_amanda,becker06_amanda} and with ICECUBE 
\cite{ahrens04,hulth06_icecube,halzen06} on its way. The detection of TeV and
higher energy neutrinos from GRB would be of great importance for
understanding these sources, as well as the acceleration mechanisms
involved. It could provide evidence for the hadronic vs. the MHD
composition of the jets, and if observed, could serve as an unabsorbed
probe of the highest redshift generation of star formation in the Universe.

\subsection{Gravitational waves}
\label{sec:gw}

The gamma-rays and the afterglows of GRB are thought to be produced 
at distances from the central engine where the plasma has become optically 
thin, $r \ge 10^{13}$ cm, which is much larger than the Schwarzschild
radius of a stellar mass black hole (or of a neutron star). Hence we have 
only very indirect information about the inner parts of the central engine
where the energy is generated.  However, in any stellar progenitor model of GRB
one expects that gravitational waves should be emitted from the immediate 
neighborhood of the central engine, and their observation should give
valuable information about its identity. Therefore, it is of interest to
study the gravitational wave emission from GRB associated with specific
progenitors. Another reason for doing this is that the present and
foreseeable sensitivity of gravitational wave  detectors is such that
for likely sources, including GRB, the detections would be
difficult, and for this reason, much effort has been devoted to the
development of data analysis techniques that can reach deep into the
detector noise.  A coincidence between a gravitational wave signal and a
gamma-ray signal would greatly enhance the statistical significance of the
detection of the gravitational wave signal \cite{finn00, kp93}.
It is therefore of interest to examine the gravitational wave signals
expected from various  specific GRB progenitors that have been recently
discussed, and based on  current astrophysical models, to consider the
range of rates and strains expected in each case, for comparison with
the LIGO sensitivity. A general reference is \cite{van05_book},
which also discusses GRB-related sources of gravitational waves.

Regardless of whether they are associated with GRBs, binary compact
object mergers (NS-NS, NS-BH, BH-BH, BH-WD, BH-Helium star etc.)
\cite{thorn87,phinney91,cutler93,kp93,ruffert97,janka99,fwhd99,kobmes02} and
stellar core-collapses\cite{rmr98,fryer02,davies02,kobmes02,van01b,van02,van03}
have been studied as potential gravitational wave (GW) sources. 
These events are also leading candidates for being GRB progenitors, and
a coincidence between a GW signal and a gamma-ray signal would greatly 
enhance the statistical significance of the former\cite{finn00}.
A binary coalescence process can be divided into three phases: in-spiral, 
merger, and ring-down\cite{fh98,kobmes02}. For collapsars, a rapidly rotating 
core could lead to development of a bar and to fragmentation instabilities 
which would produce similar GW signals as in the binary merger scenarios,
although a larger uncertainty is involved. The GW frequencies of
various phases cover the $10-10^3$ Hz band which is relevant for the
Laser Interferometer Gravitational-wave Observatory (LIGO) \cite{ligo_site} 
and other related detectors such as VIRGO \cite{virgo_site}, GEO600 
\cite{geo600_site} and TAMA300 \cite{tama300_site}. Because of the faint 
nature of the typical GW strain, only
nearby sources (e.g. within $\sim 200$ Mpc for NS-NS and NS-BH
mergers, and within $\sim 30$ Mpc for collapsars)\cite{kobmes02} have
strong enough signals to be detectable by LIGO-II. When event rates
are taken into account\cite{fwh99,bbr02}, order of magnitude estimates
indicate that after one-year operation of the  advanced LIGO, one 
event for the in-spiral chirp signal of the NS-NS or NS-BH merger, and 
probably one collapsar event (subject to uncertainties), would be
detected\cite{kobmes02}. Other binary merger scenarios such as BH-WD and
BH-Helium star mergers are unlikely to be detectable\cite{kobmes02}, and
they are also unfavored as sources of GRBs according to other
arguments\cite{npk01}.

A time-integrated GW luminosity of the order of a solar rest mass 
($\siml 10^{54}$ erg) is predicted from merging
NS-NS and NS-BH models \cite{kp93,ruffert98,nakamura00}, while the
luminosity from collapsar models is more model-dependent, but expected
to be lower (\cite{fryerwh01,muellergw01}; c.f. \cite{van01b}).
Specific estimates have been made of the GW strains from some of the
most widely discussed current GRB progenitor stellar systems \cite{kobmes02}.
The expected detection rates of gravitational wave events with LIGO from 
compact binary mergers, in coincidence with GRBs, has been estimated by 
\cite{finn00,finn03}.  If some fraction of GRBs are produced by DNS  or 
NS-BH mergers, the gravitational wave chirp signal of the in-spiral
phase should be detectable by the advanced LIGO within one year, associated
with the GRB electromagnetic signal. One also expects signals from
the black hole ring-down phase, as well as the possible contribution of a bar
configuration from gravitational instability in the accretion disk following
tidal disruption or infall in GRB scenarios.

\begin{figure}[ht]
\begin{center}
\begin{minipage}[t]{\figuresize}
\epsfxsize=\boxsize
\epsfbox{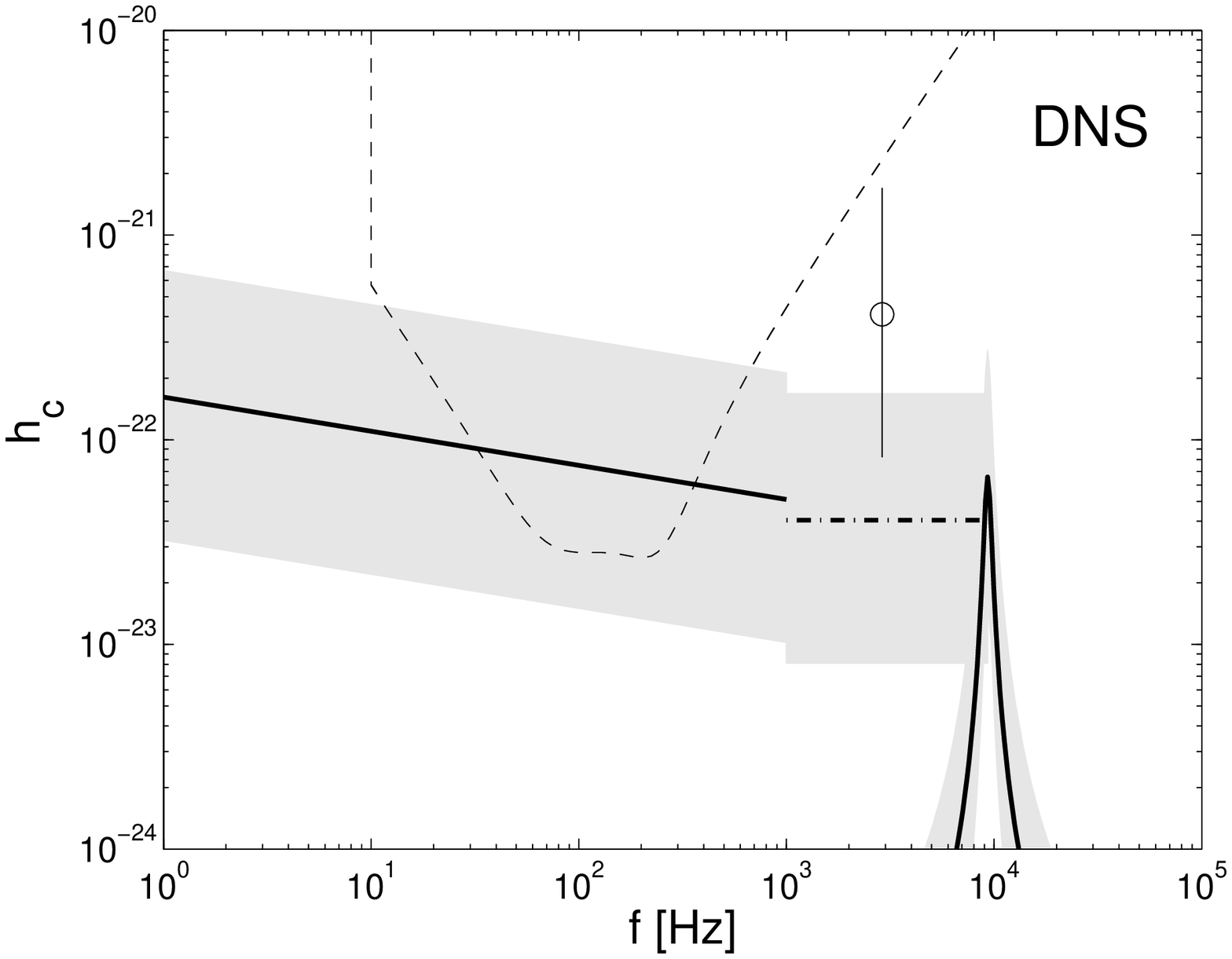}
\end{minipage}
\hspace{15mm}
\begin{minipage}[t]{\figuresize}
\epsfxsize=\boxsize
\epsfbox{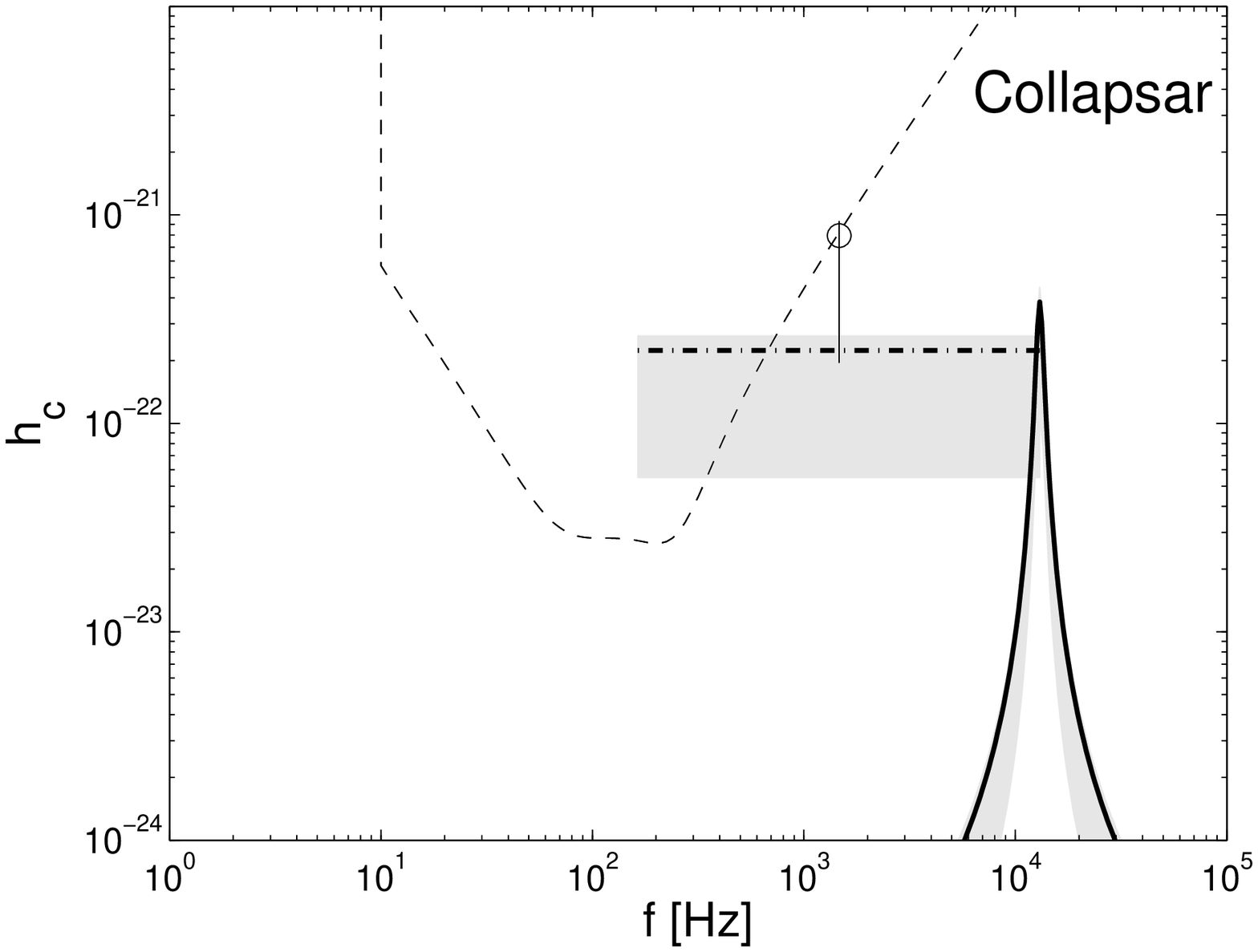}
\end{minipage}
\end{center}
\caption{Gravitational wave strain from a double neutron merger (left) and
a collapsar (right) compared to advanced LIGO sensitivity \cite{kobmes02}.}
\label{fig:gw}
\end{figure}

The most promising GW-GRB candidates in terms of detections per year are 
the DNS and BH-NS mergers \cite{kobmes02} (Fig. \ref{fig:gw}), based on 
assumed mean distances from the formation rates estimated by \cite{fwh99}. 
More recent rate estimates are in \cite{van05_book}, and rates incorporating 
new information relating to Swift short GRB  detections are in 
\cite{belcz06_sho,nakar06_sho}.
Other binary progenitor scenarios, such as black hole -- Helium star and black 
hole -- white dwarf merger GRB progenitors are unlikely to be detectable, due 
to the low estimates obtained for the maximum non-axisymmetrical perturbations.
%

For the massive rotating stellar collapse (collapsar) scenario of GRB, 
the non-axisymmetrical perturbations are very uncertain, but may be strong
\cite{davies02,fryer02,van02}, and the estimated formation rates are
much higher than for other progenitors \cite{fryer02,belcz02}, with
typically lower mean distances to the Earth. For such long GRB the rate
estimates must incorporate the beaming correction \cite{van05_book}.
This type of scenario is of special interest, since it has the most 
observational support from GRB afterglow observations. For collapsars, in 
the absence of detailed numerical 3D
calculations specifically aimed at GRB progenitors, estimates were made
\cite{kobmes02} of the strongest signals that might be expected in the case of
bar instabilities occurring in the accretion disk around the resulting
black hole, and in the maximal version of the recently proposed
fragmentation scenario of the infalling cores.
Although the waveforms of the gravitational waves produced in the
break-up, merger and/or bar instability phase of collapsars are not
known, a cross-correlation technique can be used making use of two
co-aligned detectors. Under these assumptions, collapsar GRB models
would be expected to be marginally detectable as gravitational wave
sources by the advanced LIGO within one year of observations.
Figure \ref{fig:gw} depicts the characteristic GW strains for the 
double neutrons star merger and  the collapsar model. 

Other calculations of massive stellar collapse GRB \cite{van02,van03}
take into account MHD effects in the disk and BH.
More general studies of massive stellar core collapse event gravitational
wave  emission are presented in \cite{fryer04_collgw}, considering both
core collapse SN and the progenitors of long GRB.

In the case of binaries the matched filtering technique can be used, while
for sources such as collapsars, where the wave forms are uncertain, the
simultaneous detection by two elements of a gravitational wave interferometer,
coupled with electromagnetic simultaneous detection, provides a possible
detection technique.  Specific detection estimates have been made 
\cite{kobmes02,van05_book} for both the compact binary scenarios 
and the collapsar scenarios,

Both the compact merger and the collapsar models have in common
a high angular rotation rate, and observations provide evidence
for jet collimation of the photon emission, with properties depending on 
the polar angle, which may also be of relevance for X-ray flashes. 
Calculations have been made \cite{kobmes03b} of the gravitational wave
emission and its polarization as a function of angle expected
from such sources.  The GRB progenitors emit $l=m=2$ gravitational
waves, which are circularly polarized on the polar axis, while the $+$
polarization dominates on the equatorial plane. Recent GRB studies suggest
that the wide variation in the apparent luminosity of GRBs are caused by
differences in the viewing angle, or possibly also in the jet opening angle.
Since GRB jets are launched along the polar axis of GRB progenitors,
correlations among the apparent luminosity of GRBs ($L_\gamma(\theta)\propto
\theta^{-2}$ and the amplitude as well as the degree of linear polarization $P$
degree of the gravitational waves are expected, $P\propto \theta^{4}\propto
L_\gamma^{-2}$.  At a viewing angle larger than the jet opening angle 
$\theta_j$ the GRB $\gamma$-ray emission may not be detected. However, in such 
cases an ``orphan'' (see, e.g. \cite{hudec04_orphan,zou06_orphan,rau06_orphan}) 
long-wavelength afterglow could be observed, which would be preceded by a 
pulse of gravitational waves with a significant linearly polarized component.
As the jet slows down and reaches $\gamma \sim \theta_j^{-1}$, the jet begins
to expand laterally, and its electromagnetic radiation begins to be observable
over increasingly wider viewing angles. Since the opening angle increases as
$\sim\gamma^{-1} \propto t^{1/2}$,  at a viewing angle $\theta > \theta_j$,
the orphan afterglow begins to be observed (or peaks) at a time $t_{p}
\propto \theta^2$ after the detection of the gravitational wave burst.
The polarization degree and the peak time should be correlated as
$P\propto t_p^2$.

Gravitational wave burst searches are underway with LIGO. The results from the
third science run \cite{abbot06_ligo} searched for sub-second bursts in the 
frequency range 100-1100 Hz for which no waveform model is assumed, with a 
sensitivity in terms of the root-sum-square (rss) strain amplitude of 
$h_{rss} \sim 10^{-20}$ Hz$^{-1/2}$. No gravitational-wave signals were detected 
in the eight days of analyzed data for this run. The search continues, as
LIGO continues to be upgraded towards it ultimate target sensitivity.

\bigskip\bigskip
\noindent
{\it Acknowledgements:} Research partially supported through NSF AST 0307376 and
NASA NAG5-13286. I am grateful for useful comments from two referees, as well as
from P. Kumar, N. Gehrels, L. Gou, E. Ramirez-Ruiz, S. Razzaque, M.J. Rees, 
X.Y. Wang and B. Zhang.

\end{document}